\documentclass[longauth]{aa}

\usepackage{graphicx}
\usepackage{natbib}
\usepackage{scalerel}
\usepackage{float}

\usepackage[table]{xcolor}

\usepackage{txfonts}
\usepackage[pdfencoding=auto,psdextra]{hyperref}
\hypersetup{
    colorlinks=true,
    linkcolor=blue,
    filecolor=magenta,      
    urlcolor=blue,
    citecolor=blue
}
\usepackage[nameinlink,capitalise]{cleveref}
\crefname{section}{Sect.}{Sects.}
\Crefname{section}{Section}{Sections}
\crefname{figure}{Fig.}{Figs.}
\Crefname{figure}{Figure}{Figures}
\crefname{equation}{Eq.}{Eqs.}
\Crefname{equation}{Equation}{Equations}
\crefname{table}{Table}{Tables}
\crefname{appendix}{Appendix}{Appendices}

\makeatletter
\renewcommand*\aa@pageof{, page \thepage{} of \pageref*{LastPage}}
\makeatother

\usepackage[utf8]{inputenc}

\usepackage[switch, modulo]{lineno}
\usepackage{euclid}

\usepackage{orcidlink}
\newcommand{\orcid}[1]{\orcidlink{#1}} 

\begin{document}
%
% Put the title of your paper here:
%
\title{Euclid Quick Data Release (Q1)} \subtitle{Optical and near-infrared identification and classification of \\point-like X-ray selected sources}

%%%% Version Wednesday 27th of August 2025 02:50:28 PM UT												
%%%% Please do not edit the author list -- contact ECEB Bureau for changes
%% if already defined in aa.cls: comment, or use renewcommand	
\author{Euclid Collaboration: W.~Roster\orcid{0000-0002-9149-6528}\thanks{\email{wroster@mpe.mpg.de}}\inst{\ref{aff1}}
\and M.~Salvato\orcid{0000-0001-7116-9303}\inst{\ref{aff1}}
\and J.~Buchner\orcid{0000-0003-0426-6634}\inst{\ref{aff1}}
\and R.~Shirley\orcid{0000-0002-1114-0135}\inst{\ref{aff1}}
\and E.~Lusso\orcid{0000-0003-0083-1157}\inst{\ref{aff2},\ref{aff3}}
\and H.~Landt\orcid{0000-0001-8391-6900}\inst{\ref{aff4}}
\and G.~Zamorani\orcid{0000-0002-2318-301X}\inst{\ref{aff5}}
\and M.~Siudek\orcid{0000-0002-2949-2155}\inst{\ref{aff6},\ref{aff7}}
\and B.~Laloux\orcid{0000-0001-9996-9732}\inst{\ref{aff8},\ref{aff1}}
\and T.~Matamoro~Zatarain\orcid{0009-0007-2976-293X}\inst{\ref{aff9}}
\and F.~Ricci\orcid{0000-0001-5742-5980}\inst{\ref{aff10},\ref{aff11}}
\and S.~Fotopoulou\orcid{0000-0002-9686-254X}\inst{\ref{aff9}}
\and A.~Ferr\'e-Mateu\orcid{0000-0002-6411-220X}\inst{\ref{aff12},\ref{aff13}}
\and X.~Lopez~Lopez\orcid{0009-0008-5194-5908}\inst{\ref{aff14},\ref{aff5}}
\and N.~Aghanim\orcid{0000-0002-6688-8992}\inst{\ref{aff15}}
\and B.~Altieri\orcid{0000-0003-3936-0284}\inst{\ref{aff16}}
\and A.~Amara\inst{\ref{aff17}}
\and S.~Andreon\orcid{0000-0002-2041-8784}\inst{\ref{aff18}}
\and N.~Auricchio\orcid{0000-0003-4444-8651}\inst{\ref{aff5}}
\and H.~Aussel\orcid{0000-0002-1371-5705}\inst{\ref{aff19}}
\and C.~Baccigalupi\orcid{0000-0002-8211-1630}\inst{\ref{aff20},\ref{aff21},\ref{aff22},\ref{aff23}}
\and M.~Baldi\orcid{0000-0003-4145-1943}\inst{\ref{aff24},\ref{aff5},\ref{aff25}}
\and A.~Balestra\orcid{0000-0002-6967-261X}\inst{\ref{aff26}}
\and S.~Bardelli\orcid{0000-0002-8900-0298}\inst{\ref{aff5}}
\and P.~Battaglia\orcid{0000-0002-7337-5909}\inst{\ref{aff5}}
\and A.~Biviano\orcid{0000-0002-0857-0732}\inst{\ref{aff21},\ref{aff20}}
\and A.~Bonchi\orcid{0000-0002-2667-5482}\inst{\ref{aff27}}
\and E.~Branchini\orcid{0000-0002-0808-6908}\inst{\ref{aff28},\ref{aff29},\ref{aff18}}
\and M.~Brescia\orcid{0000-0001-9506-5680}\inst{\ref{aff30},\ref{aff8}}
\and J.~Brinchmann\orcid{0000-0003-4359-8797}\inst{\ref{aff31},\ref{aff32}}
\and S.~Camera\orcid{0000-0003-3399-3574}\inst{\ref{aff33},\ref{aff34},\ref{aff35}}
\and G.~Ca\~nas-Herrera\orcid{0000-0003-2796-2149}\inst{\ref{aff36},\ref{aff37},\ref{aff38}}
\and V.~Capobianco\orcid{0000-0002-3309-7692}\inst{\ref{aff35}}
\and C.~Carbone\orcid{0000-0003-0125-3563}\inst{\ref{aff39}}
\and J.~Carretero\orcid{0000-0002-3130-0204}\inst{\ref{aff40},\ref{aff41}}
\and S.~Casas\orcid{0000-0002-4751-5138}\inst{\ref{aff42}}
\and M.~Castellano\orcid{0000-0001-9875-8263}\inst{\ref{aff11}}
\and G.~Castignani\orcid{0000-0001-6831-0687}\inst{\ref{aff5}}
\and S.~Cavuoti\orcid{0000-0002-3787-4196}\inst{\ref{aff8},\ref{aff43}}
\and K.~C.~Chambers\orcid{0000-0001-6965-7789}\inst{\ref{aff44}}
\and A.~Cimatti\inst{\ref{aff45}}
\and C.~Colodro-Conde\inst{\ref{aff12}}
\and G.~Congedo\orcid{0000-0003-2508-0046}\inst{\ref{aff46}}
\and C.~J.~Conselice\orcid{0000-0003-1949-7638}\inst{\ref{aff47}}
\and L.~Conversi\orcid{0000-0002-6710-8476}\inst{\ref{aff48},\ref{aff16}}
\and Y.~Copin\orcid{0000-0002-5317-7518}\inst{\ref{aff49}}
\and F.~Courbin\orcid{0000-0003-0758-6510}\inst{\ref{aff50},\ref{aff51}}
\and H.~M.~Courtois\orcid{0000-0003-0509-1776}\inst{\ref{aff52}}
\and M.~Cropper\orcid{0000-0003-4571-9468}\inst{\ref{aff53}}
\and A.~Da~Silva\orcid{0000-0002-6385-1609}\inst{\ref{aff54},\ref{aff55}}
\and H.~Degaudenzi\orcid{0000-0002-5887-6799}\inst{\ref{aff56}}
\and G.~De~Lucia\orcid{0000-0002-6220-9104}\inst{\ref{aff21}}
\and A.~M.~Di~Giorgio\orcid{0000-0002-4767-2360}\inst{\ref{aff57}}
\and C.~Dolding\orcid{0009-0003-7199-6108}\inst{\ref{aff53}}
\and H.~Dole\orcid{0000-0002-9767-3839}\inst{\ref{aff15}}
\and F.~Dubath\orcid{0000-0002-6533-2810}\inst{\ref{aff56}}
\and C.~A.~J.~Duncan\orcid{0009-0003-3573-0791}\inst{\ref{aff47}}
\and X.~Dupac\inst{\ref{aff16}}
\and S.~Dusini\orcid{0000-0002-1128-0664}\inst{\ref{aff58}}
\and S.~Escoffier\orcid{0000-0002-2847-7498}\inst{\ref{aff59}}
\and M.~Fabricius\orcid{0000-0002-7025-6058}\inst{\ref{aff1},\ref{aff60}}
\and M.~Farina\orcid{0000-0002-3089-7846}\inst{\ref{aff57}}
\and R.~Farinelli\inst{\ref{aff5}}
\and F.~Faustini\orcid{0000-0001-6274-5145}\inst{\ref{aff27},\ref{aff11}}
\and S.~Ferriol\inst{\ref{aff49}}
\and F.~Finelli\orcid{0000-0002-6694-3269}\inst{\ref{aff5},\ref{aff61}}
\and P.~Fosalba\orcid{0000-0002-1510-5214}\inst{\ref{aff62},\ref{aff7}}
\and N.~Fourmanoit\orcid{0009-0005-6816-6925}\inst{\ref{aff59}}
\and M.~Frailis\orcid{0000-0002-7400-2135}\inst{\ref{aff21}}
\and E.~Franceschi\orcid{0000-0002-0585-6591}\inst{\ref{aff5}}
\and S.~Galeotta\orcid{0000-0002-3748-5115}\inst{\ref{aff21}}
\and K.~George\orcid{0000-0002-1734-8455}\inst{\ref{aff60}}
\and B.~Gillis\orcid{0000-0002-4478-1270}\inst{\ref{aff46}}
\and C.~Giocoli\orcid{0000-0002-9590-7961}\inst{\ref{aff5},\ref{aff25}}
\and J.~Gracia-Carpio\inst{\ref{aff1}}
\and B.~R.~Granett\orcid{0000-0003-2694-9284}\inst{\ref{aff18}}
\and A.~Grazian\orcid{0000-0002-5688-0663}\inst{\ref{aff26}}
\and F.~Grupp\inst{\ref{aff1},\ref{aff60}}
\and S.~Gwyn\orcid{0000-0001-8221-8406}\inst{\ref{aff63}}
\and S.~V.~H.~Haugan\orcid{0000-0001-9648-7260}\inst{\ref{aff64}}
\and W.~Holmes\inst{\ref{aff65}}
\and I.~M.~Hook\orcid{0000-0002-2960-978X}\inst{\ref{aff66}}
\and F.~Hormuth\inst{\ref{aff67}}
\and A.~Hornstrup\orcid{0000-0002-3363-0936}\inst{\ref{aff68},\ref{aff69}}
\and P.~Hudelot\inst{\ref{aff70}}
\and K.~Jahnke\orcid{0000-0003-3804-2137}\inst{\ref{aff71}}
\and M.~Jhabvala\inst{\ref{aff72}}
\and E.~Keih\"anen\orcid{0000-0003-1804-7715}\inst{\ref{aff73}}
\and S.~Kermiche\orcid{0000-0002-0302-5735}\inst{\ref{aff59}}
\and A.~Kiessling\orcid{0000-0002-2590-1273}\inst{\ref{aff65}}
\and B.~Kubik\orcid{0009-0006-5823-4880}\inst{\ref{aff49}}
\and M.~K\"ummel\orcid{0000-0003-2791-2117}\inst{\ref{aff60}}
\and M.~Kunz\orcid{0000-0002-3052-7394}\inst{\ref{aff74}}
\and H.~Kurki-Suonio\orcid{0000-0002-4618-3063}\inst{\ref{aff75},\ref{aff76}}
\and Q.~Le~Boulc'h\inst{\ref{aff77}}
\and A.~M.~C.~Le~Brun\orcid{0000-0002-0936-4594}\inst{\ref{aff78}}
\and D.~Le~Mignant\orcid{0000-0002-5339-5515}\inst{\ref{aff79}}
\and S.~Ligori\orcid{0000-0003-4172-4606}\inst{\ref{aff35}}
\and P.~B.~Lilje\orcid{0000-0003-4324-7794}\inst{\ref{aff64}}
\and V.~Lindholm\orcid{0000-0003-2317-5471}\inst{\ref{aff75},\ref{aff76}}
\and I.~Lloro\orcid{0000-0001-5966-1434}\inst{\ref{aff80}}
\and G.~Mainetti\orcid{0000-0003-2384-2377}\inst{\ref{aff77}}
\and D.~Maino\inst{\ref{aff81},\ref{aff39},\ref{aff82}}
\and E.~Maiorano\orcid{0000-0003-2593-4355}\inst{\ref{aff5}}
\and O.~Mansutti\orcid{0000-0001-5758-4658}\inst{\ref{aff21}}
\and S.~Marcin\inst{\ref{aff83}}
\and O.~Marggraf\orcid{0000-0001-7242-3852}\inst{\ref{aff84}}
\and M.~Martinelli\orcid{0000-0002-6943-7732}\inst{\ref{aff11},\ref{aff85}}
\and N.~Martinet\orcid{0000-0003-2786-7790}\inst{\ref{aff79}}
\and F.~Marulli\orcid{0000-0002-8850-0303}\inst{\ref{aff14},\ref{aff5},\ref{aff25}}
\and R.~Massey\orcid{0000-0002-6085-3780}\inst{\ref{aff86}}
\and D.~C.~Masters\orcid{0000-0001-5382-6138}\inst{\ref{aff87}}
\and E.~Medinaceli\orcid{0000-0002-4040-7783}\inst{\ref{aff5}}
\and S.~Mei\orcid{0000-0002-2849-559X}\inst{\ref{aff88},\ref{aff89}}
\and M.~Melchior\inst{\ref{aff90}}
\and Y.~Mellier\inst{\ref{aff91},\ref{aff70}}
\and M.~Meneghetti\orcid{0000-0003-1225-7084}\inst{\ref{aff5},\ref{aff25}}
\and E.~Merlin\orcid{0000-0001-6870-8900}\inst{\ref{aff11}}
\and G.~Meylan\inst{\ref{aff92}}
\and A.~Mora\orcid{0000-0002-1922-8529}\inst{\ref{aff93}}
\and M.~Moresco\orcid{0000-0002-7616-7136}\inst{\ref{aff14},\ref{aff5}}
\and L.~Moscardini\orcid{0000-0002-3473-6716}\inst{\ref{aff14},\ref{aff5},\ref{aff25}}
\and R.~Nakajima\orcid{0009-0009-1213-7040}\inst{\ref{aff84}}
\and C.~Neissner\orcid{0000-0001-8524-4968}\inst{\ref{aff94},\ref{aff41}}
\and S.-M.~Niemi\inst{\ref{aff36}}
\and J.~W.~Nightingale\orcid{0000-0002-8987-7401}\inst{\ref{aff95}}
\and C.~Padilla\orcid{0000-0001-7951-0166}\inst{\ref{aff94}}
\and S.~Paltani\orcid{0000-0002-8108-9179}\inst{\ref{aff56}}
\and F.~Pasian\orcid{0000-0002-4869-3227}\inst{\ref{aff21}}
\and K.~Pedersen\inst{\ref{aff96}}
\and W.~J.~Percival\orcid{0000-0002-0644-5727}\inst{\ref{aff97},\ref{aff98},\ref{aff99}}
\and V.~Pettorino\inst{\ref{aff36}}
\and S.~Pires\orcid{0000-0002-0249-2104}\inst{\ref{aff19}}
\and G.~Polenta\orcid{0000-0003-4067-9196}\inst{\ref{aff27}}
\and M.~Poncet\inst{\ref{aff100}}
\and L.~A.~Popa\inst{\ref{aff101}}
\and L.~Pozzetti\orcid{0000-0001-7085-0412}\inst{\ref{aff5}}
\and F.~Raison\orcid{0000-0002-7819-6918}\inst{\ref{aff1}}
\and R.~Rebolo\orcid{0000-0003-3767-7085}\inst{\ref{aff12},\ref{aff102},\ref{aff13}}
\and A.~Renzi\orcid{0000-0001-9856-1970}\inst{\ref{aff103},\ref{aff58}}
\and J.~Rhodes\orcid{0000-0002-4485-8549}\inst{\ref{aff65}}
\and G.~Riccio\inst{\ref{aff8}}
\and E.~Romelli\orcid{0000-0003-3069-9222}\inst{\ref{aff21}}
\and M.~Roncarelli\orcid{0000-0001-9587-7822}\inst{\ref{aff5}}
\and R.~Saglia\orcid{0000-0003-0378-7032}\inst{\ref{aff60},\ref{aff1}}
\and Z.~Sakr\orcid{0000-0002-4823-3757}\inst{\ref{aff104},\ref{aff105},\ref{aff106}}
\and A.~G.~S\'anchez\orcid{0000-0003-1198-831X}\inst{\ref{aff1}}
\and D.~Sapone\orcid{0000-0001-7089-4503}\inst{\ref{aff107}}
\and B.~Sartoris\orcid{0000-0003-1337-5269}\inst{\ref{aff60},\ref{aff21}}
\and J.~A.~Schewtschenko\orcid{0000-0002-4913-6393}\inst{\ref{aff46}}
\and M.~Schirmer\orcid{0000-0003-2568-9994}\inst{\ref{aff71}}
\and P.~Schneider\orcid{0000-0001-8561-2679}\inst{\ref{aff84}}
\and T.~Schrabback\orcid{0000-0002-6987-7834}\inst{\ref{aff108}}
\and A.~Secroun\orcid{0000-0003-0505-3710}\inst{\ref{aff59}}
\and G.~Seidel\orcid{0000-0003-2907-353X}\inst{\ref{aff71}}
\and M.~Seiffert\orcid{0000-0002-7536-9393}\inst{\ref{aff65}}
\and S.~Serrano\orcid{0000-0002-0211-2861}\inst{\ref{aff62},\ref{aff109},\ref{aff7}}
\and P.~Simon\inst{\ref{aff84}}
\and C.~Sirignano\orcid{0000-0002-0995-7146}\inst{\ref{aff103},\ref{aff58}}
\and G.~Sirri\orcid{0000-0003-2626-2853}\inst{\ref{aff25}}
\and L.~Stanco\orcid{0000-0002-9706-5104}\inst{\ref{aff58}}
\and J.~Steinwagner\orcid{0000-0001-7443-1047}\inst{\ref{aff1}}
\and P.~Tallada-Cresp\'{i}\orcid{0000-0002-1336-8328}\inst{\ref{aff40},\ref{aff41}}
\and D.~Tavagnacco\orcid{0000-0001-7475-9894}\inst{\ref{aff21}}
\and A.~N.~Taylor\inst{\ref{aff46}}
\and I.~Tereno\inst{\ref{aff54},\ref{aff110}}
\and S.~Toft\orcid{0000-0003-3631-7176}\inst{\ref{aff111},\ref{aff112}}
\and R.~Toledo-Moreo\orcid{0000-0002-2997-4859}\inst{\ref{aff113}}
\and F.~Torradeflot\orcid{0000-0003-1160-1517}\inst{\ref{aff41},\ref{aff40}}
\and I.~Tutusaus\orcid{0000-0002-3199-0399}\inst{\ref{aff105}}
\and L.~Valenziano\orcid{0000-0002-1170-0104}\inst{\ref{aff5},\ref{aff61}}
\and J.~Valiviita\orcid{0000-0001-6225-3693}\inst{\ref{aff75},\ref{aff76}}
\and T.~Vassallo\orcid{0000-0001-6512-6358}\inst{\ref{aff60},\ref{aff21}}
\and G.~Verdoes~Kleijn\orcid{0000-0001-5803-2580}\inst{\ref{aff114}}
\and A.~Veropalumbo\orcid{0000-0003-2387-1194}\inst{\ref{aff18},\ref{aff29},\ref{aff28}}
\and Y.~Wang\orcid{0000-0002-4749-2984}\inst{\ref{aff87}}
\and J.~Weller\orcid{0000-0002-8282-2010}\inst{\ref{aff60},\ref{aff1}}
\and A.~Zacchei\orcid{0000-0003-0396-1192}\inst{\ref{aff21},\ref{aff20}}
\and F.~M.~Zerbi\inst{\ref{aff18}}
\and I.~A.~Zinchenko\orcid{0000-0002-2944-2449}\inst{\ref{aff60}}
\and E.~Zucca\orcid{0000-0002-5845-8132}\inst{\ref{aff5}}
\and V.~Allevato\orcid{0000-0001-7232-5152}\inst{\ref{aff8}}
\and M.~Ballardini\orcid{0000-0003-4481-3559}\inst{\ref{aff115},\ref{aff116},\ref{aff5}}
\and M.~Bolzonella\orcid{0000-0003-3278-4607}\inst{\ref{aff5}}
\and E.~Bozzo\orcid{0000-0002-8201-1525}\inst{\ref{aff56}}
\and C.~Burigana\orcid{0000-0002-3005-5796}\inst{\ref{aff117},\ref{aff61}}
\and R.~Cabanac\orcid{0000-0001-6679-2600}\inst{\ref{aff105}}
\and A.~Cappi\inst{\ref{aff5},\ref{aff118}}
\and D.~Di~Ferdinando\inst{\ref{aff25}}
\and J.~A.~Escartin~Vigo\inst{\ref{aff1}}
\and L.~Gabarra\orcid{0000-0002-8486-8856}\inst{\ref{aff119}}
\and M.~Huertas-Company\orcid{0000-0002-1416-8483}\inst{\ref{aff12},\ref{aff6},\ref{aff120},\ref{aff121}}
\and J.~Mart\'{i}n-Fleitas\orcid{0000-0002-8594-569X}\inst{\ref{aff93}}
\and S.~Matthew\orcid{0000-0001-8448-1697}\inst{\ref{aff46}}
\and N.~Mauri\orcid{0000-0001-8196-1548}\inst{\ref{aff45},\ref{aff25}}
\and R.~B.~Metcalf\orcid{0000-0003-3167-2574}\inst{\ref{aff14},\ref{aff5}}
\and A.~Pezzotta\orcid{0000-0003-0726-2268}\inst{\ref{aff122},\ref{aff1}}
\and M.~P\"ontinen\orcid{0000-0001-5442-2530}\inst{\ref{aff75}}
\and C.~Porciani\orcid{0000-0002-7797-2508}\inst{\ref{aff84}}
\and I.~Risso\orcid{0000-0003-2525-7761}\inst{\ref{aff123}}
\and V.~Scottez\inst{\ref{aff91},\ref{aff124}}
\and M.~Sereno\orcid{0000-0003-0302-0325}\inst{\ref{aff5},\ref{aff25}}
\and M.~Tenti\orcid{0000-0002-4254-5901}\inst{\ref{aff25}}
\and M.~Viel\orcid{0000-0002-2642-5707}\inst{\ref{aff20},\ref{aff21},\ref{aff23},\ref{aff22},\ref{aff125}}
\and M.~Wiesmann\orcid{0009-0000-8199-5860}\inst{\ref{aff64}}
\and Y.~Akrami\orcid{0000-0002-2407-7956}\inst{\ref{aff126},\ref{aff127}}
\and I.~T.~Andika\orcid{0000-0001-6102-9526}\inst{\ref{aff128},\ref{aff129}}
\and S.~Anselmi\orcid{0000-0002-3579-9583}\inst{\ref{aff58},\ref{aff103},\ref{aff130}}
\and M.~Archidiacono\orcid{0000-0003-4952-9012}\inst{\ref{aff81},\ref{aff82}}
\and F.~Atrio-Barandela\orcid{0000-0002-2130-2513}\inst{\ref{aff131}}
\and C.~Benoist\inst{\ref{aff118}}
\and K.~Benson\inst{\ref{aff53}}
\and D.~Bertacca\orcid{0000-0002-2490-7139}\inst{\ref{aff103},\ref{aff26},\ref{aff58}}
\and M.~Bethermin\orcid{0000-0002-3915-2015}\inst{\ref{aff132}}
\and L.~Bisigello\orcid{0000-0003-0492-4924}\inst{\ref{aff26}}
\and A.~Blanchard\orcid{0000-0001-8555-9003}\inst{\ref{aff105}}
\and L.~Blot\orcid{0000-0002-9622-7167}\inst{\ref{aff133},\ref{aff130}}
\and H.~B\"ohringer\orcid{0000-0001-8241-4204}\inst{\ref{aff1},\ref{aff134},\ref{aff135}}
\and M.~L.~Brown\orcid{0000-0002-0370-8077}\inst{\ref{aff47}}
\and S.~Bruton\orcid{0000-0002-6503-5218}\inst{\ref{aff136}}
\and A.~Calabro\orcid{0000-0003-2536-1614}\inst{\ref{aff11}}
\and F.~Caro\inst{\ref{aff11}}
\and C.~S.~Carvalho\inst{\ref{aff110}}
\and T.~Castro\orcid{0000-0002-6292-3228}\inst{\ref{aff21},\ref{aff22},\ref{aff20},\ref{aff125}}
\and F.~Cogato\orcid{0000-0003-4632-6113}\inst{\ref{aff14},\ref{aff5}}
\and A.~R.~Cooray\orcid{0000-0002-3892-0190}\inst{\ref{aff137}}
\and O.~Cucciati\orcid{0000-0002-9336-7551}\inst{\ref{aff5}}
\and S.~Davini\orcid{0000-0003-3269-1718}\inst{\ref{aff29}}
\and F.~De~Paolis\orcid{0000-0001-6460-7563}\inst{\ref{aff138},\ref{aff139},\ref{aff140}}
\and G.~Desprez\orcid{0000-0001-8325-1742}\inst{\ref{aff114}}
\and A.~D\'iaz-S\'anchez\orcid{0000-0003-0748-4768}\inst{\ref{aff141}}
\and J.~J.~Diaz\inst{\ref{aff6},\ref{aff12}}
\and S.~Di~Domizio\orcid{0000-0003-2863-5895}\inst{\ref{aff28},\ref{aff29}}
\and J.~M.~Diego\orcid{0000-0001-9065-3926}\inst{\ref{aff142}}
\and A.~Enia\orcid{0000-0002-0200-2857}\inst{\ref{aff24},\ref{aff5}}
\and Y.~Fang\inst{\ref{aff60}}
\and A.~G.~Ferrari\orcid{0009-0005-5266-4110}\inst{\ref{aff25}}
\and A.~Finoguenov\orcid{0000-0002-4606-5403}\inst{\ref{aff75}}
\and A.~Fontana\orcid{0000-0003-3820-2823}\inst{\ref{aff11}}
\and A.~Franco\orcid{0000-0002-4761-366X}\inst{\ref{aff139},\ref{aff138},\ref{aff140}}
\and K.~Ganga\orcid{0000-0001-8159-8208}\inst{\ref{aff88}}
\and J.~Garc\'ia-Bellido\orcid{0000-0002-9370-8360}\inst{\ref{aff126}}
\and T.~Gasparetto\orcid{0000-0002-7913-4866}\inst{\ref{aff21}}
\and V.~Gautard\inst{\ref{aff143}}
\and E.~Gaztanaga\orcid{0000-0001-9632-0815}\inst{\ref{aff7},\ref{aff62},\ref{aff144}}
\and F.~Giacomini\orcid{0000-0002-3129-2814}\inst{\ref{aff25}}
\and F.~Gianotti\orcid{0000-0003-4666-119X}\inst{\ref{aff5}}
\and G.~Gozaliasl\orcid{0000-0002-0236-919X}\inst{\ref{aff145},\ref{aff75}}
\and M.~Guidi\orcid{0000-0001-9408-1101}\inst{\ref{aff24},\ref{aff5}}
\and C.~M.~Gutierrez\orcid{0000-0001-7854-783X}\inst{\ref{aff146}}
\and A.~Hall\orcid{0000-0002-3139-8651}\inst{\ref{aff46}}
\and W.~G.~Hartley\inst{\ref{aff56}}
\and S.~Hemmati\orcid{0000-0003-2226-5395}\inst{\ref{aff147}}
\and C.~Hern\'andez-Monteagudo\orcid{0000-0001-5471-9166}\inst{\ref{aff13},\ref{aff12}}
\and H.~Hildebrandt\orcid{0000-0002-9814-3338}\inst{\ref{aff148}}
\and J.~Hjorth\orcid{0000-0002-4571-2306}\inst{\ref{aff96}}
\and J.~J.~E.~Kajava\orcid{0000-0002-3010-8333}\inst{\ref{aff149},\ref{aff150}}
\and Y.~Kang\orcid{0009-0000-8588-7250}\inst{\ref{aff56}}
\and V.~Kansal\orcid{0000-0002-4008-6078}\inst{\ref{aff151},\ref{aff152}}
\and D.~Karagiannis\orcid{0000-0002-4927-0816}\inst{\ref{aff115},\ref{aff153}}
\and K.~Kiiveri\inst{\ref{aff73}}
\and C.~C.~Kirkpatrick\inst{\ref{aff73}}
\and S.~Kruk\orcid{0000-0001-8010-8879}\inst{\ref{aff16}}
\and J.~Le~Graet\orcid{0000-0001-6523-7971}\inst{\ref{aff59}}
\and L.~Legrand\orcid{0000-0003-0610-5252}\inst{\ref{aff154},\ref{aff155}}
\and M.~Lembo\orcid{0000-0002-5271-5070}\inst{\ref{aff115},\ref{aff116}}
\and F.~Lepori\orcid{0009-0000-5061-7138}\inst{\ref{aff156}}
\and G.~Leroy\orcid{0009-0004-2523-4425}\inst{\ref{aff4},\ref{aff86}}
\and G.~F.~Lesci\orcid{0000-0002-4607-2830}\inst{\ref{aff14},\ref{aff5}}
\and J.~Lesgourgues\orcid{0000-0001-7627-353X}\inst{\ref{aff42}}
\and L.~Leuzzi\orcid{0009-0006-4479-7017}\inst{\ref{aff14},\ref{aff5}}
\and T.~I.~Liaudat\orcid{0000-0002-9104-314X}\inst{\ref{aff157}}
\and A.~Loureiro\orcid{0000-0002-4371-0876}\inst{\ref{aff158},\ref{aff159}}
\and J.~Macias-Perez\orcid{0000-0002-5385-2763}\inst{\ref{aff160}}
\and G.~Maggio\orcid{0000-0003-4020-4836}\inst{\ref{aff21}}
\and M.~Magliocchetti\orcid{0000-0001-9158-4838}\inst{\ref{aff57}}
\and F.~Mannucci\orcid{0000-0002-4803-2381}\inst{\ref{aff2}}
\and R.~Maoli\orcid{0000-0002-6065-3025}\inst{\ref{aff161},\ref{aff11}}
\and C.~J.~A.~P.~Martins\orcid{0000-0002-4886-9261}\inst{\ref{aff162},\ref{aff31}}
\and L.~Maurin\orcid{0000-0002-8406-0857}\inst{\ref{aff15}}
\and M.~Miluzio\inst{\ref{aff16},\ref{aff163}}
\and P.~Monaco\orcid{0000-0003-2083-7564}\inst{\ref{aff164},\ref{aff21},\ref{aff22},\ref{aff20}}
\and C.~Moretti\orcid{0000-0003-3314-8936}\inst{\ref{aff23},\ref{aff125},\ref{aff21},\ref{aff20},\ref{aff22}}
\and G.~Morgante\inst{\ref{aff5}}
\and K.~Naidoo\orcid{0000-0002-9182-1802}\inst{\ref{aff144}}
\and A.~Navarro-Alsina\orcid{0000-0002-3173-2592}\inst{\ref{aff84}}
\and S.~Nesseris\orcid{0000-0002-0567-0324}\inst{\ref{aff126}}
\and F.~Passalacqua\orcid{0000-0002-8606-4093}\inst{\ref{aff103},\ref{aff58}}
\and K.~Paterson\orcid{0000-0001-8340-3486}\inst{\ref{aff71}}
\and L.~Patrizii\inst{\ref{aff25}}
\and A.~Pisani\orcid{0000-0002-6146-4437}\inst{\ref{aff59},\ref{aff165}}
\and D.~Potter\orcid{0000-0002-0757-5195}\inst{\ref{aff156}}
\and S.~Quai\orcid{0000-0002-0449-8163}\inst{\ref{aff14},\ref{aff5}}
\and M.~Radovich\orcid{0000-0002-3585-866X}\inst{\ref{aff26}}
\and S.~Sacquegna\orcid{0000-0002-8433-6630}\inst{\ref{aff138},\ref{aff139},\ref{aff140}}
\and M.~Sahl\'en\orcid{0000-0003-0973-4804}\inst{\ref{aff166}}
\and D.~B.~Sanders\orcid{0000-0002-1233-9998}\inst{\ref{aff44}}
\and E.~Sarpa\orcid{0000-0002-1256-655X}\inst{\ref{aff23},\ref{aff125},\ref{aff22}}
\and A.~Schneider\orcid{0000-0001-7055-8104}\inst{\ref{aff156}}
\and D.~Sciotti\orcid{0009-0008-4519-2620}\inst{\ref{aff11},\ref{aff85}}
\and E.~Sellentin\inst{\ref{aff167},\ref{aff38}}
\and F.~Shankar\orcid{0000-0001-8973-5051}\inst{\ref{aff168}}
\and L.~C.~Smith\orcid{0000-0002-3259-2771}\inst{\ref{aff169}}
\and K.~Tanidis\orcid{0000-0001-9843-5130}\inst{\ref{aff119}}
\and G.~Testera\inst{\ref{aff29}}
\and R.~Teyssier\orcid{0000-0001-7689-0933}\inst{\ref{aff165}}
\and S.~Tosi\orcid{0000-0002-7275-9193}\inst{\ref{aff28},\ref{aff123}}
\and A.~Troja\orcid{0000-0003-0239-4595}\inst{\ref{aff103},\ref{aff58}}
\and M.~Tucci\inst{\ref{aff56}}
\and C.~Valieri\inst{\ref{aff25}}
\and A.~Venhola\orcid{0000-0001-6071-4564}\inst{\ref{aff170}}
\and D.~Vergani\orcid{0000-0003-0898-2216}\inst{\ref{aff5}}
\and G.~Verza\orcid{0000-0002-1886-8348}\inst{\ref{aff171}}
\and P.~Vielzeuf\orcid{0000-0003-2035-9339}\inst{\ref{aff59}}
\and A.~Viitanen\orcid{0000-0001-9383-786X}\inst{\ref{aff73},\ref{aff11}}
\and N.~A.~Walton\orcid{0000-0003-3983-8778}\inst{\ref{aff169}}
\and E.~Soubrie\orcid{0000-0001-9295-1863}\inst{\ref{aff15}}
\and D.~Scott\orcid{0000-0002-6878-9840}\inst{\ref{aff172}}}
										   
%%%% please do not edit the affiliation list -- contact ECEB Bureau for changes
\institute{Max Planck Institute for Extraterrestrial Physics, Giessenbachstr. 1, 85748 Garching, Germany\label{aff1}
\and
INAF-Osservatorio Astrofisico di Arcetri, Largo E. Fermi 5, 50125, Firenze, Italy\label{aff2}
\and
Dipartimento di Fisica e Astronomia, Universit\`{a} di Firenze, via G. Sansone 1, 50019 Sesto Fiorentino, Firenze, Italy\label{aff3}
\and
Department of Physics, Centre for Extragalactic Astronomy, Durham University, South Road, Durham, DH1 3LE, UK\label{aff4}
\and
INAF-Osservatorio di Astrofisica e Scienza dello Spazio di Bologna, Via Piero Gobetti 93/3, 40129 Bologna, Italy\label{aff5}
\and
Instituto de Astrof\'isica de Canarias (IAC); Departamento de Astrof\'isica, Universidad de La Laguna (ULL), 38200, La Laguna, Tenerife, Spain\label{aff6}
\and
Institute of Space Sciences (ICE, CSIC), Campus UAB, Carrer de Can Magrans, s/n, 08193 Barcelona, Spain\label{aff7}
\and
INAF-Osservatorio Astronomico di Capodimonte, Via Moiariello 16, 80131 Napoli, Italy\label{aff8}
\and
School of Physics, HH Wills Physics Laboratory, University of Bristol, Tyndall Avenue, Bristol, BS8 1TL, UK\label{aff9}
\and
Department of Mathematics and Physics, Roma Tre University, Via della Vasca Navale 84, 00146 Rome, Italy\label{aff10}
\and
INAF-Osservatorio Astronomico di Roma, Via Frascati 33, 00078 Monteporzio Catone, Italy\label{aff11}
\and
Instituto de Astrof\'{\i}sica de Canarias, V\'{\i}a L\'actea, 38205 La Laguna, Tenerife, Spain\label{aff12}
\and
Universidad de La Laguna, Departamento de Astrof\'{\i}sica, 38206 La Laguna, Tenerife, Spain\label{aff13}
\and
Dipartimento di Fisica e Astronomia "Augusto Righi" - Alma Mater Studiorum Universit\`a di Bologna, via Piero Gobetti 93/2, 40129 Bologna, Italy\label{aff14}
\and
Universit\'e Paris-Saclay, CNRS, Institut d'astrophysique spatiale, 91405, Orsay, France\label{aff15}
\and
ESAC/ESA, Camino Bajo del Castillo, s/n., Urb. Villafranca del Castillo, 28692 Villanueva de la Ca\~nada, Madrid, Spain\label{aff16}
\and
School of Mathematics and Physics, University of Surrey, Guildford, Surrey, GU2 7XH, UK\label{aff17}
\and
INAF-Osservatorio Astronomico di Brera, Via Brera 28, 20122 Milano, Italy\label{aff18}
\and
Universit\'e Paris-Saclay, Universit\'e Paris Cit\'e, CEA, CNRS, AIM, 91191, Gif-sur-Yvette, France\label{aff19}
\and
IFPU, Institute for Fundamental Physics of the Universe, via Beirut 2, 34151 Trieste, Italy\label{aff20}
\and
INAF-Osservatorio Astronomico di Trieste, Via G. B. Tiepolo 11, 34143 Trieste, Italy\label{aff21}
\and
INFN, Sezione di Trieste, Via Valerio 2, 34127 Trieste TS, Italy\label{aff22}
\and
SISSA, International School for Advanced Studies, Via Bonomea 265, 34136 Trieste TS, Italy\label{aff23}
\and
Dipartimento di Fisica e Astronomia, Universit\`a di Bologna, Via Gobetti 93/2, 40129 Bologna, Italy\label{aff24}
\and
INFN-Sezione di Bologna, Viale Berti Pichat 6/2, 40127 Bologna, Italy\label{aff25}
\and
INAF-Osservatorio Astronomico di Padova, Via dell'Osservatorio 5, 35122 Padova, Italy\label{aff26}
\and
Space Science Data Center, Italian Space Agency, via del Politecnico snc, 00133 Roma, Italy\label{aff27}
\and
Dipartimento di Fisica, Universit\`a di Genova, Via Dodecaneso 33, 16146, Genova, Italy\label{aff28}
\and
INFN-Sezione di Genova, Via Dodecaneso 33, 16146, Genova, Italy\label{aff29}
\and
Department of Physics "E. Pancini", University Federico II, Via Cinthia 6, 80126, Napoli, Italy\label{aff30}
\and
Instituto de Astrof\'isica e Ci\^encias do Espa\c{c}o, Universidade do Porto, CAUP, Rua das Estrelas, PT4150-762 Porto, Portugal\label{aff31}
\and
Faculdade de Ci\^encias da Universidade do Porto, Rua do Campo de Alegre, 4150-007 Porto, Portugal\label{aff32}
\and
Dipartimento di Fisica, Universit\`a degli Studi di Torino, Via P. Giuria 1, 10125 Torino, Italy\label{aff33}
\and
INFN-Sezione di Torino, Via P. Giuria 1, 10125 Torino, Italy\label{aff34}
\and
INAF-Osservatorio Astrofisico di Torino, Via Osservatorio 20, 10025 Pino Torinese (TO), Italy\label{aff35}
\and
European Space Agency/ESTEC, Keplerlaan 1, 2201 AZ Noordwijk, The Netherlands\label{aff36}
\and
Institute Lorentz, Leiden University, Niels Bohrweg 2, 2333 CA Leiden, The Netherlands\label{aff37}
\and
Leiden Observatory, Leiden University, Einsteinweg 55, 2333 CC Leiden, The Netherlands\label{aff38}
\and
INAF-IASF Milano, Via Alfonso Corti 12, 20133 Milano, Italy\label{aff39}
\and
Centro de Investigaciones Energ\'eticas, Medioambientales y Tecnol\'ogicas (CIEMAT), Avenida Complutense 40, 28040 Madrid, Spain\label{aff40}
\and
Port d'Informaci\'{o} Cient\'{i}fica, Campus UAB, C. Albareda s/n, 08193 Bellaterra (Barcelona), Spain\label{aff41}
\and
Institute for Theoretical Particle Physics and Cosmology (TTK), RWTH Aachen University, 52056 Aachen, Germany\label{aff42}
\and
INFN section of Naples, Via Cinthia 6, 80126, Napoli, Italy\label{aff43}
\and
Institute for Astronomy, University of Hawaii, 2680 Woodlawn Drive, Honolulu, HI 96822, USA\label{aff44}
\and
Dipartimento di Fisica e Astronomia "Augusto Righi" - Alma Mater Studiorum Universit\`a di Bologna, Viale Berti Pichat 6/2, 40127 Bologna, Italy\label{aff45}
\and
Institute for Astronomy, University of Edinburgh, Royal Observatory, Blackford Hill, Edinburgh EH9 3HJ, UK\label{aff46}
\and
Jodrell Bank Centre for Astrophysics, Department of Physics and Astronomy, University of Manchester, Oxford Road, Manchester M13 9PL, UK\label{aff47}
\and
European Space Agency/ESRIN, Largo Galileo Galilei 1, 00044 Frascati, Roma, Italy\label{aff48}
\and
Universit\'e Claude Bernard Lyon 1, CNRS/IN2P3, IP2I Lyon, UMR 5822, Villeurbanne, F-69100, France\label{aff49}
\and
Institut de Ci\`{e}ncies del Cosmos (ICCUB), Universitat de Barcelona (IEEC-UB), Mart\'{i} i Franqu\`{e}s 1, 08028 Barcelona, Spain\label{aff50}
\and
Instituci\'o Catalana de Recerca i Estudis Avan\c{c}ats (ICREA), Passeig de Llu\'{\i}s Companys 23, 08010 Barcelona, Spain\label{aff51}
\and
UCB Lyon 1, CNRS/IN2P3, IUF, IP2I Lyon, 4 rue Enrico Fermi, 69622 Villeurbanne, France\label{aff52}
\and
Mullard Space Science Laboratory, University College London, Holmbury St Mary, Dorking, Surrey RH5 6NT, UK\label{aff53}
\and
Departamento de F\'isica, Faculdade de Ci\^encias, Universidade de Lisboa, Edif\'icio C8, Campo Grande, PT1749-016 Lisboa, Portugal\label{aff54}
\and
Instituto de Astrof\'isica e Ci\^encias do Espa\c{c}o, Faculdade de Ci\^encias, Universidade de Lisboa, Campo Grande, 1749-016 Lisboa, Portugal\label{aff55}
\and
Department of Astronomy, University of Geneva, ch. d'Ecogia 16, 1290 Versoix, Switzerland\label{aff56}
\and
INAF-Istituto di Astrofisica e Planetologia Spaziali, via del Fosso del Cavaliere, 100, 00100 Roma, Italy\label{aff57}
\and
INFN-Padova, Via Marzolo 8, 35131 Padova, Italy\label{aff58}
\and
Aix-Marseille Universit\'e, CNRS/IN2P3, CPPM, Marseille, France\label{aff59}
\and
Universit\"ats-Sternwarte M\"unchen, Fakult\"at f\"ur Physik, Ludwig-Maximilians-Universit\"at M\"unchen, Scheinerstrasse 1, 81679 M\"unchen, Germany\label{aff60}
\and
INFN-Bologna, Via Irnerio 46, 40126 Bologna, Italy\label{aff61}
\and
Institut d'Estudis Espacials de Catalunya (IEEC),  Edifici RDIT, Campus UPC, 08860 Castelldefels, Barcelona, Spain\label{aff62}
\and
Herzberg Astronomy and Astrophysics Research Centre, 5071 W. Saanich Rd. Victoria, BC, V9E 2E7, Canada\label{aff63}
\and
Institute of Theoretical Astrophysics, University of Oslo, P.O. Box 1029 Blindern, 0315 Oslo, Norway\label{aff64}
\and
Jet Propulsion Laboratory, California Institute of Technology, 4800 Oak Grove Drive, Pasadena, CA, 91109, USA\label{aff65}
\and
Department of Physics, Lancaster University, Lancaster, LA1 4YB, UK\label{aff66}
\and
Felix Hormuth Engineering, Goethestr. 17, 69181 Leimen, Germany\label{aff67}
\and
Technical University of Denmark, Elektrovej 327, 2800 Kgs. Lyngby, Denmark\label{aff68}
\and
Cosmic Dawn Center (DAWN), Denmark\label{aff69}
\and
Institut d'Astrophysique de Paris, UMR 7095, CNRS, and Sorbonne Universit\'e, 98 bis boulevard Arago, 75014 Paris, France\label{aff70}
\and
Max-Planck-Institut f\"ur Astronomie, K\"onigstuhl 17, 69117 Heidelberg, Germany\label{aff71}
\and
NASA Goddard Space Flight Center, Greenbelt, MD 20771, USA\label{aff72}
\and
Department of Physics and Helsinki Institute of Physics, Gustaf H\"allstr\"omin katu 2, University of Helsinki, 00014 Helsinki, Finland\label{aff73}
\and
Universit\'e de Gen\`eve, D\'epartement de Physique Th\'eorique and Centre for Astroparticle Physics, 24 quai Ernest-Ansermet, CH-1211 Gen\`eve 4, Switzerland\label{aff74}
\and
Department of Physics, P.O. Box 64, University of Helsinki, 00014 Helsinki, Finland\label{aff75}
\and
Helsinki Institute of Physics, Gustaf H{\"a}llstr{\"o}min katu 2, University of Helsinki, 00014 Helsinki, Finland\label{aff76}
\and
Centre de Calcul de l'IN2P3/CNRS, 21 avenue Pierre de Coubertin 69627 Villeurbanne Cedex, France\label{aff77}
\and
Laboratoire d'etude de l'Univers et des phenomenes eXtremes, Observatoire de Paris, Universit\'e PSL, Sorbonne Universit\'e, CNRS, 92190 Meudon, France\label{aff78}
\and
Aix-Marseille Universit\'e, CNRS, CNES, LAM, Marseille, France\label{aff79}
\and
SKA Observatory, Jodrell Bank, Lower Withington, Macclesfield, Cheshire SK11 9FT, UK\label{aff80}
\and
Dipartimento di Fisica "Aldo Pontremoli", Universit\`a degli Studi di Milano, Via Celoria 16, 20133 Milano, Italy\label{aff81}
\and
INFN-Sezione di Milano, Via Celoria 16, 20133 Milano, Italy\label{aff82}
\and
University of Applied Sciences and Arts of Northwestern Switzerland, School of Computer Science, 5210 Windisch, Switzerland\label{aff83}
\and
Universit\"at Bonn, Argelander-Institut f\"ur Astronomie, Auf dem H\"ugel 71, 53121 Bonn, Germany\label{aff84}
\and
INFN-Sezione di Roma, Piazzale Aldo Moro, 2 - c/o Dipartimento di Fisica, Edificio G. Marconi, 00185 Roma, Italy\label{aff85}
\and
Department of Physics, Institute for Computational Cosmology, Durham University, South Road, Durham, DH1 3LE, UK\label{aff86}
\and
Infrared Processing and Analysis Center, California Institute of Technology, Pasadena, CA 91125, USA\label{aff87}
\and
Universit\'e Paris Cit\'e, CNRS, Astroparticule et Cosmologie, 75013 Paris, France\label{aff88}
\and
CNRS-UCB International Research Laboratory, Centre Pierre Bin\'etruy, IRL2007, CPB-IN2P3, Berkeley, USA\label{aff89}
\and
University of Applied Sciences and Arts of Northwestern Switzerland, School of Engineering, 5210 Windisch, Switzerland\label{aff90}
\and
Institut d'Astrophysique de Paris, 98bis Boulevard Arago, 75014, Paris, France\label{aff91}
\and
Institute of Physics, Laboratory of Astrophysics, Ecole Polytechnique F\'ed\'erale de Lausanne (EPFL), Observatoire de Sauverny, 1290 Versoix, Switzerland\label{aff92}
\and
Aurora Technology for European Space Agency (ESA), Camino bajo del Castillo, s/n, Urbanizacion Villafranca del Castillo, Villanueva de la Ca\~nada, 28692 Madrid, Spain\label{aff93}
\and
Institut de F\'{i}sica d'Altes Energies (IFAE), The Barcelona Institute of Science and Technology, Campus UAB, 08193 Bellaterra (Barcelona), Spain\label{aff94}
\and
School of Mathematics, Statistics and Physics, Newcastle University, Herschel Building, Newcastle-upon-Tyne, NE1 7RU, UK\label{aff95}
\and
DARK, Niels Bohr Institute, University of Copenhagen, Jagtvej 155, 2200 Copenhagen, Denmark\label{aff96}
\and
Waterloo Centre for Astrophysics, University of Waterloo, Waterloo, Ontario N2L 3G1, Canada\label{aff97}
\and
Department of Physics and Astronomy, University of Waterloo, Waterloo, Ontario N2L 3G1, Canada\label{aff98}
\and
Perimeter Institute for Theoretical Physics, Waterloo, Ontario N2L 2Y5, Canada\label{aff99}
\and
Centre National d'Etudes Spatiales -- Centre spatial de Toulouse, 18 avenue Edouard Belin, 31401 Toulouse Cedex 9, France\label{aff100}
\and
Institute of Space Science, Str. Atomistilor, nr. 409 M\u{a}gurele, Ilfov, 077125, Romania\label{aff101}
\and
Consejo Superior de Investigaciones Cientificas, Calle Serrano 117, 28006 Madrid, Spain\label{aff102}
\and
Dipartimento di Fisica e Astronomia "G. Galilei", Universit\`a di Padova, Via Marzolo 8, 35131 Padova, Italy\label{aff103}
\and
Institut f\"ur Theoretische Physik, University of Heidelberg, Philosophenweg 16, 69120 Heidelberg, Germany\label{aff104}
\and
Institut de Recherche en Astrophysique et Plan\'etologie (IRAP), Universit\'e de Toulouse, CNRS, UPS, CNES, 14 Av. Edouard Belin, 31400 Toulouse, France\label{aff105}
\and
Universit\'e St Joseph; Faculty of Sciences, Beirut, Lebanon\label{aff106}
\and
Departamento de F\'isica, FCFM, Universidad de Chile, Blanco Encalada 2008, Santiago, Chile\label{aff107}
\and
Universit\"at Innsbruck, Institut f\"ur Astro- und Teilchenphysik, Technikerstr. 25/8, 6020 Innsbruck, Austria\label{aff108}
\and
Satlantis, University Science Park, Sede Bld 48940, Leioa-Bilbao, Spain\label{aff109}
\and
Instituto de Astrof\'isica e Ci\^encias do Espa\c{c}o, Faculdade de Ci\^encias, Universidade de Lisboa, Tapada da Ajuda, 1349-018 Lisboa, Portugal\label{aff110}
\and
Cosmic Dawn Center (DAWN)\label{aff111}
\and
Niels Bohr Institute, University of Copenhagen, Jagtvej 128, 2200 Copenhagen, Denmark\label{aff112}
\and
Universidad Polit\'ecnica de Cartagena, Departamento de Electr\'onica y Tecnolog\'ia de Computadoras,  Plaza del Hospital 1, 30202 Cartagena, Spain\label{aff113}
\and
Kapteyn Astronomical Institute, University of Groningen, PO Box 800, 9700 AV Groningen, The Netherlands\label{aff114}
\and
Dipartimento di Fisica e Scienze della Terra, Universit\`a degli Studi di Ferrara, Via Giuseppe Saragat 1, 44122 Ferrara, Italy\label{aff115}
\and
Istituto Nazionale di Fisica Nucleare, Sezione di Ferrara, Via Giuseppe Saragat 1, 44122 Ferrara, Italy\label{aff116}
\and
INAF, Istituto di Radioastronomia, Via Piero Gobetti 101, 40129 Bologna, Italy\label{aff117}
\and
Universit\'e C\^{o}te d'Azur, Observatoire de la C\^{o}te d'Azur, CNRS, Laboratoire Lagrange, Bd de l'Observatoire, CS 34229, 06304 Nice cedex 4, France\label{aff118}
\and
Department of Physics, Oxford University, Keble Road, Oxford OX1 3RH, UK\label{aff119}
\and
Universit\'e PSL, Observatoire de Paris, Sorbonne Universit\'e, CNRS, LERMA, 75014, Paris, France\label{aff120}
\and
Universit\'e Paris-Cit\'e, 5 Rue Thomas Mann, 75013, Paris, France\label{aff121}
\and
INAF - Osservatorio Astronomico di Brera, via Emilio Bianchi 46, 23807 Merate, Italy\label{aff122}
\and
INAF-Osservatorio Astronomico di Brera, Via Brera 28, 20122 Milano, Italy, and INFN-Sezione di Genova, Via Dodecaneso 33, 16146, Genova, Italy\label{aff123}
\and
ICL, Junia, Universit\'e Catholique de Lille, LITL, 59000 Lille, France\label{aff124}
\and
ICSC - Centro Nazionale di Ricerca in High Performance Computing, Big Data e Quantum Computing, Via Magnanelli 2, Bologna, Italy\label{aff125}
\and
Instituto de F\'isica Te\'orica UAM-CSIC, Campus de Cantoblanco, 28049 Madrid, Spain\label{aff126}
\and
CERCA/ISO, Department of Physics, Case Western Reserve University, 10900 Euclid Avenue, Cleveland, OH 44106, USA\label{aff127}
\and
Technical University of Munich, TUM School of Natural Sciences, Physics Department, James-Franck-Str.~1, 85748 Garching, Germany\label{aff128}
\and
Max-Planck-Institut f\"ur Astrophysik, Karl-Schwarzschild-Str.~1, 85748 Garching, Germany\label{aff129}
\and
Laboratoire Univers et Th\'eorie, Observatoire de Paris, Universit\'e PSL, Universit\'e Paris Cit\'e, CNRS, 92190 Meudon, France\label{aff130}
\and
Departamento de F{\'\i}sica Fundamental. Universidad de Salamanca. Plaza de la Merced s/n. 37008 Salamanca, Spain\label{aff131}
\and
Universit\'e de Strasbourg, CNRS, Observatoire astronomique de Strasbourg, UMR 7550, 67000 Strasbourg, France\label{aff132}
\and
Center for Data-Driven Discovery, Kavli IPMU (WPI), UTIAS, The University of Tokyo, Kashiwa, Chiba 277-8583, Japan\label{aff133}
\and
Ludwig-Maximilians-University, Schellingstrasse 4, 80799 Munich, Germany\label{aff134}
\and
Max-Planck-Institut f\"ur Physik, Boltzmannstr. 8, 85748 Garching, Germany\label{aff135}
\and
California Institute of Technology, 1200 E California Blvd, Pasadena, CA 91125, USA\label{aff136}
\and
Department of Physics \& Astronomy, University of California Irvine, Irvine CA 92697, USA\label{aff137}
\and
Department of Mathematics and Physics E. De Giorgi, University of Salento, Via per Arnesano, CP-I93, 73100, Lecce, Italy\label{aff138}
\and
INFN, Sezione di Lecce, Via per Arnesano, CP-193, 73100, Lecce, Italy\label{aff139}
\and
INAF-Sezione di Lecce, c/o Dipartimento Matematica e Fisica, Via per Arnesano, 73100, Lecce, Italy\label{aff140}
\and
Departamento F\'isica Aplicada, Universidad Polit\'ecnica de Cartagena, Campus Muralla del Mar, 30202 Cartagena, Murcia, Spain\label{aff141}
\and
Instituto de F\'isica de Cantabria, Edificio Juan Jord\'a, Avenida de los Castros, 39005 Santander, Spain\label{aff142}
\and
CEA Saclay, DFR/IRFU, Service d'Astrophysique, Bat. 709, 91191 Gif-sur-Yvette, France\label{aff143}
\and
Institute of Cosmology and Gravitation, University of Portsmouth, Portsmouth PO1 3FX, UK\label{aff144}
\and
Department of Computer Science, Aalto University, PO Box 15400, Espoo, FI-00 076, Finland\label{aff145}
\and
Instituto de Astrof\'\i sica de Canarias, c/ Via Lactea s/n, La Laguna 38200, Spain. Departamento de Astrof\'\i sica de la Universidad de La Laguna, Avda. Francisco Sanchez, La Laguna, 38200, Spain\label{aff146}
\and
Caltech/IPAC, 1200 E. California Blvd., Pasadena, CA 91125, USA\label{aff147}
\and
Ruhr University Bochum, Faculty of Physics and Astronomy, Astronomical Institute (AIRUB), German Centre for Cosmological Lensing (GCCL), 44780 Bochum, Germany\label{aff148}
\and
Department of Physics and Astronomy, Vesilinnantie 5, University of Turku, 20014 Turku, Finland\label{aff149}
\and
Serco for European Space Agency (ESA), Camino bajo del Castillo, s/n, Urbanizacion Villafranca del Castillo, Villanueva de la Ca\~nada, 28692 Madrid, Spain\label{aff150}
\and
ARC Centre of Excellence for Dark Matter Particle Physics, Melbourne, Australia\label{aff151}
\and
Centre for Astrophysics \& Supercomputing, Swinburne University of Technology,  Hawthorn, Victoria 3122, Australia\label{aff152}
\and
Department of Physics and Astronomy, University of the Western Cape, Bellville, Cape Town, 7535, South Africa\label{aff153}
\and
DAMTP, Centre for Mathematical Sciences, Wilberforce Road, Cambridge CB3 0WA, UK\label{aff154}
\and
Kavli Institute for Cosmology Cambridge, Madingley Road, Cambridge, CB3 0HA, UK\label{aff155}
\and
Department of Astrophysics, University of Zurich, Winterthurerstrasse 190, 8057 Zurich, Switzerland\label{aff156}
\and
IRFU, CEA, Universit\'e Paris-Saclay 91191 Gif-sur-Yvette Cedex, France\label{aff157}
\and
Oskar Klein Centre for Cosmoparticle Physics, Department of Physics, Stockholm University, Stockholm, SE-106 91, Sweden\label{aff158}
\and
Astrophysics Group, Blackett Laboratory, Imperial College London, London SW7 2AZ, UK\label{aff159}
\and
Univ. Grenoble Alpes, CNRS, Grenoble INP, LPSC-IN2P3, 53, Avenue des Martyrs, 38000, Grenoble, France\label{aff160}
\and
Dipartimento di Fisica, Sapienza Universit\`a di Roma, Piazzale Aldo Moro 2, 00185 Roma, Italy\label{aff161}
\and
Centro de Astrof\'{\i}sica da Universidade do Porto, Rua das Estrelas, 4150-762 Porto, Portugal\label{aff162}
\and
HE Space for European Space Agency (ESA), Camino bajo del Castillo, s/n, Urbanizacion Villafranca del Castillo, Villanueva de la Ca\~nada, 28692 Madrid, Spain\label{aff163}
\and
Dipartimento di Fisica - Sezione di Astronomia, Universit\`a di Trieste, Via Tiepolo 11, 34131 Trieste, Italy\label{aff164}
\and
Department of Astrophysical Sciences, Peyton Hall, Princeton University, Princeton, NJ 08544, USA\label{aff165}
\and
Theoretical astrophysics, Department of Physics and Astronomy, Uppsala University, Box 516, 751 37 Uppsala, Sweden\label{aff166}
\and
Mathematical Institute, University of Leiden, Einsteinweg 55, 2333 CA Leiden, The Netherlands\label{aff167}
\and
School of Physics \& Astronomy, University of Southampton, Highfield Campus, Southampton SO17 1BJ, UK\label{aff168}
\and
Institute of Astronomy, University of Cambridge, Madingley Road, Cambridge CB3 0HA, UK\label{aff169}
\and
Space physics and astronomy research unit, University of Oulu, Pentti Kaiteran katu 1, FI-90014 Oulu, Finland\label{aff170}
\and
Center for Computational Astrophysics, Flatiron Institute, 162 5th Avenue, 10010, New York, NY, USA\label{aff171}
\and
Department of Physics and Astronomy, University of British Columbia, Vancouver, BC V6T 1Z1, Canada\label{aff172}}

 \date{Received ; accepted}

\abstract{

To better understand the role of active galactic nuclei (AGN) in galaxy evolution, it is crucial to work with a complete and pure AGN sample. X-ray surveys are key to doing so, but their larger positional uncertainties complicate counterpart (CTP) association, further compounded by the limited availability of deep, uniform multi-wavelength ancillary data. \Euclid is revolutionising this identification effort, offering extensive coverage of nearly the entire extragalactic sky, particularly in the near-infrared bands, where AGN are more easily detected. Using the first Euclid Quick Data Release (Q1), we validated the methods for identifying and classifying \Euclid CTPs of known point-like sources from major X-ray surveys, including \XMMN, \textit{Chandra}, and eROSITA. Using Bayesian statistics, combined with machine learning (ML), as incorporated in the algorithm {\tt NWAY}, we identified the CTPs of 11\,286 X-ray sources from the three X-ray telescopes. For the large majority of 10\,194 sources, the association is unique, with the remaining $\sim$ 10\% of multi-CTP cases equally split between \XMMN and eROSITA. Six percent of the \Euclid CTPs are detected in more than one X-ray survey. We then used ML to distinguish between Galactic (8\%) and extragalactic (92\%) sources. We computed photo-$z$s using deep learning for the 9259 sources detected in the tenth data release of the DESI Legacy Survey, reaching an accuracy and a fraction of outliers of roughly 5\%. Based on their X-ray luminosities, all CTPs identified as extragalactic are classified as AGN, most of which appear as type I AGN according to their hardness ratios. With this paper, we release our catalogue, which includes identifiers, basic X-ray properties, the reliability of the associations, and additional property extensions, such as Galactic- or extragalactic classifications and photometric/spectroscopic redshifts. We also provide probabilities for sub-selecting the sample based on purity and completeness, in order to allow users to tailor the sample according to their specific needs.

}
%
% Provide up to five key words:
%
\keywords{%select up to 6 key words from the list given in
  %\url{https://www.aanda.org/for-authors/latex-issues/information-files\#pop},
Methods: statistical, data analysis; Surveys; Catalogues; Galaxies: active; X-rays: galaxies}
%
% Add short versions of title and author list for page headings
%
   \titlerunning{\Euclid\/: Identification of point-like X-ray selected sources}
   \authorrunning{Euclid Collaboration: W. Roster et al.}
   
   \maketitle
%
%-------------------------------------------------------------------
%
%
%   Start the main text of your paper here
%

\section{\label{sc:Intro}Introduction}

%###################################################

Active galactic nuclei (AGN\footnote{We use this term interchangeably to indicate both the singular and plural cases.}), which rank among the most energetic phenomena in the Universe, play a pivotal role in shaping the evolutionary trajectories of galaxies throughout cosmic time. Their activity is fundamentally driven by the accretion of matter onto supermassive black holes (SMBHs) from their surrounding environment \citep[e.g.,][]{Salpeter_1964,LB_1969,Pringle_1972,Shakura_1973,LB_1974,Kormendy_2013}. Situated at the centres of galaxies, AGN emit intense radiation across the entire electromagnetic spectrum, ranging from radio waves to $\gamma$-rays \citep[e.g.,][]{Elvis_1994,Urry_1995, Padovani_2017}. While this is particularly true for quasi-stellar objects (QSOs), it is important to note that the range of activity, emission, and interaction of AGN span a broad spectrum, with different AGN exhibiting varying levels of energy outputs \citep[e.g.,][]{Peterson_1997,Pierce2010,Povic2012, Bettoni2015}. 

Consequently, with enough coupling efficiency, this allows AGN to exert significant influence over both their host galaxies and the growth of their central SMBHs, where the interplay between AGN activity and the respective host is believed to regulate star formation and shape the morphology of galaxies. Here, energetic feedback drives powerful winds and outflows, impacting the surrounding interstellar and intergalactic medium \citep{Green_2011, cielo_2018}. These close connections are supported by fundamental scaling relations that link the SMBH mass to various galactic properties. Notable among these are the ${M}_{\textrm{BH}}$-$\sigma$ relation, which ties black hole (BH) mass to the velocity dispersion of stars in the galaxy's bulge \citep{Gebhardt2000, Ferrarese_2000}, and the ${M}_{\textrm{BH}}$-${M}_{*}$ relation, which connects it to the stellar mass of the host galaxy \citep{Magorrian_1998, Haring_2004,Gultekin_2009,Sani_2011,Shankar_2016,Suh_2020}, thus pointing to an intricate co-evolution of the systems \citep{Kormendy_2013, Heckman2014, Madau2014, Bisigello_2024}. However, the physical mechanisms driving this connection remain under debate, highlighting the need to construct a complete picture of AGN diversity and complexity to better understand galaxy formation and transformation across the history of the Universe \citep[e.g.,][]{Aird2010, Buchner2015, Georgakakis2015, Ananna2017, Morganti_2017,Delvecchio_2017,Husemann_2018,Harrison_2024}.

Despite significant advancements over the years, the current AGN census remains incomplete due to inherent biases introduced to individual samples by various multi-wavelength selection methods, as each favours specific properties of the population \citep{Messias_2014, Padovani_2017, Delvecchio_2017, Lyu_2022, Green_2024}. Comprehensive wavelength coverage is vital for identifying diverse AGN across redshift, as each method that contributes to this coverage capitalises on the dominant physical processes of AGN activity, dominating the spectral energy distribution (SED) of its host galaxy \citep{Padovani_2017}. For instance, radio observations are effective at identifying jet-dominated AGN \citep{DB_2002, Smol_2017}, while optical diagnostics \citep{Feltre_2016}, such as emission line ratios \citep[e.g., BPT diagrams as in][]{BPT_1, BPT_2, Juneau_2014}, are widely used to distinguish AGN from star-forming galaxies by probing ionisation mechanisms. Other selections include colour or variability \citep[e.g.,][]{Richards_2002, Bongiorno_2010, Bovy2012, Donley_2012, Kirkpatrick_2013,Peters_2015,Palanque_2016,Lusso_2016}. Soft X-ray selection excels at detecting unobscured accretion-powered emission \citep[e.g.,][]{Hasinger_2008, brandt_2005, Nandra_2015} and can be complemented by mid-infrared (MIR) observations \citep[e.g.,][]{Stern_2012, Assef_2013}, which can effectively penetrate dense dust clouds \citep{Hickox_2018}. Notably, the significantly higher contrast between the intrinsic brightness of accreting SMBHs and their host galaxies in X-rays, as opposed to other wavelengths, makes this selection one of the most uncontaminated, and thus it provides pure AGN samples of a larger diversity \citep{Xue_2011, Donley_2012}. To combine different observational perspectives of AGN, these X-ray samples must then be paired with ancillary multi-wavelength data.

In the past, identifying multi-wavelength counterparts (CTPs) for X-ray-selected sources was a significant hurdle in AGN characterisation and redshift estimation \citep{Salvato_2018}. While combining data from multiple wavelengths is crucial for constructing a more complete picture of AGN, the process of cross-identifying sources remains complicated. Due to the significant positional uncertainties in X-ray surveys, confidently linking most X-ray sources to a single, definitive CTP in, for example, optical, infrared, or radio surveys, is challenging, except in the brightest cases. These uncertainties, along with the uneven spatial coverage of ancillary data, further complicate the task of reliably pinpointing AGN across different surveys. Consequently, a simple coordinate match is insufficient, and more sophisticated methods are needed for accurate cross-identification. For this purpose, machine learning (ML) algorithms that integrate multi-wavelength source classification approaches have become increasingly prevalent \citep[e.g.,][]{Cavuoti_2014, Brescia_2015, Karsten_2023,Cooper_2023,Daoutis_2023,Zeraatgari_2024,Perez-diaz_2024,Mechbal_2024}.

The European Space Agency's \Euclid satellite will also help overcome this limitation due to its uniform depth and wide coverage \citep{laureijs11, EuclidSkyOverview}. At the same time, \Euclid's depth results in very high source densities, thus increasing the risk of unforced misidentification and chance alignments. Over its six-year mission lifetime, \Euclid will cover a total of $\sim$ 14\,000\,deg\textsuperscript{2} in the Euclid Wide Survey \citep[EWS,][]{Scaramella-EP1}, and it is expected to detect billions of sources. In addition to its cosmological objectives, \Euclid will play a crucial role in the detection and characterisation of AGN, as at least 10 million AGN are anticipated to be detected \citep{Bisigello_2024,EP-Lusso,EP-Selwood}. To achieve these goals, \Euclid employs two main instruments: the VISible imager \citep[VIS,][]{Desprez-EP10,EuclidSkyVIS}, covering wavelengths from 0.53 to 0.90\,\micron \,(\IE), and the Near-Infrared Spectrometer and Photometer \citep[NISP,][]{Maciaszek22,EuclidSkyNISP}. Operating in the near-infrared and with photometric measurements in three bands (\YE, \JE, and \HE), \Euclid  covers in total the wavelength range from 0.95 to 2.02\,\micron\, while also including a slitless spectrograph for the detection of emission lines.

With the first Euclid Quick Data Release (Q1), we can now begin building and testing the necessary machinery for identifying and characterising AGN previously selected in the X-rays. We aim to leverage \Euclid's depth, resolution, and wavelength coverage to provide more secure associations of these AGN. Moreover, this initial testing will form the foundation for refining our methods and ensuring accurate AGN identification as more data become available over the next years. 

%This paper
In this paper, we address the task of reliably associating X-ray sources with their \Euclid CTPs and investigate their respective NIR properties as a function of X-ray flux, among other properties. The paper is structured as follows: In \cref{sec2} we give an overview of the Q1 data. In \cref{sec3} we introduce the X-ray source catalogues used in this work, while \cref{sec4} provides an outline of the approach to identifying CTPs. In \cref{sec6}, using a sample of sources clearly characterised with the tenth Data Release of
the Legacy Survey \citep{Dey_2019}, we describe how we assign a Galactic- or extragalactic nature to our sources using \Euclid photometry. In \cref{sec7}, we compute photometric redshifts for all CTPs. \Cref{sec8} displays CTP sample properties, and in \cref{sec9} we explain the release of the catalogues and how to use them. Finally, a summary and discussion are given in \cref{sec10}.

In this paper, unless stated otherwise, we express magnitudes in the AB system \citep{oke_1983} and adopt a flat $\Lambda$CDM cosmology with $H_{0} = 70$\,km\,s\textsuperscript{-1}\,Mpc\textsuperscript{-1}, $\Omega_{\text{m}}=0.3$, and $\Omega_{\Lambda}=0.7$ to facilitate direct comparison with similar works from literature.

%###################################################

\section{Euclid Q1 fields}

%###################################################
\label{sec2}

\begin{figure*}[htbp!]
\centering
\includegraphics[angle=0,width=1.0\hsize]{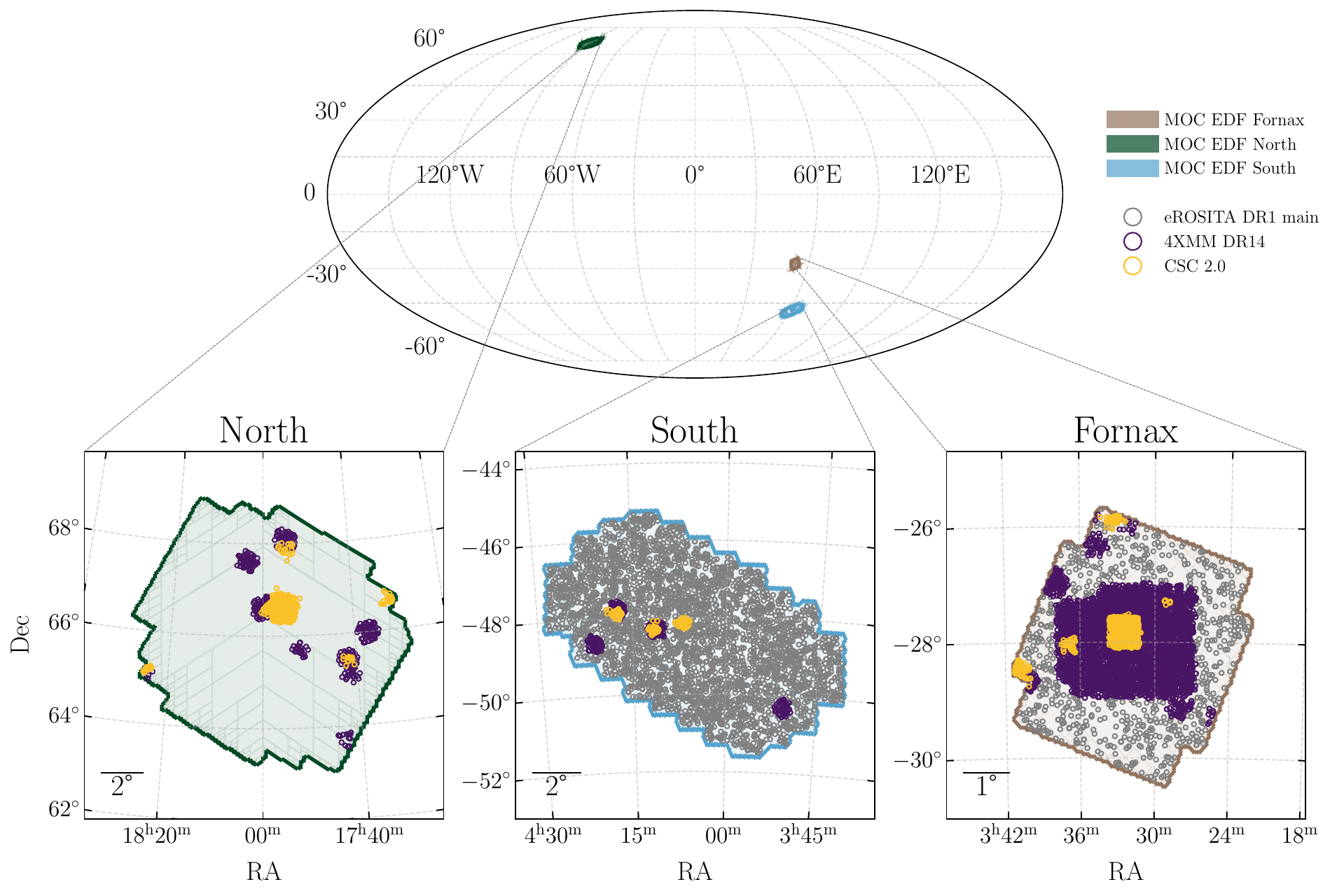}
\caption{Sky coverage of the EDF fields and corresponding X-ray source distributions. \textit{Top:} Mollweide projection of the sky indicating the locations of EDF-N, EDF-S, and EDF-F using their multi-order coverage maps (MOCs). \textit{Bottom:} Zoomed-in views of each EDF overlaid with X-ray sources from the eROSITA-DE DR1 catalogue (grey), the \XMMN 4XMM DR14 catalogue (purple), and the \textit{Chandra} Source Catalog 2.0 (orange), to illustrate the overlap between \Euclid's deep fields and existing X-ray source catalogues.}
\label{fig:1}
\end{figure*}

\Euclid Q1 marks the initial data release with a footprint area of $\sim$ 63.1\,deg\textsuperscript{2} \citep{EuclidSkyOverview}. It consists of observations making up the Euclid Deep Survey (EDS), selected for their considerable multi-wavelength coverage. However, the data is currently at the depth of the EWS, with a 5$\sigma$ depth of \IE(AB)$\,=\,$24.5 for point sources. The EDS targets three important Euclid Deep Fields (EDFs; see \cref{fig:1}): the EDF Fornax (EDF-F), EDF South (EDF-S), and the EDF North (EDF-N). For more details, we refer the reader to \cite{Q1-TP001}.

The catalogues provide comprehensive photometric information across all four \Euclid bands \citep{Q1-TP002, Q1-TP003, Q1-TP004}. In addition to the \Euclid bands, the catalogue is supplemented with ground-based photometry in the $ugriz$ optical bands, where available, sourced from various instruments depending on their access to the northern or southern extragalactic sky. Photometric measurements are provided as both template/S\'ersic model fits and fluxes extracted within apertures corresponding to radii of 1--4 times the seeing full-width at half maximum (FWHM).

%###################################################
\section{The X-ray surveys}
\label{sec3}

% Surveys and Euclid
X-ray surveys are often constrained by the flux detection limit of the most sensitive X-ray telescopes, which typically operate with small fields of view. Therefore, surveys with different observational strategies need to be combined to facilitate the characterisation of the full AGN population. Pencil-beam surveys that provide exceptionally deep observations, valuable for probing faint AGN and for characterising the low-luminosity end of the AGN population, are limited in sky coverage and, consequently, miss rare (bright or distant) sources \citep[e.g.,][]{Brusa_2010, Hsu2014,Nandra_2015,Marchesi_2016,Luo_2017,Oh_2018}. Subsequently, the data releases of \XMMN 4XMM\,DR14 \citep{Webb2020} and \textit{Chandra} Source Catalogue 2 \citep[CSC\,2.0,][]{Evans_2024} feature outstanding sensitivity for X-ray-selected sources but cover relatively small regions of the sky. 

On the other hand, wide-area surveys are crucial for complementing pointed observations and uncovering those AGN missed \citep[e.g.,][]{Ananna2017, Fotopoulou_2016}. Therefore, we incorporate data from the extended \textit{Roentgen} Survey with an Imaging Array \citep[eROSITA; ][]{Predehl2021} DR1 main sample \citep{Merloni2024}. As part of the ongoing eROSITA All-Sky Survey (eRASS), these data provide homogeneous, soft X-ray coverage of the entire extragalactic sky. While eROSITA’s positional accuracy is somewhat poorer than that of \XMMN and \textit{Chandra}, its unparalleled breadth allows us to include a significantly larger sample of sources, enhancing the statistical power of our analysis and ensuring a more comprehensive AGN census (see \cref{fig:1} and \cref{tab:1}). In the following subsections, we describe the selection criteria applied to the catalogues from each X-ray survey.

\begin{table*}
\centering
\caption[]{Sample sizes for the three X-ray catalogues at various stages.}

\begin{tabular}{lccccccr}
 \hline\hline\noalign{\vskip 1.5pt}
  X-ray catalogue & MOC & Number of sources & AND point-like & AND POSERR < limit & EDF-F & EDF-N & EDF-S \\
 (1)&(2)&(3)&(4)& (5)&(6)&(7)&(8)\\\noalign{\vskip 1.5pt}
 Unit & deg$^{2}$ & &&&&& \\
 \hline\noalign{\vskip 1.5pt}
 4XMM DR14 & 6.18 & 692\,109 & 612\,521 & 612\,504 & 4523&445&273 \\
 CSC\,2.0 & 1.31 & 317\,167 & 296\,473 & 296\,162 &1016&506&147 \\
 eROSITA DR1 main & 43.14 & 898\,812 & 879\,153 & 878\,984 &1012& 0 &3369 \\
 \hline     
    
\end{tabular}
\tablefoot{(1) Areas covered by the X-ray survey MOCs in the Q1 footprint (see \cref{fig:1}), (2) X-ray catalogue, (3) total number of X-ray sources (starting from unique source ID entries and QFs set to zero), (4) after applying the point-like source criterion, and (5) after following the application of the positional error threshold. Columns (6), (7), and (8) give the respective number of sources in each EDF.}
\label{tab:1}
\end{table*}

\subsection{XMM-Newton}
The \XMMN space-telescope provides some of the deepest and most detailed X-ray imaging available, with a spatial resolution of approximately 6\arcsec\, and positional uncertainties typically below 1--2\arcsec. The data used in this study includes both targeted and serendipitous observations, selected from the 4XMM DR14 catalogue\footnote{catalogue and data model description available at \url{https://cdsarc.cds.unistra.fr/viz-bin/cat/IX/69}}, the most recent of the XMM releases\footnote{We note that dedicated XMM observations targeting the Fornax field are ongoing. These additional data are not included in this work but may be used to refine and extend the CTPs in future.}. The catalogue offers multi-band flux measurements and quality flags. To prepare the sample, point-like sources were filtered based on their extent likelihood and positional uncertainties:  {\tt SC\_EXT\_ML}\,$<$\,10,  {\tt SC\_EXTENT}\,$<$\,10 and {\tt SC\_POSERR}\,$<$\,10\arcsec, where the threshold for the positional uncertainty is motivated by the respective distribution as a function of flux as shown in \cref{fig:flux_poserr}. This effectively reduces the computational load for cross-matching with \Euclid. To be consistent with other X-ray surveys, we computed the X-ray flux, $F_{\textrm{x}}$, for the \mbox{0.5--2\,keV} band as 
\begin{equation}
    \textrm{F$_{\textrm{X}}$} = \textrm{SC\_EP\_2\_FLUX$_{0.5-1.0\textrm{keV}}$} \, + \, \textrm{SC\_EP\_3\_FLUX$_{1.0-2.0\textrm{keV}}$}\,
\end{equation}
and its uncertainty as  
\begin{equation}
    \textrm{EF}_{\textrm{X}} = \sqrt{\textrm{SC\_EP\_2\_FLUX\_ERR}^{2} + \textrm{SC\_EP\_3\_FLUX\_ERR}^{2}}\,
\end{equation}
by combining the 0.5--1\,keV and 1--2\,keV band fluxes and errors.

\subsection{Chandra}

The \textit{Chandra} X-ray space telescope provides sub-arcsecond angular resolution, which allows for precise localisation of sources and makes it particularly effective in crowded or complex fields. Our sample incorporates sources from the CSC\,2.0 catalogue\footnote{\url{https://vizier.cds.unistra.fr/viz-bin/VizieR?-source=IX/57&-to=3}}. To subsample only point-like sources, we selected sources with the extended flag, $f_{\textrm{e}}$, set to zero. The 0.5--2\,keV X-ray flux and its uncertainty were computed as

\begin{equation}
    \textrm{F$_{\textrm{X}}$} = \textrm{Flux$_{0.5-1.2\textrm{\,keV}}$} + \textrm{Flux$_{1.2-2.0\textrm{\,keV}}$}\,,
\end{equation} 
\begin{equation}
    \textrm{EF$_{\textrm{X}}$} = \sqrt{(\textrm{B\_Fluxs} - \textrm{b\_Fluxs})^{2} + (\textrm{B\_Fluxm} - \textrm{b\_Fluxm})^{2}}\,, 
\end{equation}

\noindent with Flux$_{0.5-1.2\textrm{\,keV}}$ and Flux$_{1.2-2.0\textrm{\,keV}}$ referring to the default CSC\,2.0 soft and medium bands, respectively, while (B) and (b) represent the upper and lower 1\,$\sigma$ error margins. The positional uncertainties were determined using the 2\,$\sigma$ errors along the semi-major ($r$\textsubscript{0}) and semi-minor ($r$\textsubscript{1}) axes of the error ellipse provided in the source catalogue as

\begin{equation}
    \textrm{C\_POSERR} = \sqrt{\frac{({r}_{0}/2)^{2} + ({r}_{1}/2)^{2}}{2}}\,, 
\end{equation}
and we downsampled the CSC\,2.0 catalogue such that {\tt C\_POSERR}\,$<$\,10\arcsec to be consistent with 4XMM DR14 (see \cref{fig:flux_poserr}).

\subsection{eROSITA}

The eROSITA-DE DR1 main catalogue\footnote{\url{https://erosita.mpe.mpg.de/dr1/AllSkySurveyData_dr1/Catalogues_dr1/}} offers comprehensive coverage of the western galactic hemisphere. While it does not include the EDF-N region (see \cref{fig:1}), with a half energy width (HEW) of 26\arcsec\, and a soft X-ray energy range of 0.2--2.3\,keV, eROSITA offers a uniform data set ideal for large statistical studies. The data preparation for eROSITA included filtering for reliable point-source detections with {\tt EXT\_LIKE}\,$=$\,0. Additionally, we required that all nine quality flags (QFs) are set to zero, and applied a cut in positional uncertainty of {\tt POS\_ERR}\,$<$\,20\arcsec (refer to \cref{tab:1}). 

\begin{figure}[htbp!]
\centering
\includegraphics[angle=0,width=1.0\hsize]{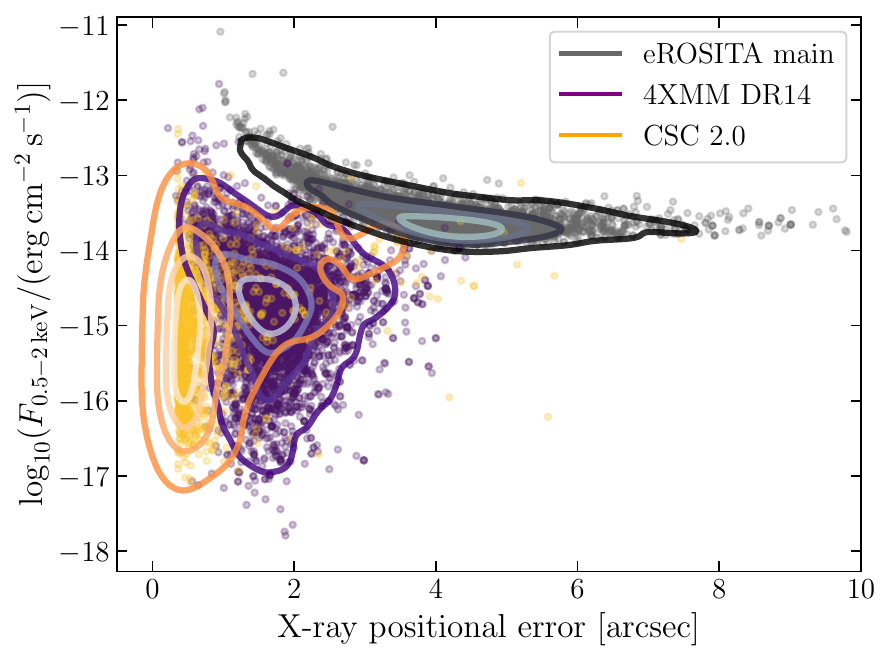}
\caption{X-ray fluxes in the 0.5--2\,keV band plotted against positional uncertainties for sources from the 4XMM DR14, CSC\,2.0, and eROSITA-DE/DR1 catalogues.}
\label{fig:flux_poserr}
\end{figure}

The choice to permit larger positional uncertainties in the eROSITA DR1 sample, as opposed to the \XMMN and \textit{Chandra} samples, is guided by the comparison of their error distributions shown in \cref{fig:flux_poserr}. This approach ensures the retention of the majority of the sample while excluding only a small number of outliers with exceptionally high positional errors. We intentionally avoid applying more restrictive criteria, such as minimum flux, signal-to-noise ratio (S/N), isolated environments, detection likelihoods, or other observational limitations, in order to prevent excluding sources that could potentially be identified by \Euclid. An overview of the sample sizes is given in \cref{tab:1}. By setting this threshold, we aim to optimise the balance between maintaining a robust sample and minimising the computational effort required for CTP identification within the defined error limits.  

\section{Identification of CTPs}
\label{sec4}

\begin{table*}
\centering
\caption[]{Overview of \Euclid features used to train the RF classifier for the photometric prior used in {\tt NWAY}.}

\begin{tabular}{ll}
\hline\hline\noalign{\vskip 1.5pt}
 Feature & Description \\\noalign{\vskip 1.5pt}
 \hline\noalign{\vskip 1.5pt}
 dered\_*\_flux& deredened template flux in \IE, \YE, \JE, and \HE \\
 colour\_* & deredened template colour for \IE$-$\YE, \IE$-$\JE, \IE$-$\HE, \YE$-$\JE, \YE$-$\HE, and \JE$-$\HE  \\
 snr\_* & S/N for template \IE, \YE, \JE, and \HE \\
 point\_like\_prob & probability between 0 and 1 that the source is point-like  \\
 vis\_det & binary flag, where $\begin{cases} 
 \textrm{source is detected in the VIS mosaic} \, (1) \,,\\ \textrm{source is only detected in the NIR mosaic } \, (0)
 \end{cases}$ \\
 \hline         
\end{tabular}
\tablefoot{Deredened fluxes are calculated by correcting for Milky Way (MW) extinction, following the procedure for obtaining intrinsic magnitudes described by \cite{EROPerseusOverview}.}
\label{tab:2}
\end{table*}

Until recently, the identification of CTPs was primarily performed using maximum likelihood matching \citep[e.g.,][]{Sutherland_1992}. This approach considers the separation between sources in the primary catalogue (e.g. X-ray) and the secondary multiwavelength catalogue, the positional uncertainties of the primary sources, and the magnitude distribution of sources in the ancillary catalogues within a certain radius from the primary sources. While this method has been applied successfully in cases with small positional errors and bright sources \citep[see e.g.,][]{Naylor_2013, Brusa_2010,Luo_2011}, its reliability decreases in situations where positional errors are large, and more than one photometric point is necessary for identifying the correct CTP or when multiple potential CTPs exist within the search radius. In such cases, positional matching often fails to provide a robust identification.

\subsection{{\tt NWAY}}
To address these challenges, Bayesian statistics needs to be invoked \citep{Budavari_2008}. With this in mind, algorithms such as {\tt NWAY} \citep{Salvato_2018}, and {\tt XMATCH} \citep{Pineau_2017} were developed. Unlike the others, {\tt NWAY} also allows for the adoption of priors in addition to positional matches between multiple catalogues. It refines its matching procedure by being able to supplement one or more features, such as colours, magnitudes, and S/N, into likelihood ratios used for CTP identification. These priors based on (anti-) correlations between source properties of the target and field populations allow {\tt NWAY} to improve the likelihood of finding true matches, boosting the accuracy significantly over methods that rely solely on positional information.

\subsubsection{The integration of priors in {\tt NWAY}}
\label{priors_in_nway}

Colour and magnitude priors are posteriors that encode knowledge from data or probabilistic distributions that describe the expected properties of true CTPs for the sources in the primary catalogue. For example, an AGN selected in X-rays is likely to have distinct colours (e.g. redder MIR colours due to dust) and magnitudes compared to stars or inactive galaxies. The difference between the methods resides then in the adoption of specific features able to distinguish an X-ray emitter (regardless of its Galactic or extragalactic nature) from a random source in the field \citep{Salvato_22}. {\tt NWAY} uses these priors to add additional dimensions to calculate a more nuanced likelihood ratio. This reduces ambiguity, especially in crowded fields or cases with large positional uncertainties.

For instance, in the ROSAT All-Sky Survey \citep[2RXS,][]{Boller_2016}, \cite{Salvato_2018} improved CTP identification by incorporating a 2D prior based on the W2 magnitude and W1$-$W2 colour from AllWISE \citep{Wright_2010}, effectively separating out AGN. In the more recent application to the eROSITA Final Equatorial-Depth Survey \citep[eFEDS,][]{Brunner2022}, \citet{Salvato_22} defined a prior that utilised not only magnitudes and colours from the Legacy Survey Data Release 10 \citep[LS10,][]{Dey_2019}, but also S/N in all bands (\textit{griz}, W1, W2, W3, and W4), as well as properties, such as proper motion, from \textit{Gaia} DR3 \citep{Vallenari_2023} when available. For further details on the formalism, we refer to the {\tt NWAY} documentation.\footnote{\url{https://github.com/JohannesBuchner/NWAY}} 

\subsubsection{Q1 prior}
\label{Q1_prior}
In this work, we have adopted the same principle as described in the previous subsection, but restricted ourselves to using only photometric features from \Euclid. To build priors based on multiple features, {\tt NWAY} can incorporate probabilities derived from ML models, such as a random forest (RF) classifier \citep[the {\tt sklearn} implementation,][]{Pedregosa_2011}. We start by utilising the training sample presented in Salvato et al., in prep., comprising secure CTPs to X-ray sources from 4XMM DR11 \citep{Webb2020} and CSC\,2.0, for training the RF model, as presented by \citep{Salvato_22}. {\tt NWAY} is then applied to identify CTPs for these X-ray sources in the EDF catalogues. A search radius of three times the positional error threshold is set around each X-ray source to include all potential EDF matches within this radius. This large search radius ensures that widely separated CTPs are not missed, even for sources with the largest positional uncertainties. The sky coverage of each X-ray sample per EDF is calculated to account for overlapping search windows. Specifically, while the coverage for the X-ray sample is given by its multi-order coverage map (MOC) or pointing, the EDF coverage was derived as
\begin{equation}
    A_{\textrm{field}} = N_{\textrm{X-ray}} \, \, \pi \, \, (\rm{POSERR})^{2} - A_{\textrm{overlap}}\,,
\end{equation}
where $N_{\textrm{X-ray}}$ corresponds to the number of X-ray sources and $A_{\textrm{overlap}}$ is the overlapping area of adjacent search windows. Following the methodology outlined in \citet{Salvato_2018}, these coverage areas are used to compute the number densities, which inform two key probabilities: \texttt{p\_any} (the probability that an X-ray source has a CTP) and \texttt{p\_i} (the probability that each EDF source within the search radius is the correct CTP). {\tt NWAY} identifies the most likely CTP for each X-ray source by selecting the EDF source with the highest \texttt{p\_i} value. This CTP is considered the primary match. To assess the reliability of this identification, the \texttt{p\_any} value is used as a confidence threshold. All sources identified as primary matches (\texttt{match\_flag}\,==\,1) with \texttt{p\_any}\,>\,0.85, form the target sample. In addition, for every X-ray source, all other EDF sources within the search radius that are not the primary match are collected into a field population, provided \texttt{p\_any}\,<\,0.1. This field population serves as a representation of non-X-ray sources, providing a comparison set for training the RF. To label the data sets, X-ray CTPs are assigned a target class of `1', while field objects are labelled as `0`. The \Euclid-only training features utilised for the RF classifier are outlined in \cref{tab:2}. At present, photometry and colours from different apertures are not included as features, as the aperture sizes vary between sources despite sharing the same designation.

\begin{figure}[htbp!]
\centering
\includegraphics[angle=0,width=1.0\hsize]{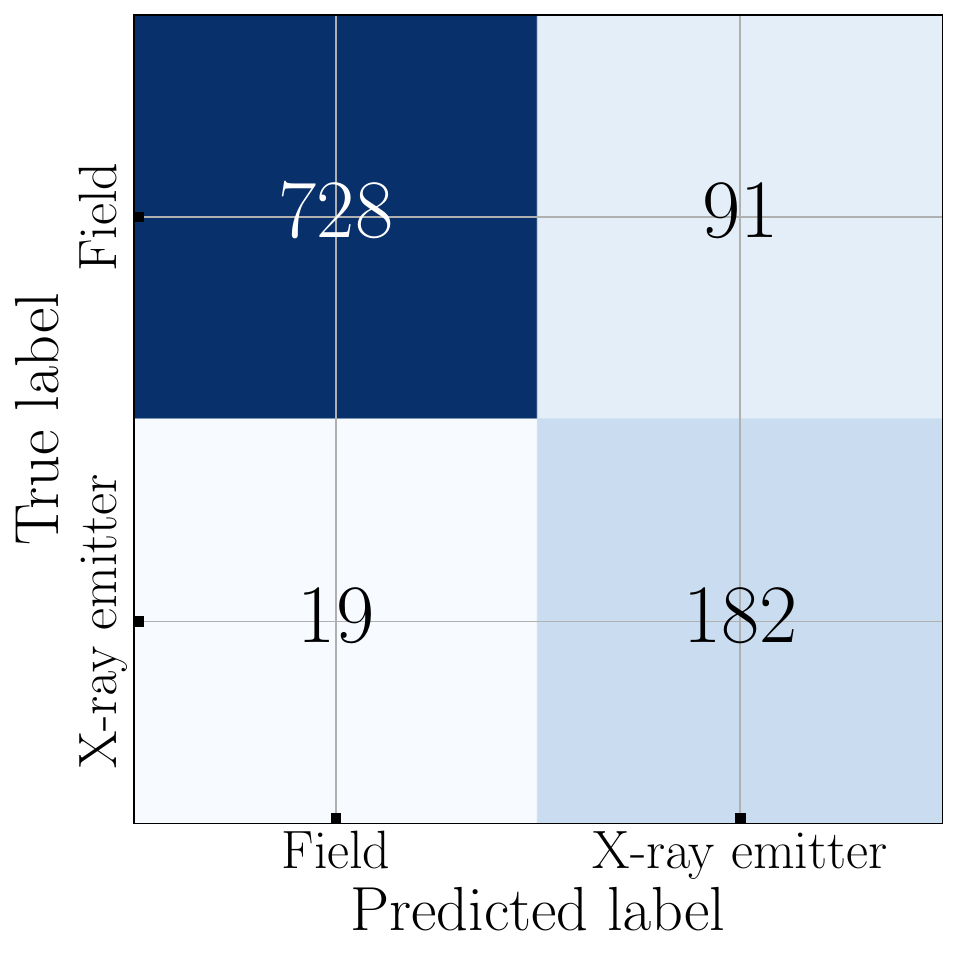}
\caption{Confusion matrix from the RF prediction on an independent test set. X-ray sources are labelled as `X-ray emitter', while field objects are labelled as `Field'. Numbers on the diagonal correspond to correctly predicted classes of TP (bottom right) and TN (top left), while those off-diagonal refer to falsely predicted classes of FP (top right) and FN (bottom left).}
\label{fig:2.1}
\end{figure}

Future data releases may standardise aperture sizes across all objects, also enhancing the RF classifier’s ability to leverage the relationship between light profiles and X-ray emission. To evaluate model performance, 15\% of the training dataset (6000 fields and 1730 target sources) is randomly set aside as a test set. The RF model is configured with 200 decision trees, with splits allowed if at least 25 samples remain in a branch. Per decision tree construction, 10 photometric features are considered, and bootstrap sampling is applied to the training set for building the ensemble of trees. Since the data set is highly imbalanced, with field objects vastly outnumbering CTPs of X-ray sources, a weighting scheme is implemented to automatically adjust the contribution of each class during training \citep{Salvato_22}. This ensures the model effectively learns to differentiate between the two classes despite the imbalance. The classifier outputs a probability score ${\textrm{X-ray}}$, the predicted likelihood of a candidate CTP being X-ray emitting. ${P}_{\textrm{X-ray}}$ is used for class prediction, where {\tt NWAY} retains candidates with ${P}_{\textrm{X-ray}} < 50\%$ if their astrometric configuration strongly supports a match. Importantly, the probability $P_{\textrm{X-ray}}$ is derived exclusively from photometric properties, independent of positional uncertainties, enabling it to be seamlessly integrated into {\tt NWAY}’s Bayesian framework, complementing astrometric priors with photometric information. 

We quantify the classification quality by comparing the input labels and predicted classes as: (i) true positive (TP, where label\,$=$\,1 and prediction\,$=$\,1); (ii) true negative (TN, where label\,$=$\,0 and prediction\,$=$\,0); (iii) false positive (FP, where label\,$=$\,0 and prediction\,$=$\,1); and (iv) false negative (FN, where label\,$=$\,1 and prediction\,$=$\,0). The model's performance is summarised in the confusion matrix shown in \cref{fig:2.1}, assessed by acquiring these values for the test set \citep[see e.g.,][]{Fotopoulou_2018}. We defined the following measures of quality for the classifier:

\begin{itemize}
    \item accuracy\,$=$\,(TP$+$TN)/(TP$+$TN$+$FP$+$FN)\, ;
    \item precision\,$=$\,TP/(TP$+$FP)\, ;
    \item recall\,$=$\,TP/(TP$+$FN)\, ;
    \item fall-out\,$=$\,FP/(TN$+$FP)\, .
\end{itemize}

The trained model achieves strong performance, with a high accuracy of 89\% and a recall of 91\%, indicating that most real X-ray emitters are correctly identified. The contamination rate remains low, with a fractional leakage of 11\%. The fall-out fraction can, in principle, be reduced by including more field objects, though at the price of reduced recall.

To better understand the model's decision-making process, we performed a detailed feature importance analysis. This revealed that while most features contribute marginally to the classification, a few stand out as particularly informative. Specifically, features related to the de-reddened \HE band, including the flux, the S/N, and respective colour combinations, headed by \YE$-$\HE, as well as the point-like probability, emerge as the most significant predictors. For a more comprehensive overview of feature contributions, we refer the reader to \cref{apdx:A1}.

\begin{figure}[htbp!]
\centering
\includegraphics[angle=0,width=1.0\hsize]{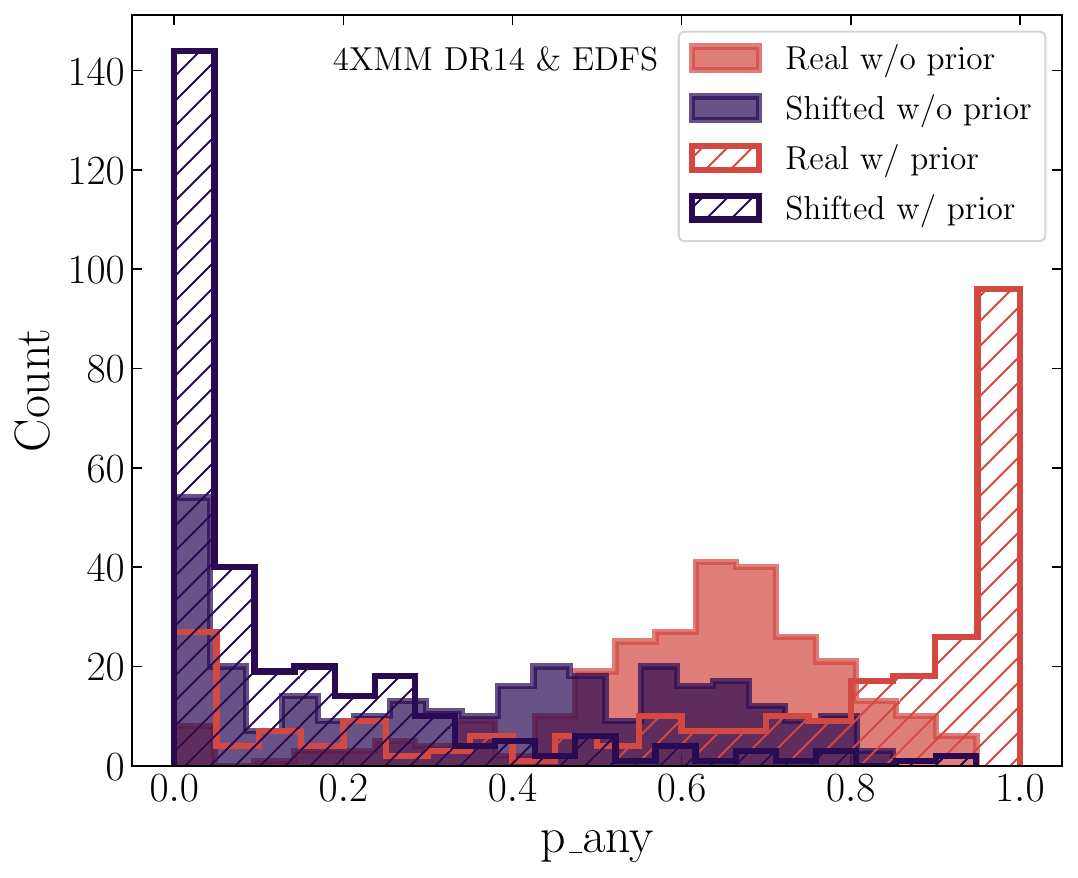}
\caption{Example of \texttt{p\_any}, the probability that an X-ray source has a \Euclid CTP, showing distributions for both random and real X-ray sources before (filled) and after (hatched) incorporating the photometric prior in {\tt NWAY}. For the random sources, the inclusion of the prior shifts the \texttt{p\_any} values to lower values, indicating that most associations are due to chance alignments. Conversely, for the real sources, the prior shifts the \texttt{p\_any} values to higher values, highlighting the increased confidence in the true matches, demonstrating the positive impact of the prior in distinguishing real CTPs from random associations.}
\label{fig:2.3}
\end{figure}

\begin{table*}
\centering
\caption[]{Breakdown of the CTP sample.}

\begin{tabular}{cccccccc}
\hline\hline\noalign{\vskip 1.5pt}
 X-ray catalogue & X-ray sources & CTPs & AND  & AND & AND & $P_{\textrm{Gal}} < 0.5$ & w/ spec-$z$ \\\noalign{\vskip 1.5pt}
 &&&$\texttt{match\_flag} == 1$&in LS10&S/N$_{r}$ >= 3& \\
 \hline\noalign{\vskip 1.5pt}
 4XMM DR14 & 5241 & 5885 & 5239   & 4107 & 4033 &5486 & 953 \\
 CSC\,2.0 & 1669 & 1712 &  1666   & 993 & 958 &1625 & 502 \\
 eROSITA DR1 & 4381 & 5048 & 4381   & 4159 & 4141 & 4375 & 321 \\
 \hline\noalign{\vskip 1.5pt} 
 Total: & 11\,291 & 12\,645 & 11\,286 & 9259 & 9132 & 11\,486 & 1776 \\
    
\end{tabular}

\tablefoot{(1) X-ray catalogue, (2) number of X-ray sources, (3) number of CTP candidates, (4) unique CTPs with \texttt{match\_flag==1}, (5) unique CTPs successfully matched to LS10, (6) unique CTPs in LS10 with S/N\textsubscript{r} $\geq$ 3. The respective number of CTP candidates that we consider to be extragalactic is given in column 7, while the number of candidates with publicly available spectroscopic redshifts is given in column 8.}

\label{tab:3}
\end{table*}

\subsection{Adding the X-ray emitter prior to {\tt NWAY}}

The a priori probability of a \Euclid source being an X-ray emitter, $P_{\textrm{X-ray}}$, is now incorporated as a prior into the {\tt NWAY} framework to enhance our CTP identification process. To visualise the positive impact of the inclusion of this prior, we plot the distribution of \texttt{p\_any} both before and after the inclusion of the X-ray emitter prior (see \cref{fig:2.3}). Incorporating the prior shifts the distribution of \texttt{p\_any} to higher values, making the identification of true CTPs more reliable. Notably, the inclusion of the prior also changes the respective CTP in about 30\% of the cases, indicating that for those sources that are indeed not the CTP after adding the prior, the \texttt{p\_any} value has been decreased. 

\begin{figure}[htbp!]
\centering
\includegraphics[angle=0,width=1.0\hsize]{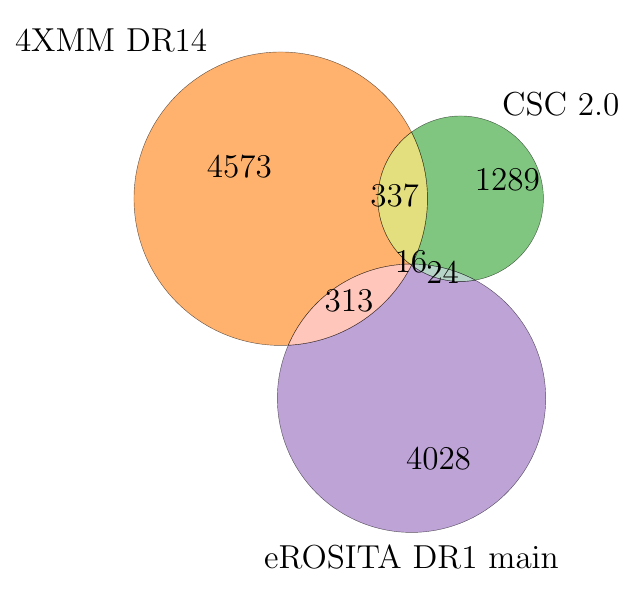}
\caption{Venn diagram illustrating the overlap of shared \Euclid CTPs among the three X-ray samples, 4XMM DR14, CSC\,2.0, and eROSITA DR1 main, as listed in \cref{tab:3}. The diagram shows the distribution of sources that are uniquely or jointly identified by the surveys, with larger overlapping areas indicating stronger agreement among the surveys.}
\label{fig:7}
\end{figure}

In total, we associated 12\,645 CTP candidates. Among these, 11\,286 sources (about $90\%$) have $\texttt{match\_flag} == 1$, indicating that they have the highest probability $\texttt{p\_i}$ to be considered reliable CTPs. The remaining candidates, with $\texttt{match\_flag} == 2$, represent secondary but plausible matches that are not ruled out entirely. We found 1092 X-ray sources with multiple CTPs, the vast majority of which are made up by \XMMN and eROSITA, with only \textit{Chandra} providing the required positional accuracy for \Euclid to have a few sources with multiple plausible CTPs. While most of the multi-CTP cases are made up of two candidates, a few cases with more than two CTPs are introduced, predominantly by eROSITA, given its larger positional uncertainty. Notably, only five X-ray sources within the Q1 footprint (see \cref{tab:1}) lack an association entirely, possibly due to boundary effects where the true CTP falls outside the region covered by the EDFs. 

With this catalogue, several scenarios emerged in terms of the relationships between CTPs and X-ray sources:
\begin{enumerate}
    \item Two X-ray entries from different surveys are assigned to a single CTP. This scenario suggests but does not confirm the possibility of a single physical source being detected by two different X-ray surveys. 
    \item Two X-ray entries from the same survey are assigned to a single CTP. Here, we can confidently state that two distinct physical sources are selected, though both are associated with the same \Euclid ID.
    \item Cases where a single \Euclid ID is linked to multiple entries from two or even all three X-ray surveys.
    \item Two X-ray entries where the same X-ray source, detected by multiple surveys, is matched to distinct \Euclid CTPs.
\end{enumerate}

Among the surveys, only eROSITA uniformly covers two of the three fields. In contrast, other surveys consist of pointed observations with differing depths and catalogues across distinct bands, meaning sources are not always detected at the same flux levels. As a result, the overlap between catalogues is complex and interpreting source matches becomes challenging. At this stage, gaps in coverage further limit the significance of overlap checks. Keeping this in mind, we find 674 pairs which either correspond to case one or two, as well as 16 triplets referring to case three, as shown in \cref{fig:7}. We deliberately choose not to quantify case four, as there is currently no definitive way to resolve it with certainty. A breakdown of the CTP sample is given in \cref{tab:3}. 

\begin{figure}[htbp!]
\centering
\includegraphics[angle=0,width=1.0\hsize]{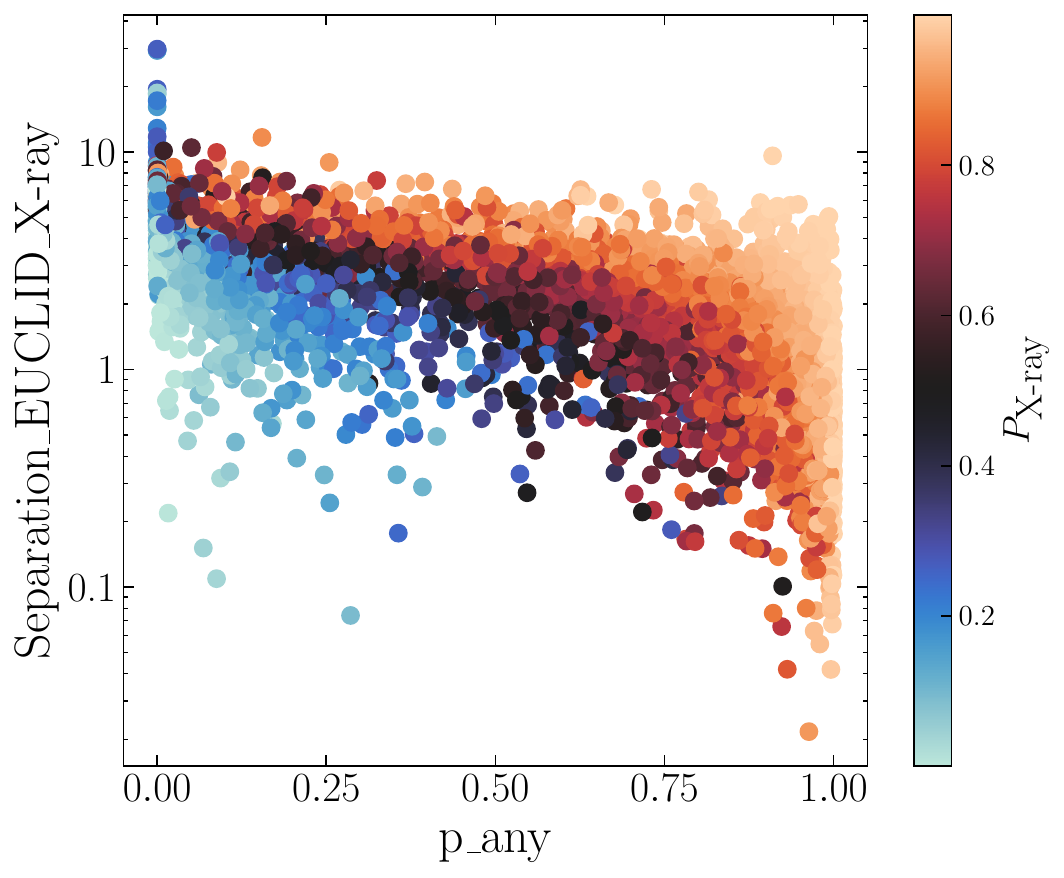}
\caption{Relationship between the separation (in arcseconds) between X-ray sources from 4XMM DR14 and their selected CTP as a function of \texttt{p\_any} colour-coded by $P_{\textrm{X-ray}}$. The plot highlights that a minimal (or very small) fraction of matches have separations above 10\arcsec, and \texttt{p\_any} is enhanced by the probability to be X-ray emitting.}
\label{fig:2.4}
\end{figure}

\begin{figure*}[htbp!]
\centering
\includegraphics[angle=0,width=1.0\hsize]{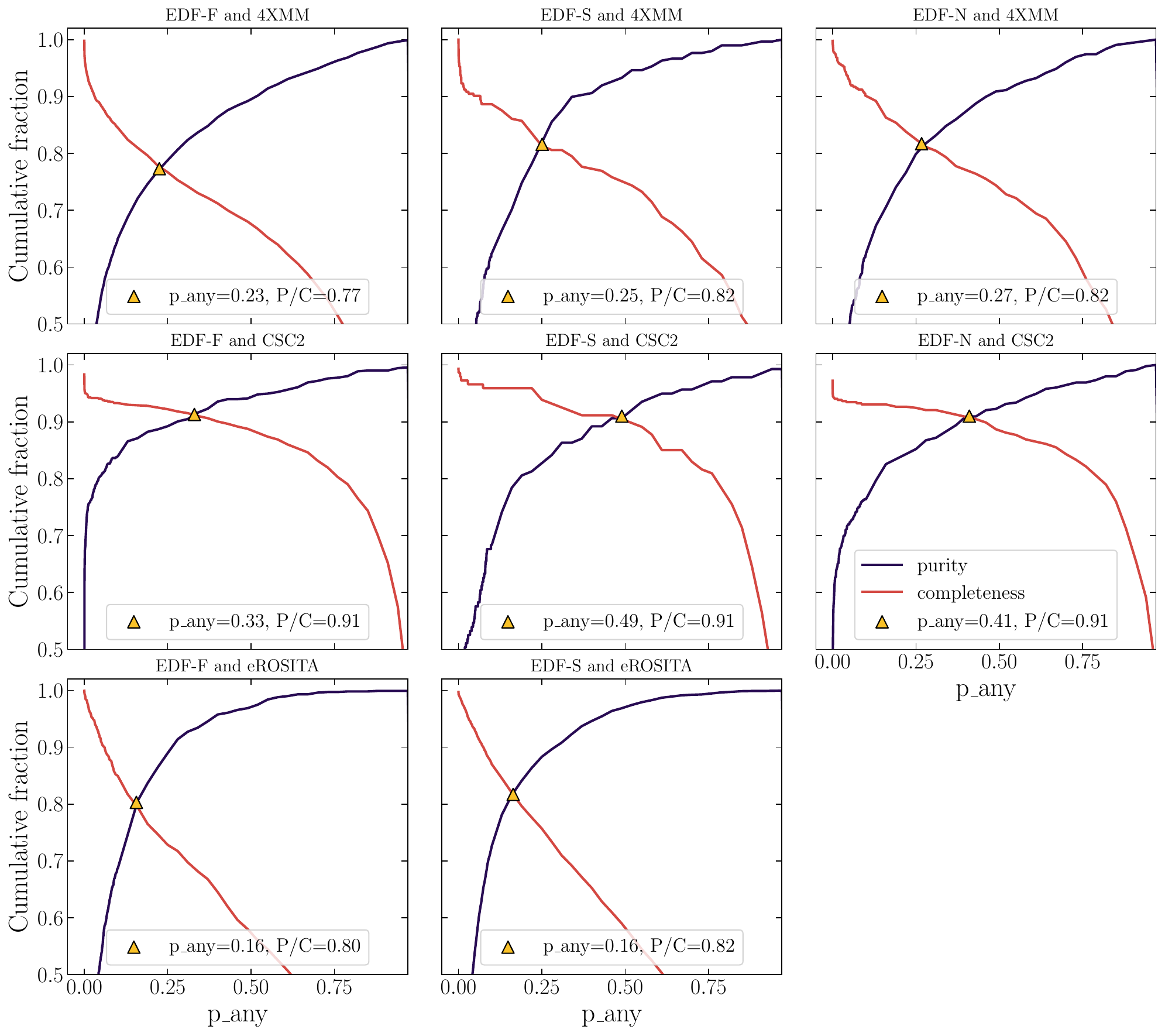}
\caption{Purity (red) and completeness (purple) as a function of \texttt{p\_any} for matches between the different EDFs and the three X-ray catalogues: 4XMM DR14 (top row), CSC\,2.0 (middle row), and eROSITA DR1 (bottom row). Based on these figures, users can adopt the \texttt{p\_any} value threshold that enhances purity or completeness, depending on their scientific preference. The intersection point serves as a threshold for the \texttt{p\_any} selection.}
\label{fig:2.6}
\end{figure*}

\subsection{Purity and completeness}

To calibrate a reliable \texttt{p\_any} cut-off, we performed the {\tt NWAY} procedure (using the same prior) on a duplicate sample of the X-ray sources, referred to as `randoms'. These randoms were generated by displacing the positions of the X-ray sources to lie beyond the search radius, ensuring that any matches with multiwavelength CTPs are purely coincidental. By comparing the distribution of \texttt{p\_any} values for both the original X-ray sample and the randoms, we could assess the impact of the prior by examining the two histograms in \cref{fig:2.3}. The figure shows how the \texttt{p\_any} value for the randoms is usually lower than for the real sample of X-ray sources, indicating that in a random position in the sky, there are no sources that have the same features as a typical X-ray emitter. For the few sources for which \texttt{p\_any} remains high, either the prior was not representative enough, or, deeper data in the future will potentially reveal a fainter X-ray source.

To further assess the impact of incorporating $P_{\textrm{X-ray}}$ into the {\tt NWAY} procedure, we compared the \texttt{p\_any} values as a function of the separation distance between the primary (X-ray) and secondary (multi-wavelength) sources. \Cref{fig:2.4} presents the separation between the 4XMM sources and the associated CTP in EDF-F as a function of \texttt{p\_any}. The sources with a higher probability of being correct also have a high probability of being X-ray emitters, even at larger separations, the latter being mostly smaller than 10\arcsec. CTPs of little separation and low \texttt{p\_any} showcase scenarios of chance alignments where objects appear close but do not show significant $P_{\textrm{X-ray}}$. Next, we focused on evaluating the purity and completeness of the CTP samples based on how they are sub-selected. In this context, we defined completeness as the cumulative fraction of sources with \texttt{p\_any} higher than a certain value and purity as the fraction of sources with \texttt{p\_any} higher than the value in the random sample. This gives a measure of how often the CTP in the real sample is associated to a CTP only by chance.

\Cref{fig:2.6} shows both the purity and completeness for each EDF and X-ray survey as well as the `sweet spot' for the best trade-off between these quantities as a function of \texttt{p\_any}. Variations in the distributions result from differences in the positional uncertainties of the X-ray surveys and the source densities within both the X-ray data and the corresponding EDF. Depending on the scientific goal, a user interested in creating a catalogue that is extremely complete or pure is hence free to select sources with a \texttt{p\_any} beyond a certain threshold. For the rest of the work, we adopt the intersection between the purity and completeness curves as the threshold. In all fields, purity and completeness are at least 80\%, with the \textit{Chandra} catalogues unsurprisingly having the most reliable associations, reaching values of about 90\%. Interestingly, the impact of the higher spatial resolution of \textit{Chandra} is clear in the EDF-F field, where Fornax, the second richest galaxy cluster of the local Universe, is located. In such a dense environment, it is more challenging to identify the correct CTP given the resolution of 4XMM and eROSITA.

\begin{figure}[htbp!]
\centering
\includegraphics[angle=0,width=1.0\hsize]{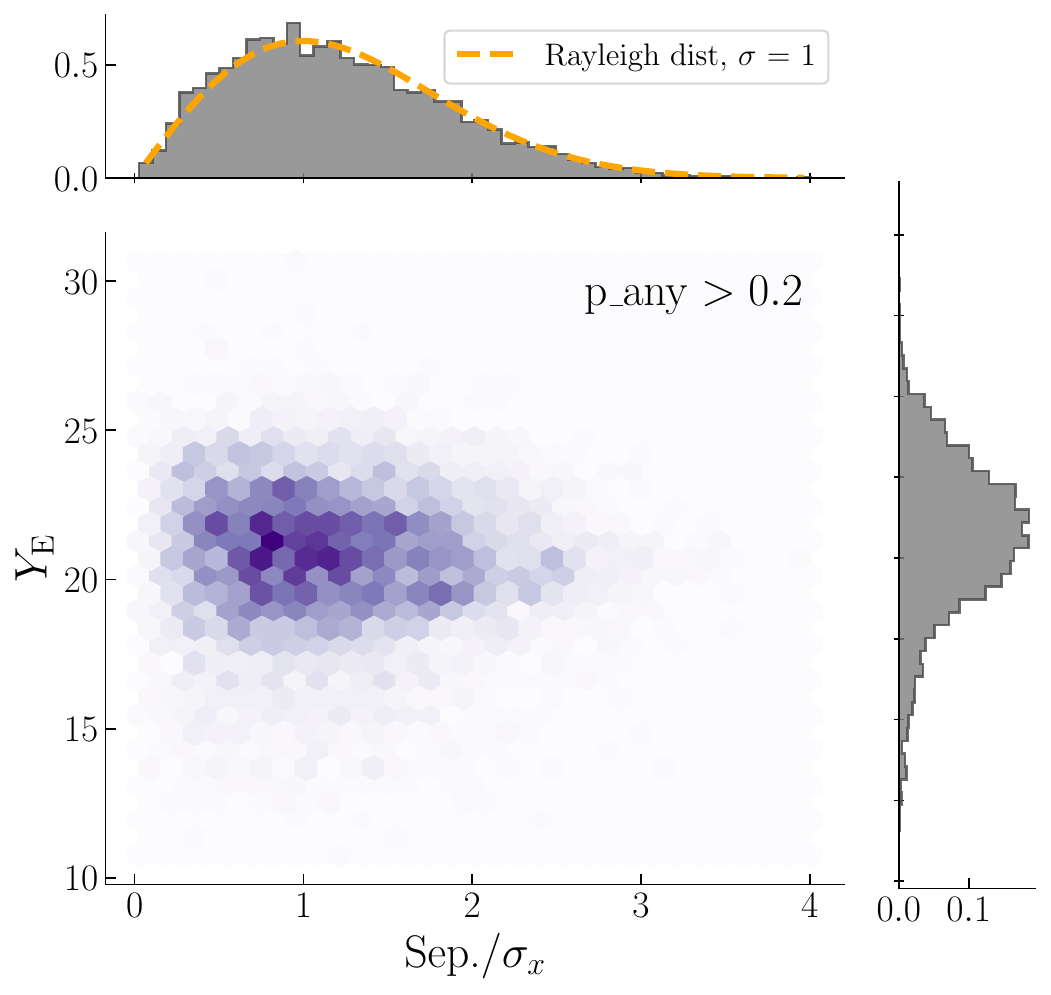}
\caption{Separation between the X-ray position and the selected CTP, where the \YE magnitude for sources with CTPs ({\tt p\_any}\,>\,0.2) is plotted against the normalised  1-dimensional positional error of the X-ray sources. The hexagonal bins are colour-coded linearly based on the count of sources within each bin. Marginal histograms show the distribution along the axes, with a linear y-axis scale. The expected 1\,$\sigma$ Rayleigh distribution for the normalised separations is overlaid in orange.}
\label{fig:11.1}
\end{figure}

To conclude this part, we assessed the stability of the method by examining the impact of RF hyperparameters such as the minimum leaf size and features per tree. Variations in these parameters mainly shift the intersection of purity and completeness curves along the \texttt{p\_any} axis, with smaller leaf sizes enhancing subclass specificity and subtly altering the intersection point.

\subsection{Separation and magnitude distribution of CTPs}
\label{Separation and magnitude distribution of CTPs}
The distribution of observed X-ray-optical separations, normalised by the X-ray positional uncertainty, is presented in \cref{fig:11.1} as a function of the CTP's \YE-band magnitude. For a quality-selected sample of \texttt{p\_any}\,$>$\,0.2, this distribution well aligns with the expected Rayleigh distribution for a scale factor of $\sigma=1$, indicating consistent behaviour with statistical expectations, demonstrating the reliability of the matching process \citep[e.g.,][]{Salvato_22}. The bulk of the sources have a mean separation of 1.\arcsec22\, and \YE\,=\,20.65.

\section{Characterisation and classification of CTPs}
\label{sec6}

Once the CTPs have been identified, it is essential to categorise the sources in order to study the underlying physical processes and populations. The primary distinction is between extragalactic sources (such as galaxies or faint and bright AGN) and galactic sources (including stars and compact objects). In the following section, we outline the methodology used for classifying the sources and the validation tests conducted to ensure the accuracy of the classification, the results of which are summarised in \cref{tab:3}.

\begin{figure}[htbp!]
\centering
\includegraphics[angle=0,width=1.0\hsize]{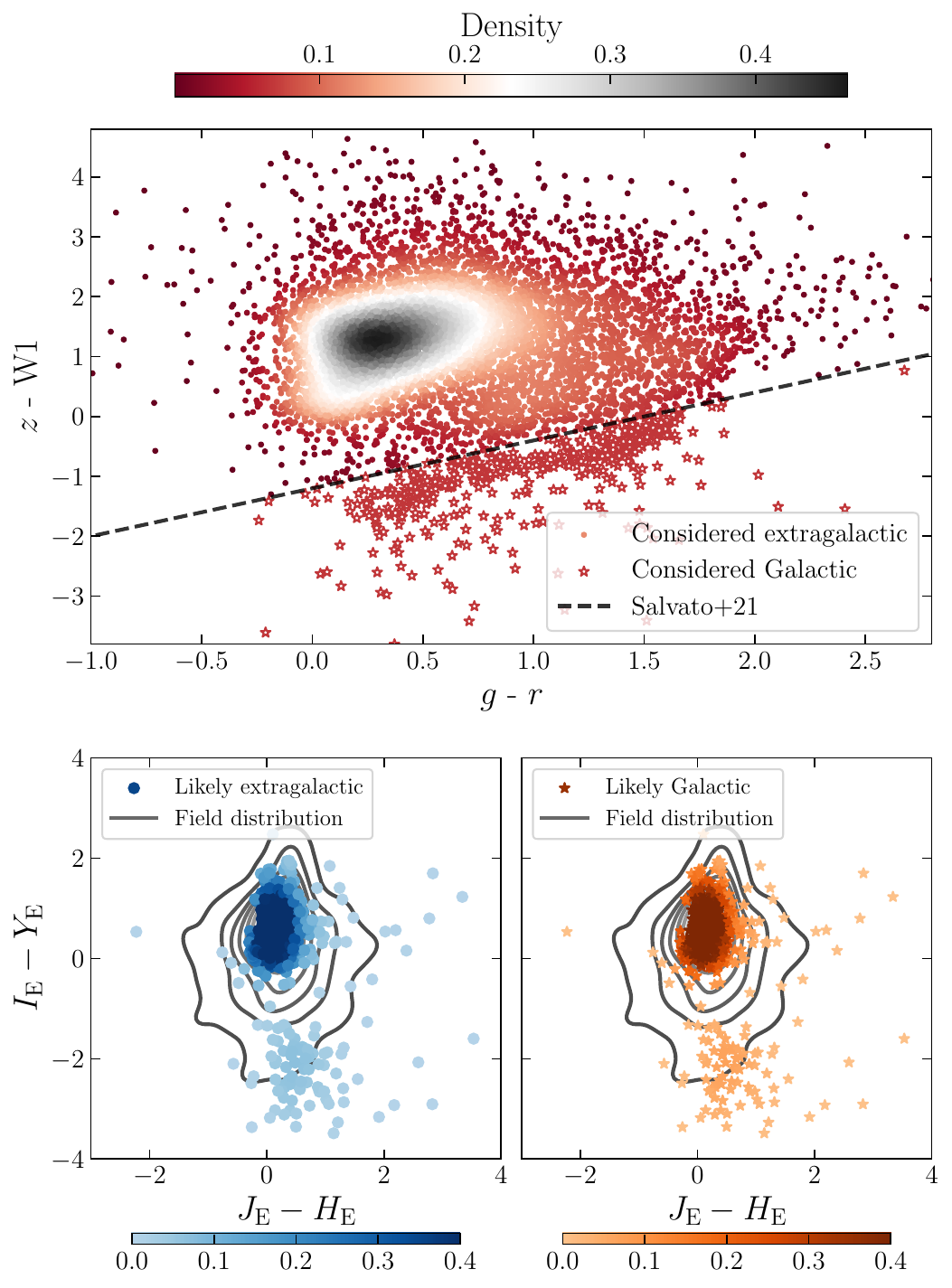}
\caption{\textit{Upper row:} Colour-colour plot showing the positional matched \Euclid CTPs to LS10 sources with `good' photometry (S/N\,>\,3), adapted from \protect\cite{Salvato_22}. This distribution is used to segment sources into Galactic and extragalactic classes. \textit{Lower row:} \Euclid colours for extragalactic (left) and Galactic (right) distributions plotted on top of the contours of \Euclid field sources. The plots confirm that \Euclid colours cannot differentiate between Galactic and extragalactic sources since both distributions are very similar.}
\label{fig:gal_exgal}
\end{figure}

\subsection{Galactic and extragalactic sources }
\label{Galactic and extragalactic sources}
The four \Euclid bands alone lack the discriminatory power required to reliably separate Galactic from extragalactic sources, given the broadness of the optical filter, where most of the differences within the SED occur. As demonstrated in \cite{Bisigello_2024}, no combination of \Euclid-only photometry achieves satisfactory fractions of purity and completeness for this task. Moreover, the ground-based photometry added to the Q1 catalogues is not completely cross-calibrated at this stage, and differences in quality and bands available differ from field to field. For this reason, we decided to use the photometry available in LS10, providing us with well-calibrated and reasonable depth in $\it{griz}$ from DECam Legacy Survey observations, including data from the Dark Energy Survey \citep[DES,][]{Abbott2016}, Beijing-Arizona Sky Survey \citep[BASS,][]{Zou17}, and the Mayall $z$-band Legacy Survey \citep[MzLS,][]{Silva2016}, as well as W[1,2,3,4] from the Near-Earth Object Wide-field Infrared Survey Explorer  \citep[NEOWISE,][]{Mainzer2011,Lang2014,Meisner2017}.
We match the 11\,286 CTPs to LS10 using a $1\arcsec$ positional tolerance and find an association for 82\% of our sources, see \cref{tab:3}. Next, we apply a cut of S/N\,>\,3 in the $g$, $r$, $z$, and W1 bands to minimise the relative impact of photometric uncertainties. The selected sources are then plotted on the $z-$W1 versus $g-r$ colour-colour plane, to distinguish between Galactic and extragalactic objects (see \cref{fig:gal_exgal}). Using the separation line defined in \cite{Salvato_22}, we preliminarily classify sources based on their location relative to the threshold. Objects above the threshold are flagged as extragalactic, while those below are classified as Galactic.

\begin{figure}[htbp!]
\centering
\includegraphics[angle=0,width=1.0\hsize]{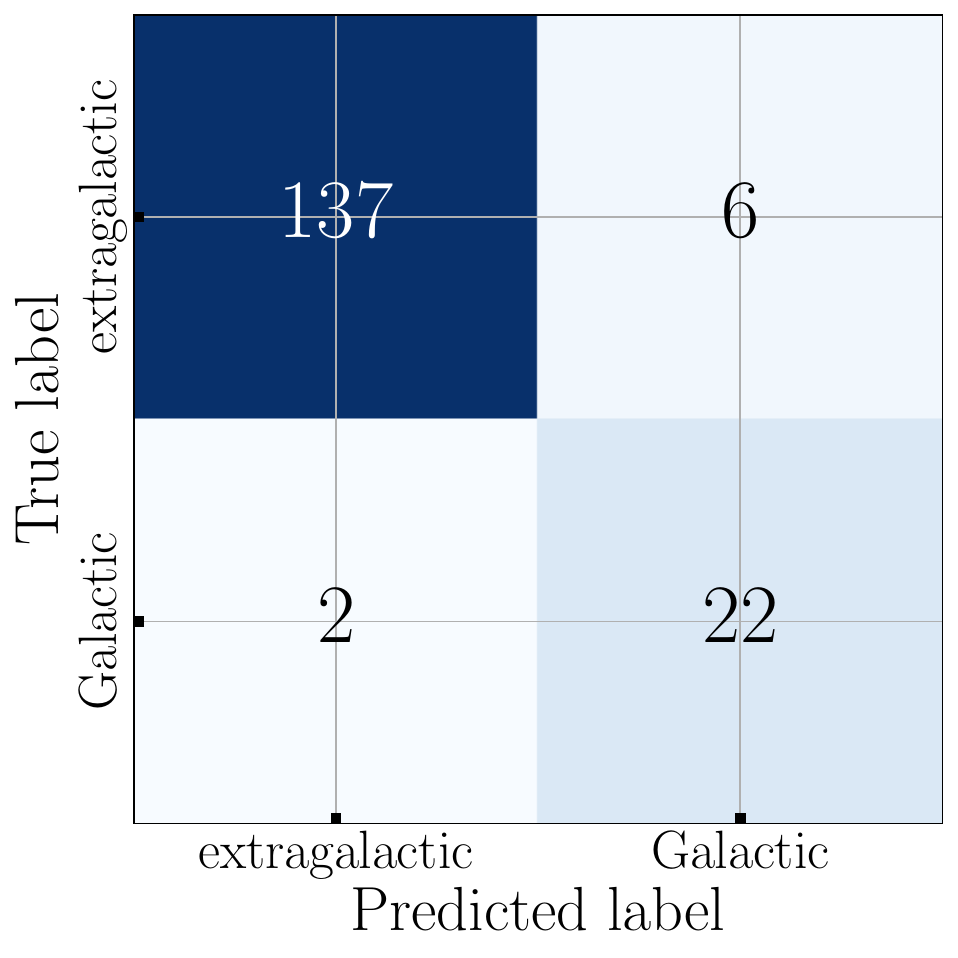}
\caption{Confusion matrix from the RF Galactic-extragalactic classifier applied to an independent test set. Sources classified as residing within the Milky Way are labelled as `Galactic', while field objects are labelled as `extragalactic'.}
\label{fig:4.1}
\end{figure}

These classifications are then used to create individual subplots (lower panels of \cref{fig:gal_exgal}), displaying the sources by their classification using only \Euclid colours. As highlighted in \cite{Bisigello_2024}, the Galactic and extragalactic sources trace almost the same feature space, confirming that \Euclid photometry alone, comprising the \IE, \YE, \JE, and \HE bands, does not provide sufficient information to reliably distinguish between the two populations.

\subsection{Probabilistic classifier approach}
\label{Q1_prior_exgal}

To overcome this limitation, we followed the methodology outlined in \cref{priors_in_nway}, training an RF classifier using the same \Euclid-only features listed in \cref{tab:2}. However, this time we labelled the sample based on the classification derived from the $z - \rm{W1}$ versus $g - r$ plot, setting Galactic objects (396) to `1` and extragalactic objects (1000) to `0`. Using the trained model, we assigned a Galactic or extragalactic label to each \Euclid CTP, providing a probabilistic classification, ${P}_{\textrm{Gal}}$, for robust downstream analyses.

\begin{figure}[t!]
\centering
\includegraphics[angle=0,width=1.0\hsize]{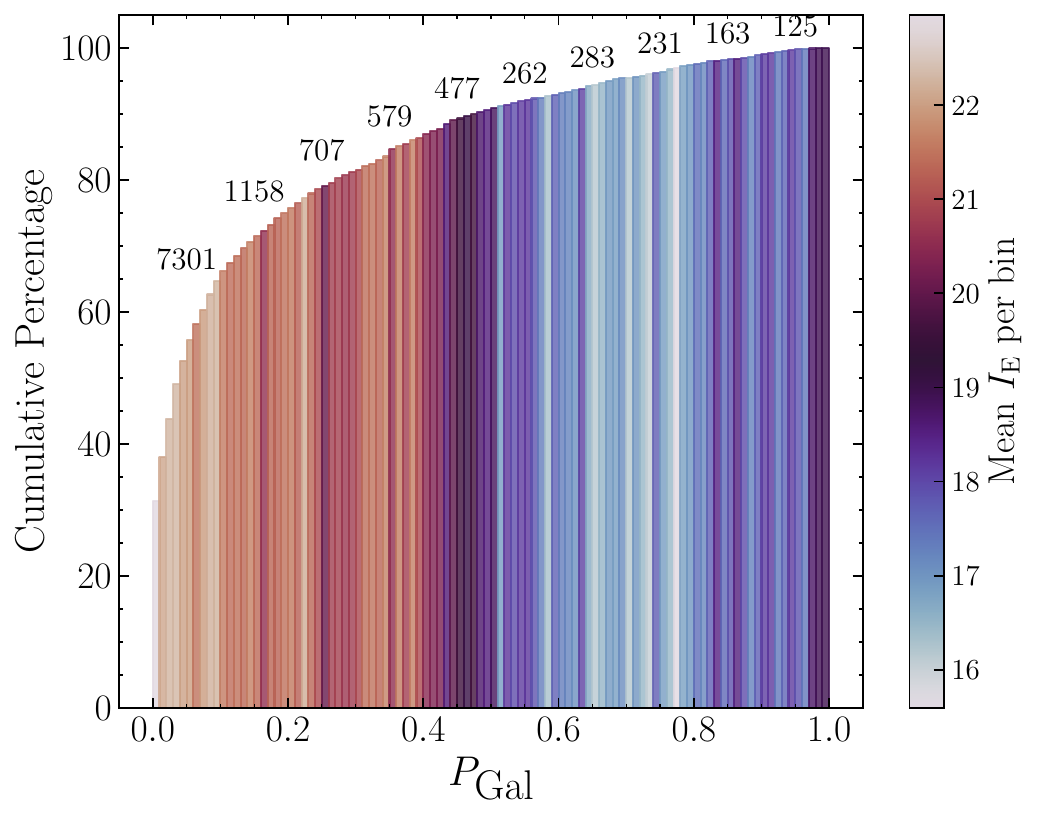}
\caption{Cumulative histograms showing the respective percentage of CTPs as a function of ${P}_{\textrm{Gal}}$. Values of ${P}_{\textrm{Gal}}$\,<\,0.5 correspond to sources classified as extragalactic, while values above 0.5 classify them as being Galactic. Only roughly 8\% of sources have ${P}_{\textrm{Gal}}$\,>\,0.5, indicating a predominantly extragalactic population. In addition, each bin of the histogram is coloured by its mean \IE magnitude, while the sum of each set of ten adjacent bins is printed on top.}
\label{fig:5.3}
\end{figure}

We present the respective confusion matrix in \cref{fig:4.1} and note a recall of 92\%, with a limited fall-out of 4\%. For more details on the feature correlations, see \cref{apdx:A1}. To further validate the RF performance, we examine the galactic probability distribution (${P}_{\textrm{Gal}}$) for the unique CTP candidates sample. Less than 10\% of sources have ${P}_{\textrm{Gal}}>0.5$, indicating effective separation of Galactic and extragalactic populations. A comparison with the \cite{Q1-SP027} classification reveals that 97\% of sources classified as Galactic there, also have ${P}_{\textrm{Gal}}>0.5$, confirming good agreement. Additionally, the colour-coding of \cref{fig:5.3} reveals that most sources classified as being Galactic make up the bright end of the sample, with the most intense colour gradient surrounding ${P}_{\textrm{Gal}} = 0.5$ and a couple of sources towards high ${P}_{\textrm{Gal}}$ interestingly appearing increasingly fainter again, probably due to different populations of stars, subject to their location in the extragalactic sky.

\begin{figure*}[htbp!]
\centering
\includegraphics[angle=0,width=1.0\hsize]{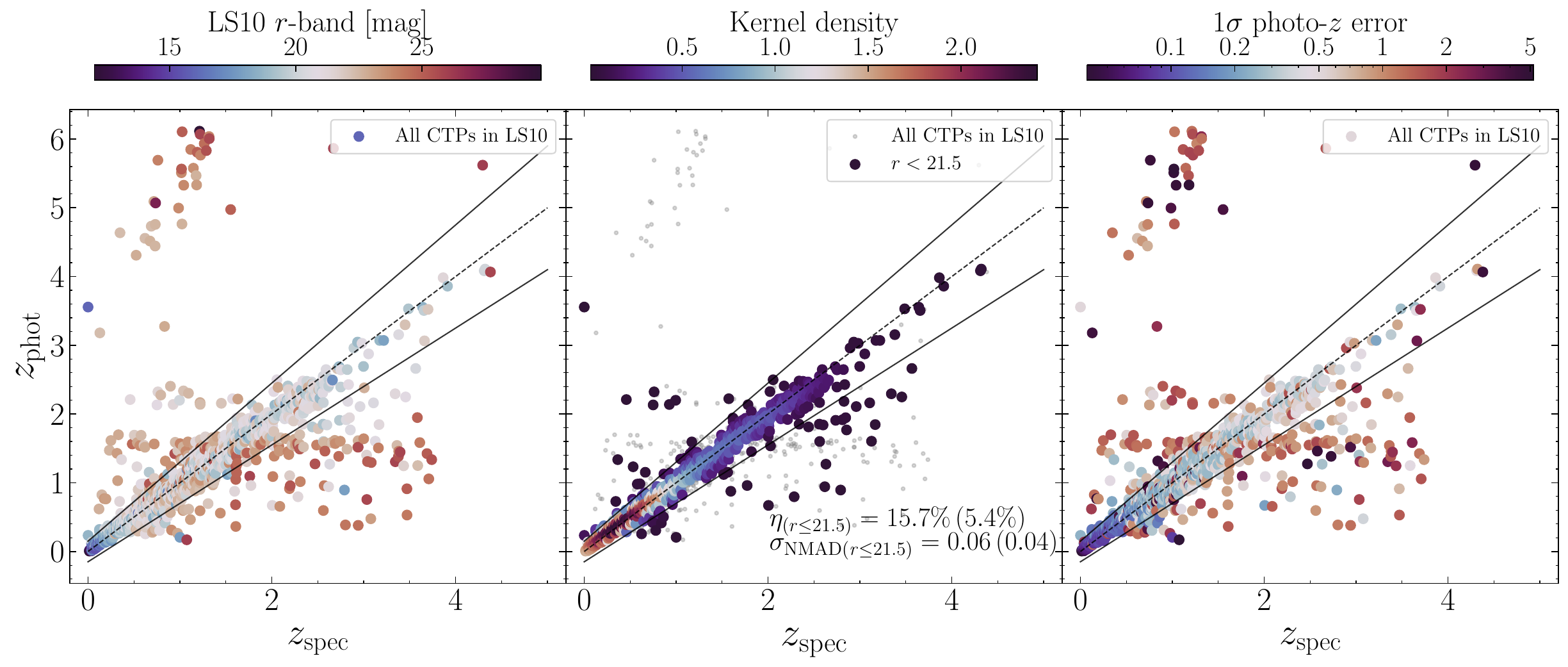}
\caption{\textit{Left}: Photo-$z$ vs spec-$z$ for all CTPs matched to LS10 colour-coded by the LS10 $r$-band magnitude. \textit{Centre}: Same distribution colour-coded according to the kernel density for sources with $r \leq 21.5$. \textit{Right}: Same distribution colour-coded by the $1\sigma$ photo-$z$ error ($z_{\rm phot,1su} - z_{\rm phot,1sl}$). Sources closely following the identity line exhibit significantly lower errors as opposed to outliers, reflecting the photometric uncertainties of faint sources in LS10.}
\label{fig:zp_zs}
\end{figure*}

\section{Redshifts}
\label{sec7}
The catalogues of CTPs we make available in this paper include spectroscopic redshifts (spec-$z$s), where available. This was done by matching their coordinates to a compilation of publicly available redshifts \citep[][Igo et al. in prep.]{kluge2024}. However, this compilation is rich in duplications; hence, the details of the cleaning are described in Sect.\,3.1 of \cite{saxena2024}. While spec-$z$s provide highly accurate distance measurements, they are inherently limited to a subset of typically brighter objects. As a result, only 16\% of the CTP sample benefits from publicly available spec-$z$ coverage (refer to \cref{tab:3}).

\subsection{Photometric redshifts}
\label{sec:photoz}
To achieve a more complete characterisation of the CTPs, photometric redshifts (photo-$z$) are essential. However, achieving reliable photo-$z$s for AGN remains significantly more challenging than for inactive galaxies, despite notable advances in recent years \citep[see][for a review]{Salvato_2019}. The current Q1 photo-$z$s \citep{Q1-TP005}, derived using {\tt{Phosphoros}} (Paltani et al. in prep.), are estimated from total fluxes and fine-tuned for inactive galaxies. This approach neglects the photometric signatures of active nuclei, whose emission can dominate the observed flux and hide key host galaxy features, crucial for breaking colour-redshift degeneracies \citep{Salvato_2019}. Instead, studies have demonstrated that incorporating aperture or pixel-based analyses can significantly enhance the accuracy of such photo-$z$ estimates. Consequently, to compute reliable photo-$z$s for AGN-dominated objects in a dedicated effort, we utilise {\tt{PICZL}} \citep{Roster24}, an ML algorithm that significantly improves photo-$z$ estimation for AGN by directly predicting redshifts from imaging, eliminating the need for manual feature extraction or combination of various surveys. Building on its predecessor, {\sc{Circlez}} \citep{saxena2024}, which analysed aperture-based photometric variations, {\tt{PICZL}} advances to pixel-level resolution, leveraging the raw images for improved precision. As a result, for sources bright enough to successfully match the LS10 catalogue, we are able to produce reliable photo-$z$s (refer to \cref{tab:3}). As in the original work, the uncertainty of the photo-$z$s in terms of 1$\sigma$ error increases with the faintness of the sources \citep[see \cref{fig:zp_zs} and][for more details]{Roster24}. 

\begin{figure}[htbp!]
\centering
\includegraphics[angle=0,width=1.0\hsize]{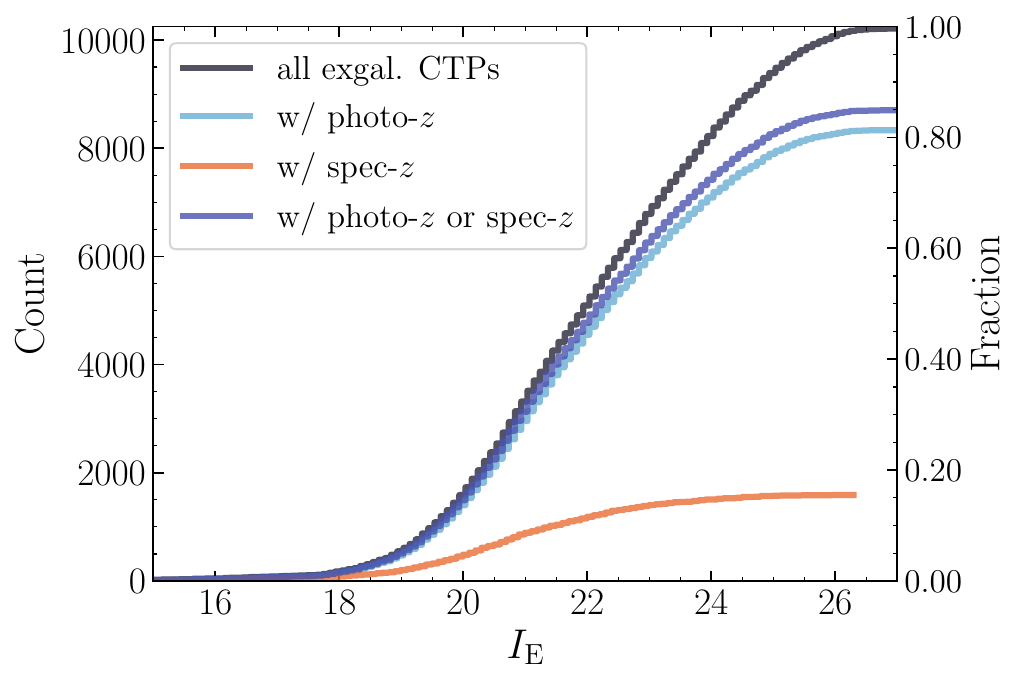}
\caption{Cumulative histogram showing the number of extragalactic CTPs according to ${P}_{\textrm{Gal}}\leq 0.5$ with spec- and/or photo-$z$ as a function of their \IE magnitude.}
\label{fig:5.1}
\end{figure}

\begin{figure*}[htbp!]
\centering
\includegraphics[angle=0,width=1.0\hsize]{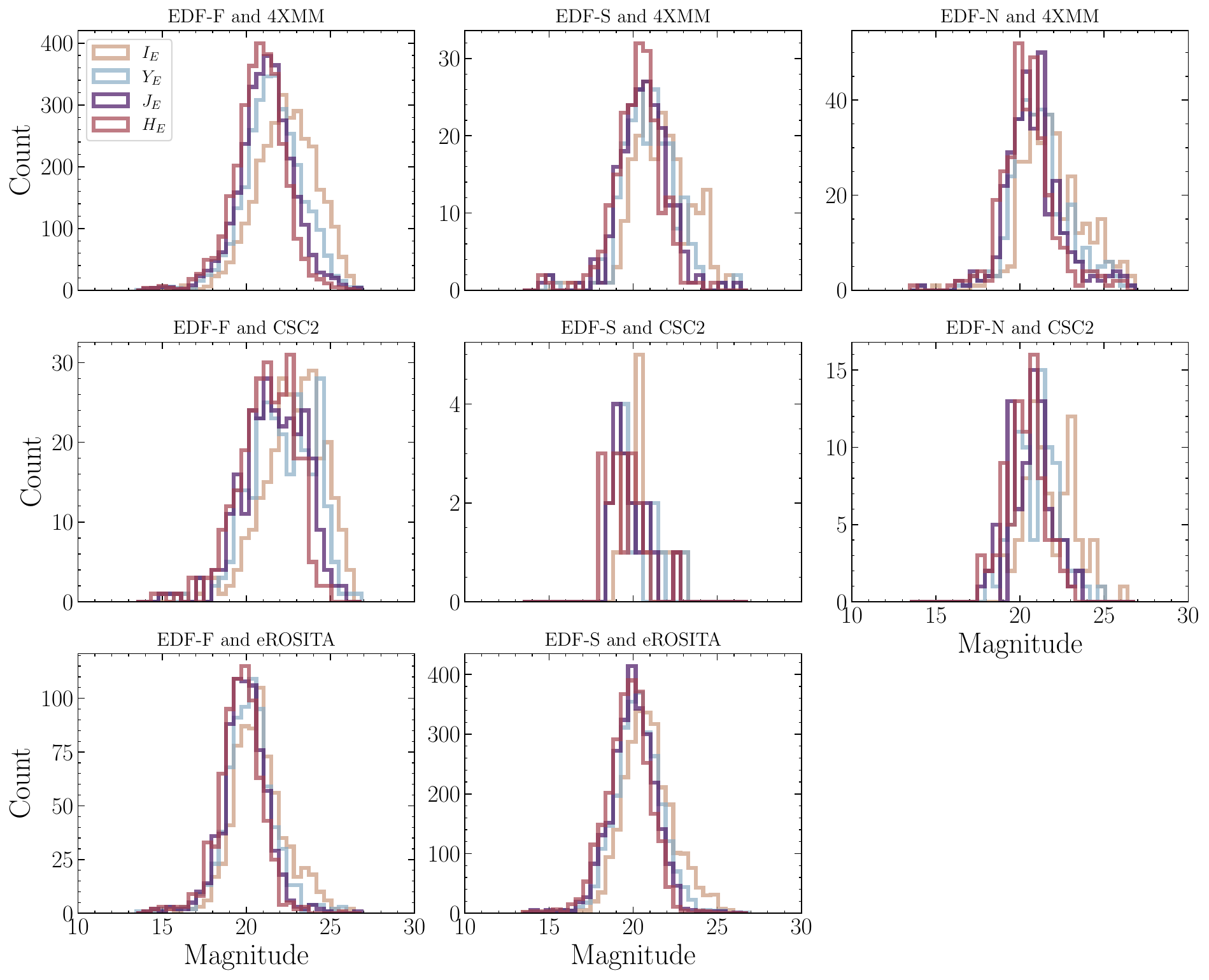}
\caption{Distribution of source counts for each combination of EDF (EDF-F, EDF-S, and EDF-N) and X-ray survey (4XMM on top row, CSC\,2.0 in the middle one, and eROSITA DR1 in the bottom panels). Each plot presents the histograms corresponding to the four \Euclid bands, depicting the count of sources as a function of magnitude.}
\label{fig:6}
\end{figure*}

This study does not incorporate \Euclid photometry in the photo-$z$ estimation, as doing so would require a dedicated pipeline, including re-training on \Euclid imaging and assembling a spectroscopic AGN training set from the Q1 field. Given the small and incomplete nature of such a sample, especially at the faint end, this effort is deferred until DR1 becomes available. An additional challenge lies in the heterogeneous photometric coverage across Q1: while \Euclid provides uniform measurements in \IE, \YE, \JE, and \HE, these are complemented by ground-based data of varying depth and filter sets, leading to inconsistencies in photo-$z$ quality. Consequently, to ensure robustness and comparability, we opted to rely on established methods, with {\tt{PICZL}} offering high completeness and reliable redshifts for $\sim$ 82\% of the X-ray sources detected in LS10\footnote{Future {\tt{PICZL}} versions will make use \Euclid imaging, possibly combined with LSST \citep{Ivezi2019}.} (see \cref{fig:5.1}). Nonetheless, we recognise the potential of \Euclid photometry to enhance our performance, particularly for faint sources and at high redshifts \citep{Graham_2020}. As illustrated in the left panel of \cref{fig:zp_zs}, numerous faint sources are dispersed horizontally around $z_{\rm{phot}} \sim 1.6$, as no strong spectral breaks can be captured by the Legacy filters around this redshift \citep{Roster24}. Access to \Euclid’s NIR bands \YE, \JE, and \HE will enable more accurate redshift recovery in this range.

For all EDF CTPs of an X-ray source detected in LS10, {\tt{PICZL}} provides full redshift posterior distributions (PDZ), derived from de-reddened calibrated images in the $g$, $r$, $i$, and $z$ bands, supplemented by $W1$, $W2$, $W3$, and $W4$ catalogue-based aperture photometry without imposing any S/N cuts. The method is agnostic to the morphological classification of the source, applying equally to extended and point-like objects. As output, {\tt{PICZL}} provides the dominant mode of the PDZ as a point prediction, along with uncertainties quantified as 1$\sigma$ confidence intervals. Notably, we compute photo-$z$s also for those CTPs matched to LS10, which we believe to be Galactic. This approach accounts for the inherent uncertainty in the probabilities produced by our model and classification threshold. To ensure flexibility for users, we include both classifications and photo-$z$ estimates in the released catalogue (see \cref{sec9}). This allows users to tailor their selection criteria to their preferences while retaining access to the full set of photo-$z$ values.

\subsection{Photo-\textit{z} quality}

To evaluate the reliability of our photo-$z$ estimates, we employed standard metrics commonly implemented in the literature, including (a) the accuracy $\sigma_{\textrm{NMAD}}$ as a measure of the scatter between the prediction and truth values of the sample, which is defined as
\begin{equation}
    \sigma_{\rm{NMAD}} = 1.4826 \, \, \textrm{median} \, \biggr[\frac{|z_{\textrm{phot}}-z_{\textrm{spec}}|}{(1+z_{\textrm{spec}})}\biggr] \,;
\end{equation}
and (b) the fraction of outliers, $\eta$, that quantifies the proportion of sources whose redshift estimates deviate significantly from their corresponding spectroscopic redshift. Specifically, this is defined as the fraction of objects for which \begin{equation}
    \eta = \frac{|z_{\textrm{phot}}-z_{\textrm{spec}}|}{(1+z_{\textrm{spec}})}>0.15 \, .
\end{equation} 
 
We evaluated the performance of the photo-$z$s by comparing them to the spec-$z$s, as shown in the middle panel of \cref{fig:zp_zs}. For the subsample of bright sources with $r$\,<\,21.5, corresponding to those that have smaller photometric errors within our dataset, we find a fraction of outliers of 5.4\%, with a $\sigma_{\rm{NMAD}}$ of roughly 0.04 as opposed to $\eta = 15.7\%$ and $\sigma_{\rm{NMAD}} = 0.06$ observed for the entire sample. The catastrophic outliers at $z_{\rm phot}>4$ appear as faint ($r$\,>\,24), galaxy-dominated objects with larger photometric errors in LS10 and highly degenerate PDZs. In scientific studies involving photo-$z$s, it is crucial to account for such uncertainties and in turn, photo-$z$ algorithms should provide reliable error estimates. {\tt{PICZL}} achieves this effectively as shown in the right panel of \cref{fig:zp_zs}, with small errors closely following the identity line and larger ones capturing photometric uncertainties. Despite occasional higher uncertainties, point estimates remain robust, ensuring that for applications such as luminosity functions, the entire PDZs remain valuable.

In \cref{fig:5.1}, we present the cumulative distribution of CTP \IE band magnitudes along with the corresponding fractions of sources for which spec- and/or and photo-$z$s from {\tt{PICZL}} are available. The figure demonstrates that redshift completeness is very high up to around $\sim \IE \leq 22$, beyond which the availability of reliable photo-$z$ estimates begins to decline. In addition to spectroscopic selection effects, this is due to fainter sources which, while clearly detected in \Euclid, often lack sufficient optical coverage or exhibit photometric uncertainties too large to yield robust photo-$z$ values. Future releases of {\tt PICZL}, adapted to directly process \Euclid imaging and incorporating the deeper and more complete photometry expected from the Legacy Survey of Space and Time \citep[LSST;][]{Ivezi2019}, will markedly enhance photo-$z$ performance for fainter AGN. 

We compared our photo-$z$ estimates with those from \cite{Duncan_2022}, who combined ML and SED fitting to derive redshifts for all extragalactic sources in the 8th data release of the Legacy Survey (LS8). In EDFN, where LS10 photometry corresponds roughly to that of LS8, {\tt PICZL} achieves a slight improvement in $\sigma_{\rm NMAD}$ (from 0.10 to 0.08) and a notable reduction in $\eta$ (from 25\% to 14\%), without applying magnitude cuts. In the two additional fields covered by LS10-South \citep[][]{zenteno2025}, where {\tt PICZL} also benefits from deeper WISE and added i-band photometry, the improvements are even more pronounced, with $\sigma_{\rm NMAD}$ improving from 0.09 to 0.05, and $\eta$ dropping from 23\% to 12\%. Likewise, while the photo-$z$s from \citet{Zhou2021,Zhou2023}, computed with LS10 photometry, perform extremely well for inactive galaxies, a trend similar to \citet{Duncan_2022} is observed for AGN, where the accuracy is known to decline\footnote{with $\eta$ reaching up to $\sim$33\% in some subsets (priv. comm.)}. For high-redshift ($z$ > 3) X-ray selected AGN, we compared our estimates to those of \citet{Pouliasis_2025}, computed using LS10 supplemented with near-infrared photometry from the VISTA Hemisphere Survey \citep[VHS,][]{mcmahon_2013}. Of the 45 sources in common, only two have spec-$z$ for which both photo-$z$ agree, preventing any meaningful additional comparison. Instead, we can assess our photo-$z$ quality for X-ray selected AGN at the faint end by benchmarking them against studies that employ significantly deeper and broader photometric datasets.

\begin{figure}[b!]
\centering
\includegraphics[angle=0,width=1.0\hsize]{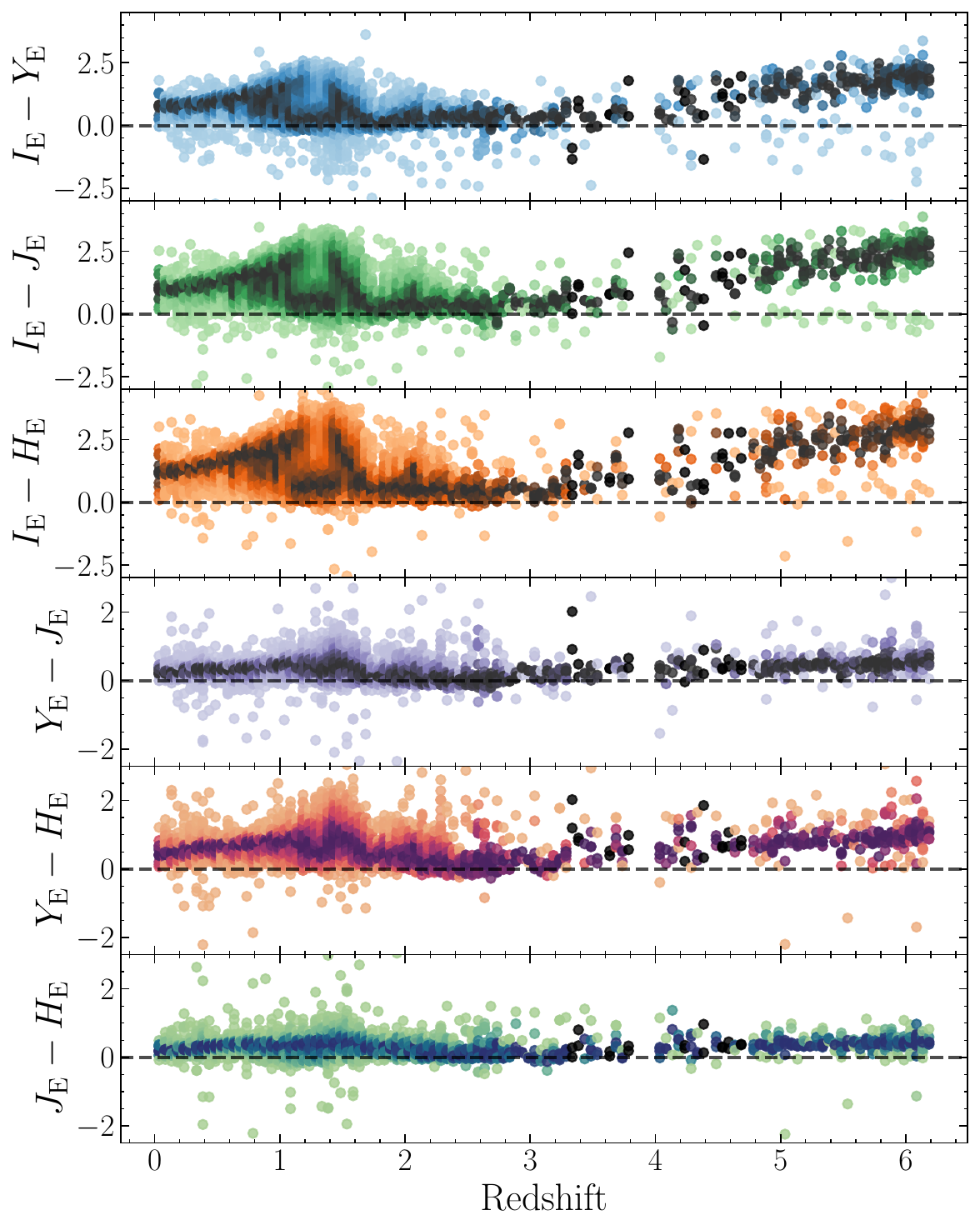}
\caption{Colour-redshift relation for all \Euclid band combinations. Each panel is divided into redshift windows, with the density of points within each bin colour-coded along the dimension of the y-axis, illustrating spectral features in the SED.}
\label{fig:11}
\end{figure}

\begin{figure*}[t!]
\centering
\includegraphics[angle=0,width=1.0\hsize]{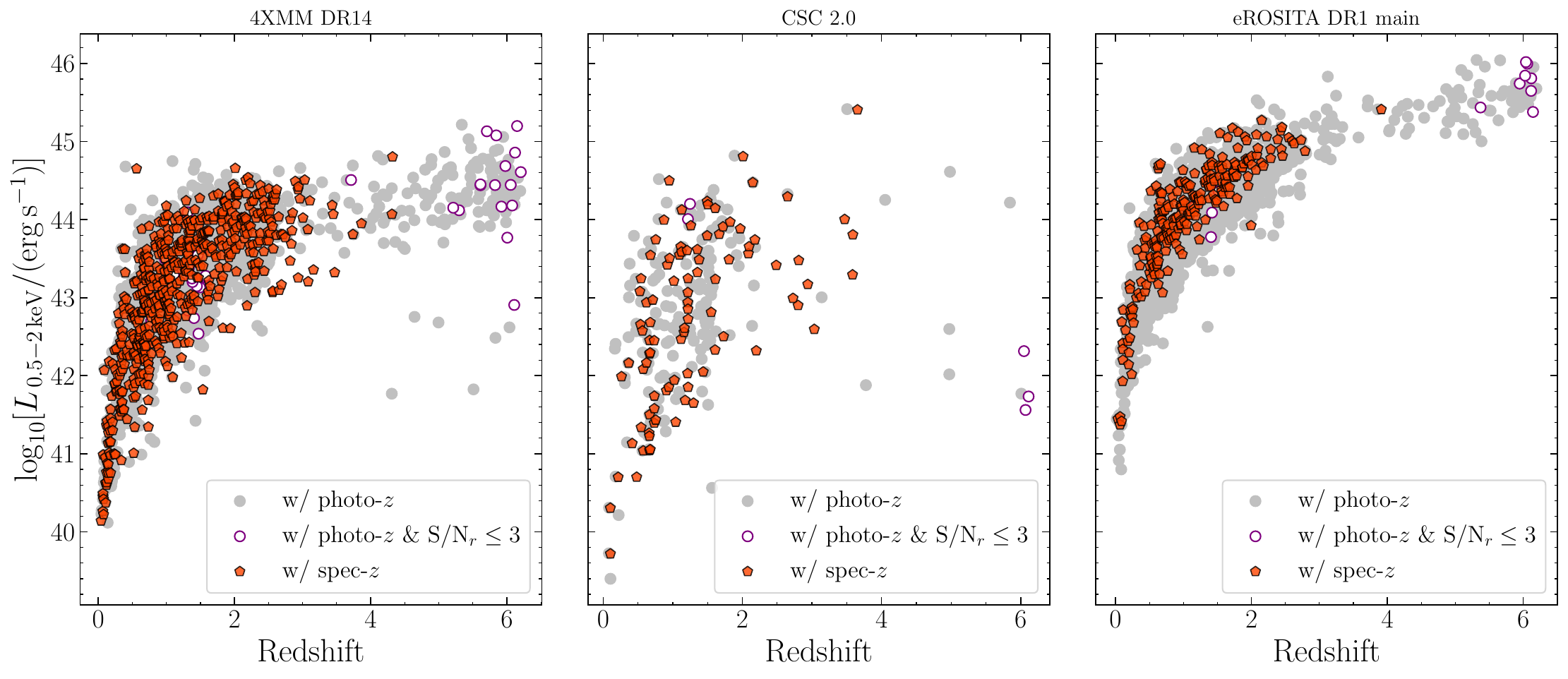}
\caption{X-ray luminosity as a function of redshift for the different surveys: 4XMM DR14 (left), CSC\,2.0 (centre), and eROSITA DR1 main, (right). Grey circles represent sources with photometric redshifts, while orange pentagons indicate sources with spectroscopic redshifts. Less reliable photometric redshifts of S/N$_{r}$\,$\leq 3$ in LS10 are highlighted by a purple outline.}
\label{fig:9}
\end{figure*}

\section{Properties of the CTP sample}
\label{sec8}
Having successfully selected, classified, and vetted our CTPs, we now turn to an analysis of their properties. From this point forward, we refer to CTPs as those sources meeting the criteria of ${P}_{\textrm{Gal}} \leq 0.5$ and \texttt{p\_any}$ > 0.05$. This refined selection yields a total of 9294 sources from the initial 11\,286 entries having \texttt{match\_flag}=1.

Building on this approach, we note that a similar comparison has already been presented in \citet{Roster24}, where {\tt PICZL} photo-$z$s were evaluated against the XMM-SERVS catalogues from \citet{Chen2018} and \citet{Ni2021}. Here, we extend this strategy by comparing our photo-$z$ results with the 7 Ms Source Catalog from the $Chandra$ Deep Field-South  \citep[CDFS,][]{Luo_2017}. This catalogue is particularly well-suited for faint-end validation, as it includes photo-$z$s derived using high-quality multi-band data, largely informed by \citet{Hsu2014}. Utilising up to 38 intermediate and broad bands spanning the ultraviolet \citep[\textit{GALEX,}][]{Martin_2005} to the mid-infrared \citep[\textit{Spitzer}/IRAC,][]{Cardamone_2010}, their work has resulted in one of the most comprehensive SED-fitting datasets for faint X-ray-selected sources to date. To ensure a proper one-to-one comparison between sources, we match our CTP catalogue to that of \citet{Luo_2017} using the respective CTP coordinates. We recover the same CTPs across the full sample, with 97.5\% of sources showing offsets below 0\farcs8, with the largest observed separation being 1\farcs8. Indicating robust matches, this consistency highlights the overall quality of our CTPs that can be attributed to the effectiveness of the \Euclid-based prior used with {\tt NWAY}, providing a solid foundation for future applications. For the 400 sources with spec-$z$s matched between this work and \citet{Luo_2017}, we compare the two datasets by evaluating $\eta$ and $\sigma_{\rm NMAD}$ as a function of $r$-band magnitude from the Wide Field Imager \citep[WFI;][]{Giavalisco_2004}. As depicted in \cref{fig:A5}, {\tt PICZL} achieves a lower fraction of outliers than \citet{Luo_2017} for $r \lesssim$ 22.3, though at larger $\sigma_{\rm NMAD}$ due to using only four shallower optical and WISE mid-IR bands. However, even up to $r \lesssim 24 \,(25)$ {\tt PICZL} extrapolation results in outlier fractions of $\eta \approx 7\%\,(13\%)$, which is lower compared to, for example, eROSITA / eFEDS \citep{Salvato_22}, where the fraction is $\eta = 14.1$\%, despite using many more and deeper bands. Given the overall quality of \cite{Luo_2017}, we decided to extend redshift completeness in this work by incorporating 378 unique photo-$z$s from the 7 Ms Source Catalog.

\Cref{fig:6} illustrates the magnitude distribution of our CTPs, grouped by the match between EDFs and the corresponding X-ray survey (4XMM DR14, CSC\,2.0, and eROSITA DR1, respectively). The 4XMM and CSC2 distributions show their ability to detect fainter X-ray sources, which correlates with the identification of fainter optical CTPs. In contrast, the eROSITA DR1 distributions exhibit brighter median magnitudes, consistent with the shallowness of the survey (see \cref{fig:flux_poserr}).
Additionally, the figure suggests a possible bimodal distribution at the faint end in the optical band for both 4XMM and CSC\,2.0, at least in EDF-N and EDF-S. The majority of sources in the second, faint peak of the distribution show X-ray fluxes $F_{(0.5-2\,\textrm{keV})} \leq 10^{-14.5} \, \textrm{erg} \, \textrm{cm} ^{-2} \, \textrm{s}^{-1}$, with no obvious drop off in \texttt{p\_any}, and the distribution will be studied further in other papers.

\begin{figure}[t!]
\centering
\includegraphics[angle=0,width=1.0\hsize]{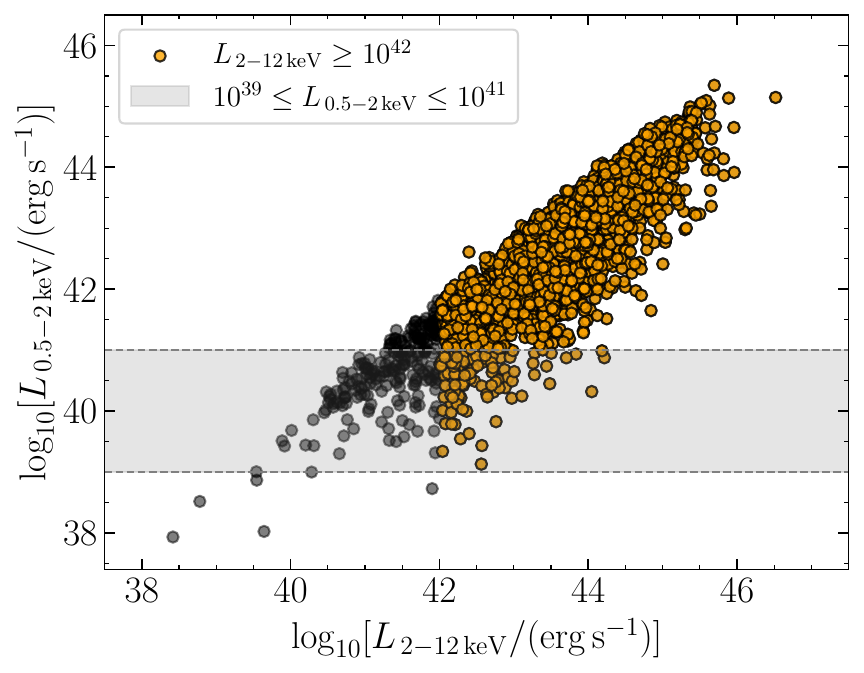}
\caption{Scatter plot of soft X-ray luminosities vs hard X-ray luminosities for sources in the 4XMM-DR14 sample. Sources classified as AGN based on their hard-band luminosities are highlighted with orange markers. Using this classification, we establish a corresponding threshold in soft-band luminosity.}
\label{fig:15}
\end{figure}

\subsection{Colour-redshift relation}
Redshifts, spectroscopic and photometric, are available for 81\% of the CTPs, allowing us to analyse their \Euclid colours as a function of redshift, as shown in \cref{fig:11}. The colour-coded kernel density is plotted along the {\it y}-axis dimension within each bin, providing a visual trace of how the physical colours of the CTP sample vary as a function of redshift.

A notable feature in the data is the detection of the Balmer break at $z \simeq 1.3$, particularly prominent in colours including the \IE filter. Unsurprisingly, NIR-only colour combinations display relatively few discernible features \citep[compare e.g., figure 7 in][]{Temple_2021}. As iterated before, \Euclid filters alone lack the fidelity to catch variations in the SED of sources, a caveat that we will be able to overcome in combination with upcoming surveys, especially LSST. 

\begin{figure*}[htbp!]
\centering
\includegraphics[angle=0,width=1.0\hsize]{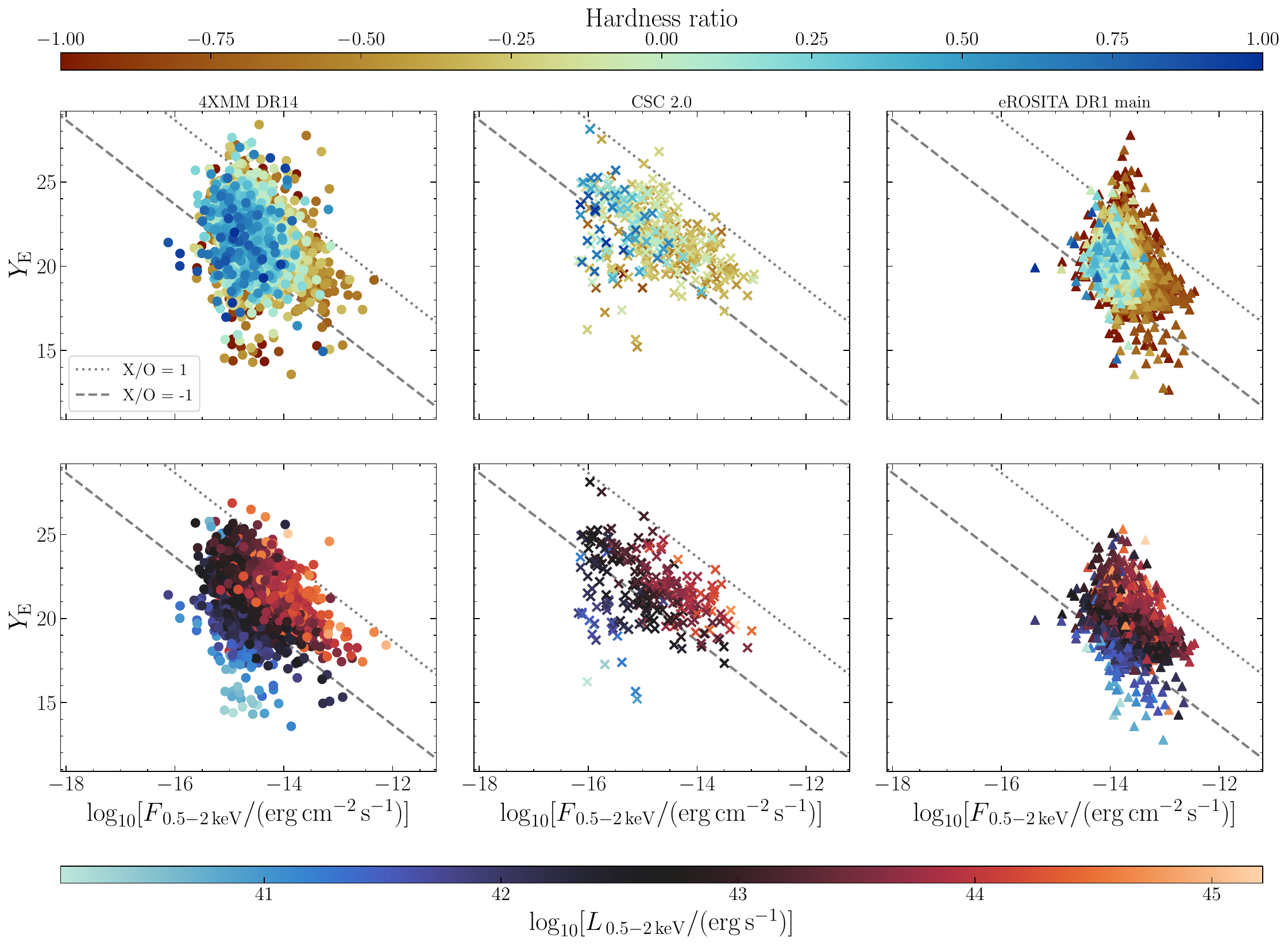}
\caption{Distribution of the \YE magnitude as a function of soft X-ray flux for the three catalogues studied in this work: 4XMM DR14 (left), CSC\,2.0 (centre), and eROSITA DR1 main (right). In each subplot, the dashed lines indicate the |X/O|\,=\,1 trend, with sources lying between the lines (|X/O|\,<\,1) usually being identified as AGN. The scatter points are colour-coded by their respective hardness ratio (top) and X-ray luminosity (bottom), when available. The markers are plotted in order of increasing hardness ratio.
}
\label{fig:8}
\end{figure*}

\subsection{X-ray luminosities}
For the rest of the analysis, we restrict the CTP sample to those  7345 extragalactic sources with a soft X-ray flux S/N$_{\rm{x}}$\,$\geq$\,2 in order to reduce potential spurious X-ray detections. In \cref{fig:9}, we observe a clear distinction between pointed observations taken with \XMMN and \textit{Chandra} at various depths versus the homogenous all-sky coverage provided by eROSITA. The complete coverage of eROSITA results in a tighter correlation between luminosity and redshift, despite the presence of a few points away from the general trend, with S/N$_{\rm{x}} < 3$. Another reason for their offset from the general trend could be erroneous photo-$z$. On the other hand, \XMMN and \textit{Chandra} catalogues include a mixture of pointed observations, resulting in a less smooth distribution between redshift and luminosity.

In all three panels, we indicate with open circles sources with low S/N$_{r}$, presumably representing examples of increased redshift uncertainties. Notably, most of the results shown focus on objects with redshifts of $z<4$. Sources with higher redshifts, while present, should be approached with caution due to potential inaccuracies. However, the future availability of LSST photometry will enable the exploration of this parameter regime by probing fainter sources with smaller errors, thus improving the reliability of redshift estimates for these populations.

In studies such as that of  \citet{Georgakakis_2011}, AGN are typically defined as sources with X-ray luminosities exceeding $L_{(2-10\,\textrm{keV})} \geq 10^{42}$$ \, \textrm{erg} \, \textrm{s}^{-1}$, since normal galaxies may be an important source of contamination below this limit \citep{Lehmer_2012,Aird_2015}. However, it is important to note that this threshold was established based on pencil-beam surveys, suggesting that it should be revisited now that wide-area surveys and deep optical data are accessible. X-ray luminosities of AGN in the soft X-ray band (0.5--2\,keV) can vary significantly, depending on the AGN type and redshift. In literature, values can range from $L_{(0.5-2\,\textrm{keV})} \geq 10^{41} - 10^{46} \, \textrm{erg} \, \textrm{s}^{-1}$ or even less when considering low-luminosity AGN \citep{Hasinger_2005,Ebrero_2009,Civano_2012,Kawamuro_2013}. According to \cite{Georgakakis_2007}, for example, X-ray detected starburst galaxy candidates, though more likely in case of \textit{Chandra} as opposed to eROSITA \citep{kyritsis2024}, can represent about 20\% of the X-ray source population in the luminosity interval $L_{(2 - 10\,\textrm{keV})} \geq 10^{40}-10^{42} \, \textrm{erg} \, \textrm{s}^{-1}$. This fraction is also likely to be an upper limit because some of the galaxy candidates may turn out to be low-luminosity AGN, such that distinguishing between these sources can be challenging. Consequently, with sources from \XMMN, where both soft and hard (2--12\,keV)\footnote{We assume that at zero order, the luminosities between (2--10) and (2--12) keV are approximately the same.} band fluxes are available, we show that, to retain all sources in this parent catalogue that are classified as AGN using the hard band limit, a soft-band luminosity threshold of $L_{(0.5-2\,\textrm{keV})} \geq 10^{39} \, \textrm{erg} \, \textrm{s}^{-1}$ is required (refer to \cref{fig:15}). Given the results of both all-sky and pencil beam surveys, at soft luminosities in the range $10^{39}-10^{41}\, \textrm{erg} \, \textrm{s}^{-1}$, AGN still appear to represent the dominant population. Nonetheless, future studies are required to determine the nature of these low-luminosity AGN at soft energies. Applying this more inclusive threshold defined in \cref{fig:15} to our extragalactic CTP sample, all of the 6653\,(91\%) sources that we were able to compute X-ray luminosities for, meet the AGN definition criterion, 49 of which have $L_{(0.5-2\,\textrm{keV})} < 10^{41} \, \textrm{erg} \, \textrm{s}^{-1}$.

\subsection{AGN type}
In \cref{fig:8}, we plot the respective \YE-band magnitudes as a function of the soft X-ray fluxes for all CTPs classified as extragalactic. We add the |X/O|\,=\,1 lines \citep{Maccacaro1988}, which confine the typical AGN locus \citep[see also e.g.,][]{Salvato2011,Civano_2012}, defined as

\begin{equation}
    \textrm{X/O} =
 \logten(F_{\textrm{x}}/F_{\textrm{opt}}) =  \logten\paren{\frac{F_x}{{\rm erg\,cm^{-2}\,s^{-1}}}} \, \frac{m_{\textrm{opt}}}{2.5} +C\,,
\end{equation}

\noindent where $F_{\textrm{x}}$ is the X-ray flux in a given energy range, $m_{\textrm{opt}}$ and $C$ correspond to the magnitude and a constant which both depend on the specific filter used in the optical observations. For $\YE$ we find $C = 5.53$. To further assess the AGN type of our CTPs, we colour-code the plots in the bottom row of \cref{fig:8} by the X-ray luminosity and the top row by hardness ratio \citep{Hasinger_2005,Cappelluti_2007,Salvato_2009}, defined as

\begin{equation}
    \textrm{HR} = \frac{\rm{Count\,rate}_{\rm{hard}}-\rm{Count\,rate}_{\textrm{soft}}}{\rm{Count\,rate}_{\textrm{hard}}+\rm{Count\,rate}_{\textrm{soft}}}\,, 
\end{equation}
\noindent where soft corresponds to the 0.5--2.0\,keV detection band, and hard refers to 2.0--12.0\,keV PN thin for 4XMM DR14, 2.0--7.0\,keV ACIS-I for \textit{Chandra}, and 2.0--5.0\,keV for eROSITA DR1 respectively, to account for each instrument's flux sensitivity. This ensures as much consistency as possible across the different surveys, to make the different instruments somewhat comparable.  A negative HR value indicates soft, unobscured sources (type I), and a positive HR value suggests harder, likely obscured sources (type II). We examine the HR distributions for all three telescopes to assess differences in their observed AGN populations. However, a non-negligible fraction of sources lack hard-band detections: approximately 47.6\% for eROSITA, 14.5\% for \XMMN, and 7.3\% for $Chandra$. These non-detections, especially in the case of eROSITA, naturally shift the medians of the true HR distributions and introduce selection biases. As shown in \cref{fig:A6}, we find that the eROSITA HR distribution median is $\tilde{\mathrm{HR}} \approx -0.88$, while that of $Chandra$ and (\XMMN) show a similar tendency towards harder values with medians of $\tilde{\mathrm{HR}} \approx -0.33\,(-0.39)$. These HR medians can then be used to compute `effective' photon indices $\Gamma_{\rm{eff}}$, which we find to be $\Gamma_{\rm{eff}} = 2.18$ for eROSITA, $\Gamma_{\rm{eff}} = 1.65$ for $Chandra$\footnote{Adopting CHANDRA-Cycle 11 as a representative benchmark, as it reflects an intermediate instrumental state within the $\sim$ 20-year span of observations included in CSC 2.0, accounting for the effects of detector degradation over time.} and $\Gamma = 1.51$ for \XMMN, assuming negligible column density. As expected, the greater sensitivity of \XMMN and $Chandra$ at higher energies allows them to detect harder and possibly mildly obscured AGNs that are missed by eROSITA. Conversely, eROSITA tends to detect brighter AGN which exhibit softer X-ray spectra. As the markers in \cref{fig:8} are plotted in order of increasing HR, many softer sources in the background are being covered up and predominantly fall within the X/O boundaries, classifying them as AGN, with differently weighted populations visible for every X-ray survey. As observed for \XMMN and $Chandra$, this includes a population of sources with $F_{(0.5-2\,\textrm{keV})} < 10^{-14} \, \textrm{erg} \, \textrm{cm}^{-2}\,\textrm{s}^{-1}$ lying below the X/O boundaries, potentially corresponding to starburst galaxies, low-luminosity AGN, or obscured AGN. However, the majority of these sources show S/N$_{\rm{x}}$\,$\leq$\,3 in the soft X-ray band, suggesting that a few of these may be spurious detections. It is also plausible that these sources are intrinsically variable, suggesting that they were observed in different states of appearance. Overall, our distribution follows the findings of \citet{Merloni_2014}, which show that only approximately 80\% of AGN selected by their SED, spectral features and HR, match their expected type I/II class.

\section{Release of the catalogue}
\label{sec9}
The catalogue containing the CTP properties of the EDF point-like X-ray sources is only available in electronic form at the CDS via anonymous ftp to \url{cdsarc.u-strasbg.fr} (130.79.128.5) or via \url{http://cdsweb.u-strasbg.fr/cgi-bin/qcat?J/A+A/}. A detailed description of the columns and their content is provided in \cref{apdx:C}. The catalogue includes unique identifiers for all relevant surveys as well as the basic X-ray and \Euclid properties (columns 1--14). For the full list of available columns, please refer to the original catalogues introduced in \cref{sec3} and, for example,  \citet{Q1-TP002,Q1-TP003,Q1-TP004}. Columns 15--24 present the results of the CTP association, followed by key parameters derived from {\tt NWAY}. Photometric data from LS10 are reported in columns 25--27. Spectroscopic information, when available, is included in columns 28--32. Lastly, the photometric redshift parameters from the {\tt{PICZL}} algorithm, as well as quantities computed with such, are listed in columns 33--39.

\section{Conclusions and outlook}
\label{sec10}
In this paper, we have presented the procedure adopted for the identification, classification, and study of the \Euclid CTPs of the point-like sources selected from \XMMN, \textit{Chandra}, and eROSITA X-ray, detected in Q1. We summarise the most important results in the following.

\begin{itemize}
    \item Utilising {\tt NWAY} \citep{Salvato_2018} supplemented by a prior on  X-ray emission based on properties of a 4XMM DR11 and CSC\,2.0 training sample, we achieved highly accurate CTP identifications while relying on \Euclid photometry only (see \cref{sec4}). By randomising the X-ray position and repeating the same procedure for the association, we quantified for each X-ray source the probability of chance associations. Users will be able to create subsamples of CTPs of desired purity and completeness by applying more restrictive or inclusive \texttt{p\_any} cuts ( probability of an X-ray source having a CTP; see \cref{fig:2.6}). 

    \item Altogether, we have identified 12\,645 CTPs, with only about 10\% of the sources having more than a single possible CTP. These are mostly in the \XMMN and eROSITA surveys, due to their larger positional uncertainties compared to \textit{Chandra}. We attribute this performance to the availability of extensive X-ray detected samples with secure CTPs and deep, homogenised multiwavelength photometry spanning optical to MIR wavelengths enabling robust SED construction for training. Further expansion to broader multiwavelength coverage will improve the quality of CTP identification, particularly for \Euclid DR1.

    \item We trained an RF algorithm to classify objects as Galactic or extragalactic using solely \Euclid photometry in \cref{sec6}. We previously trained another RF on a sample of secure Galactic and extragalactic sources detected in LS10, using $g-r$ over $z-$W1 thresholds, as presented in \citet{Salvato_22}, effectively breaking the degeneracy in classification that arises when relying on \Euclid photometry alone \citep[see][]{Bisigello_2024}. We validated these classifications by comparing to \cite{Q1-SP027}. 

    \item We also added spectroscopic redshifts for 16\% of the CTP sample using publicly available data. For the 82\% of the sample detected in LS10, we computed photo-$z$ using {\tt{PICZL}} \citep{Roster24}, reaching up to $z \approx 6$. However at these high redshifts, the values are associated with large uncertainties. With all of these redshifts, we investigated the colour-redshift relation of our sample in addition to computing their X-ray luminosities (refer to \cref{sec7}). We find that all of our CTPs pass our soft X-ray luminosity threshold to be considered an AGN, with more of them appearing  as bright type I sources rather than type II.

    \item  As the primary outcome of this paper, we present the CTP catalogue, which includes a wealth of information such as IDs, basic X-ray properties, {\tt NWAY} outputs, classification probabilities, and redshift data (see \cref{sec9}). By providing probabilities for all applicable sample characteristics, we ensure users have the flexibility to make informed selections tailored to their specific scientific goals, without being constrained by predefined thresholds or assumptions.
\end{itemize}

\begin{figure}[htbp!]
\centering
\includegraphics[angle=0,width=1.0\hsize]{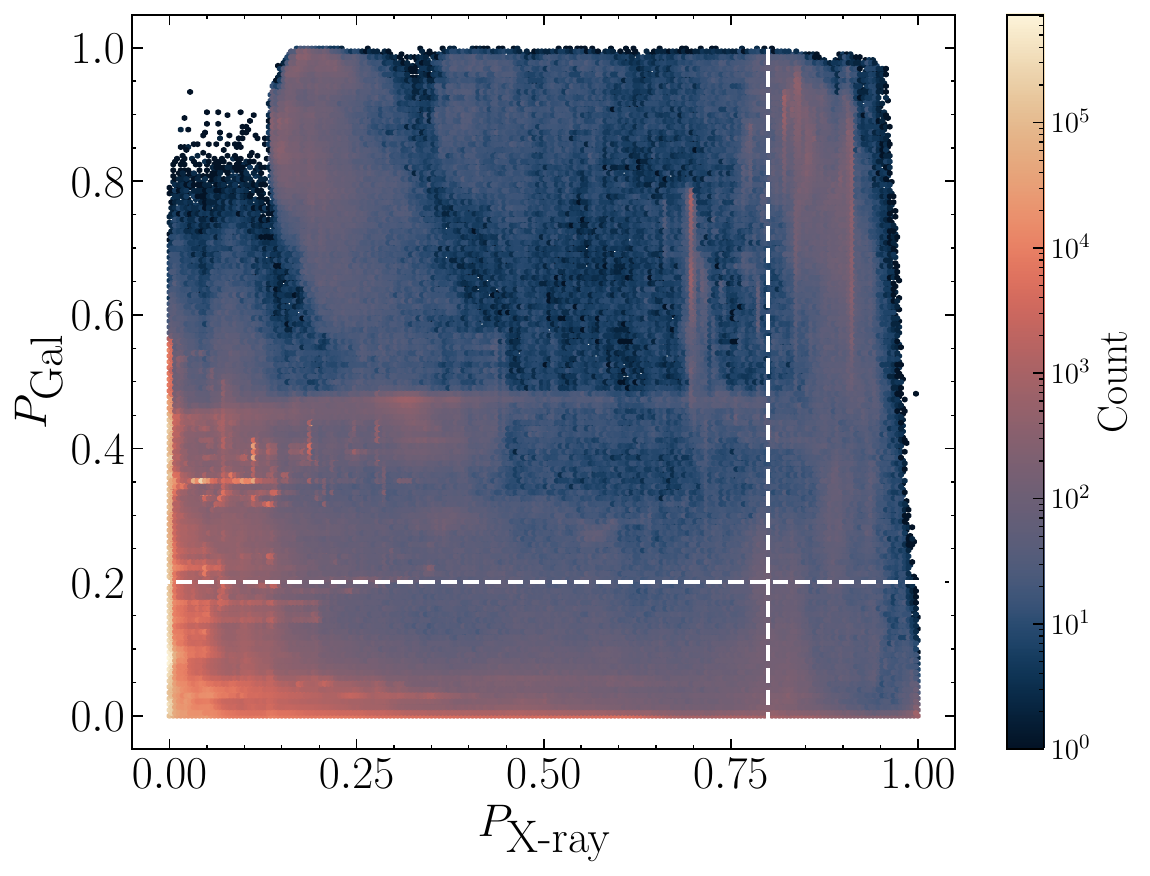}
\caption{Hexbin plot illustrating the relationship between the probability of being an X-ray emitter and the probability of being Galactic for all sources in Q1. The plot is colour-coded by the logarithmic kernel density. Orientation lines are included at an X-ray probability of 0.8 and a Galactic probability of 0.2.
}
\label{fig:13}
\end{figure}

The RF models used for the identification of the CTPs and their classification as Galactic or extragalactic, can be applied blindly to the entire Q1 catalogue. For each source, they provide the probability of being an X-ray emitter as well as an extragalactic source, despite not having been detected by any of the three X-ray surveys yet. Consequently, as illustrated in \cref{fig:13}, limiting the selection of Q1 sources to those with $P_{\textrm{X-ray}} > 0.8 $ and $P_{\textrm{Gal}} > 0.2 $ while excluding \Euclid sources classified as CTPs in this study yields a final sample of 135\,303 objects. Notably, 73 of these objects have been successfully cross-matched geometrically with the little red dots (LRDs) sample introduced by \cite{Q1-SP011}, where the matches predominantly occupy the bright end of the LRD magnitude distribution. A forthcoming paper will explore the multi-wavelength nature of the likely extragalactic high $P_{\textrm{X-ray}}$ sources that remain undetected in current X-ray surveys.

\begin{acknowledgements}
 
WR and MS acknowledge DLR support (Foerderkennzeichen 50002207). This work has benefited from the support of Royal Society Research Grant RGS{\textbackslash}R1\textbackslash231450. This research was supported by the International Space Science Institute (ISSI) in Bern, through ISSI International Team project N. 23-573 ``Active Galactic Nuclei in Next Generation Surveys''.
BL and FR acknowledge the support from the INAF Large Grant "AGN and Euclid: a close entanglement" Ob. Fu. 01.05.23.01.14. This research has made use of data obtained from the Chandra Source Catalog provided by the Chandra X-ray Center (CXC), and the 4XMM XMM-Newton serendipitous source catalogue compiled by the XMM-Newton Survey Science Centre consortium. This work is based on data from eROSITA, the soft X-ray instrument aboard SRG, a joint Russian-German science mission supported by the Russian Space Agency (Roskosmos), in the interests of the Russian Academy of Sciences represented by its Space Research Institute (IKI), and the Deutsches Zentrum für Luft- und Raumfahrt (DLR). The SRG spacecraft was built by Lavochkin Association (NPOL) and its subcontractors, and is operated by NPOL with support from the Max Planck Institute for Extraterrestrial Physics (MPE). The development and construction of the eROSITA X-ray instrument was led by MPE, with contributions from the Dr. Karl Remeis Observatory Bamberg \& ECAP (FAU Erlangen-Nuernberg), the University of Hamburg Observatory, the Leibniz Institute for Astrophysics Potsdam (AIP), and the Institute for Astronomy and Astrophysics of the University of Tübingen, with the support of DLR and the Max Planck Society. The Argelander Institute for Astronomy of the University of Bonn and the Ludwig Maximilians Universität Munich also participated in the science preparation for eROSITA. Based on data from UNIONS, a scientific collaboration using
three Hawaii-based telescopes: CFHT, Pan-STARRS, and Subaru
\url{www.skysurvey.cc}\, and data from the Dark Energy Camera (DECam) on the Blanco 4-m Telescope
at CTIO in Chile \url{https://www.darkenergysurvey.org}\,. This work uses results from the ESA mission {\it Gaia},
whose data are being processed by the Gaia Data Processing and
Analysis Consortium \url{https://www.cosmos.esa.int/gaia}\,. The authors thank Pietro Baldini for valuable additional feedback and discussion.

  \AckEC\,\AckQone\,\AckCosmoHub
\end{acknowledgements}

%
% Here comes the reference list, generated via bibtex from
% the bibfile AandA.bib
%
\bibliography{all_bib}

%
% Now you can add appendices.
\begin{appendix}
  %\onecolumn %If you don't want single column for the Appendix, please
             %comment this out
\section{RF parameters\label{apdx:A1}}

Understanding which features are most relevant to our model estimates is crucial for predictive accuracy. An ML model operating as a black box may produce predictions, but the lack of transparency raises questions about the reliability of the outcomes. This not only hinders immediate interpretability but also complicates post-analyses. The ML branch investigating the role each feature plays in the model's predictions is commonly referred to as `feature importance’. To this end, we compute SHapley Additive exPlanations \citep[SHAP;][]{lundberg2017unified, rozemberczki2022shapley, Chen_2022} values, based on a game-theoretic approach, by calculating the partial relevance for each feature. Based on a conditional expectation function, features are assigned importance values given their impact on test sample predictions. For a total achievable value $v$, SHAP values $\phi$ are computed via 

\begin{equation}
    \phi_{f}(v) = \frac{1}{p} \sum_{S} \frac{[v\,(S \cup \{f\}) - v\,(S)]}{\begin{pmatrix}
p-1\\
k\,(S)
\end{pmatrix}}\,.
\end{equation}

\noindent Given a given feature $f$, the summation is taken over all subsets 
$S$ of the feature set $F = {1,2,3,\dots,p}$, which one can construct after excluding $f$. The size of $S$ is given by $k(S)$ while $v\,(S)$ and $v\,(S\cup\{f\})$ refer to the value achieved by per subset before and after $f$ joins $S$, respectively. This ensures that the contribution of each feature to a prediction collectively adds up to the value of the prediction as
\begin{equation}
    \sum_{f=1}^{f=p} \phi_{f}(v) = v\,(F)\,.
\end{equation}

\noindent However, achieving optimal accuracy in large modern data sets often involves complex models such as ensembles or deep-learning frameworks. Since it would be unfeasible to compute contributions across all possibilities of the feature space, we approximate SHAP values through Deep SHAP \citep{rozemberczki2022shapley, Chen_2022}, which efficiently combines values computed for smaller network components into comprehensive values for the entire network, avoiding the need for heuristic choices in linearising components. The sign of SHAP values denotes whether a feature contributes to an increase or decrease in the predictive output, while the magnitude of the SHAP value quantifies the extent of the feature's influence on the prediction. Features with SHAP values close to zero for individual sources can be considered negligible or non-influential in this scenario. As a consequence, a feature can be seen as most influential if it exerts the most influence on the output across all predictions \citep{Dey2022}.

\begin{figure}[htbp!]
\centering
\includegraphics[angle=0,width=1\hsize]{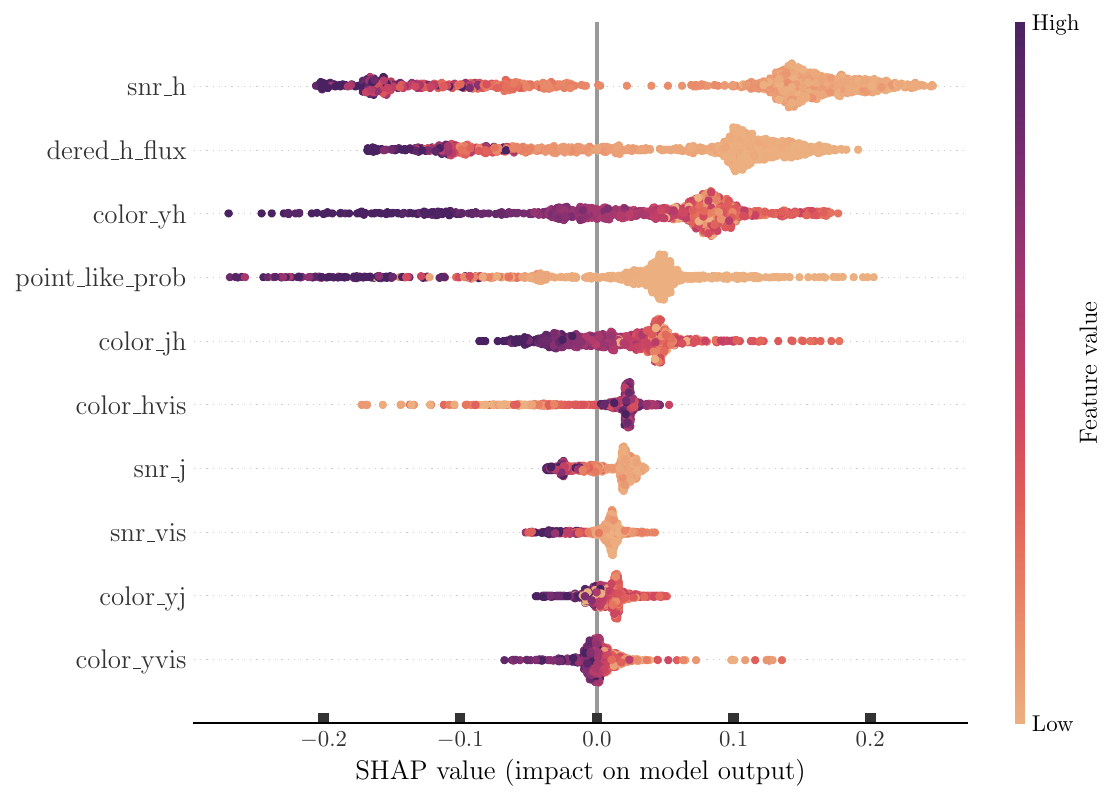}
\caption{`Beeswarm’ plot showing the SHAP values for the features used in the RF model introduced in \cref{Q1_prior}. Each point represents a SHAP value for a specific feature and data instance, illustrating the impact of that feature on the model's predictions. Features are sorted by their mean absolute SHAP values, with the most influential features appearing at the top. The colour gradient indicates the relative feature value (i.e. low to high), providing insight into how feature magnitude correlates with its effect on predictions.}
\label{fig:A2}
\end{figure}

\begin{figure}[htbp!]
\centering
\includegraphics[angle=0,width=1\hsize]{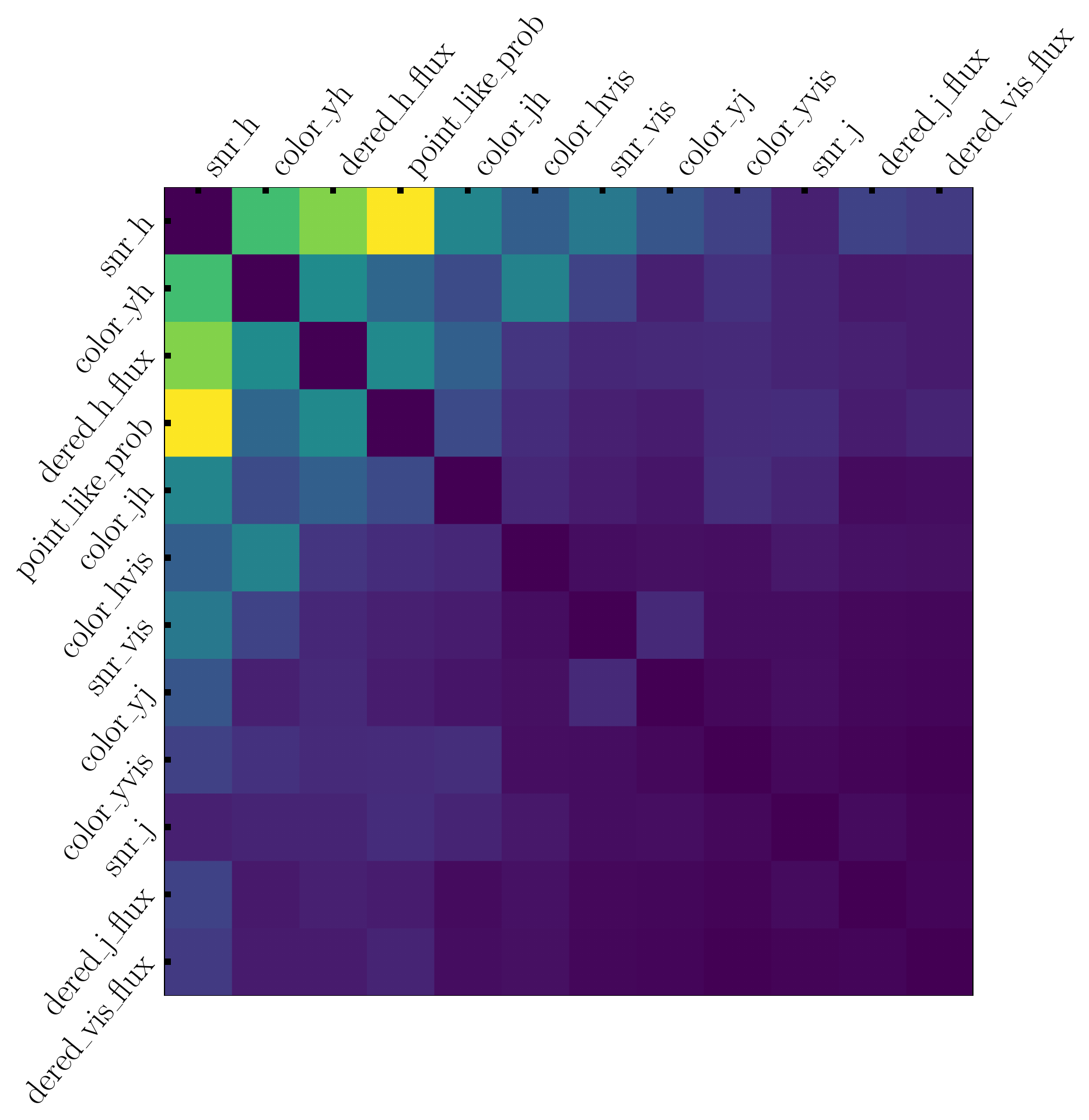}
\caption{Correlation matrix displaying the pairwise Pearson correlation coefficients between the features used in the RF model introduced in \cref{Q1_prior}. The colour intensity indicates the strength and direction of the correlations, with positive correlations shown in warm colours (yellow) and negative correlations in cool colours (blue). This visualisation helps identify interdependencies among features, which can inform feature selection and the interpretation of model performance.}
\label{fig:A3}
\end{figure}

\newpage

\begin{figure}[htbp!]
\centering
\includegraphics[angle=0,width=1\hsize]{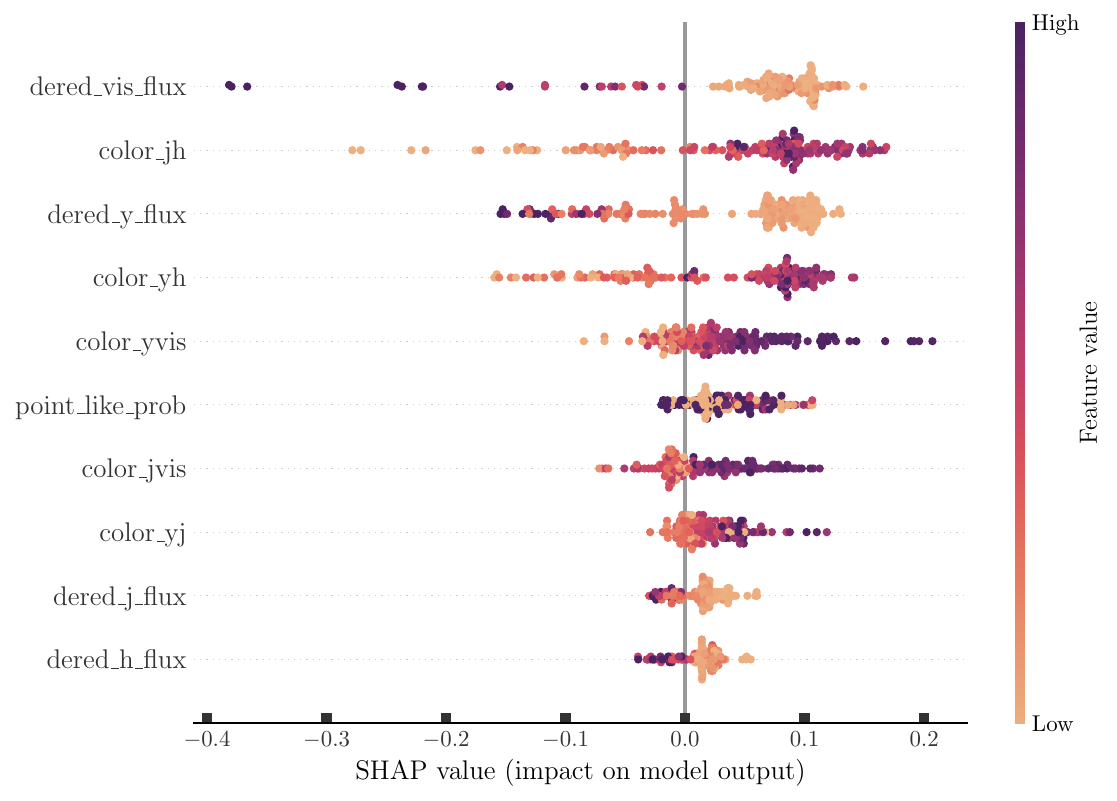}
\caption{`Beeswarm’ plot showing the SHAP values for the features used in the RF model introduced in \cref{Q1_prior_exgal}. Each point represents a SHAP value for a specific feature and data instance, illustrating the impact of that feature on the model's predictions. Features are sorted by their mean absolute SHAP values, with the most influential features appearing at the top. The colour gradient indicates the relative feature value (i.e. low to high), providing insight into how feature magnitude correlates with its effect on predictions.}
\label{fig:A4}
\end{figure}

\section{Photometric redshift comparison\label{apdx:A2}}

\begin{figure}[H]
\centering
\includegraphics[angle=0,width=1\hsize]{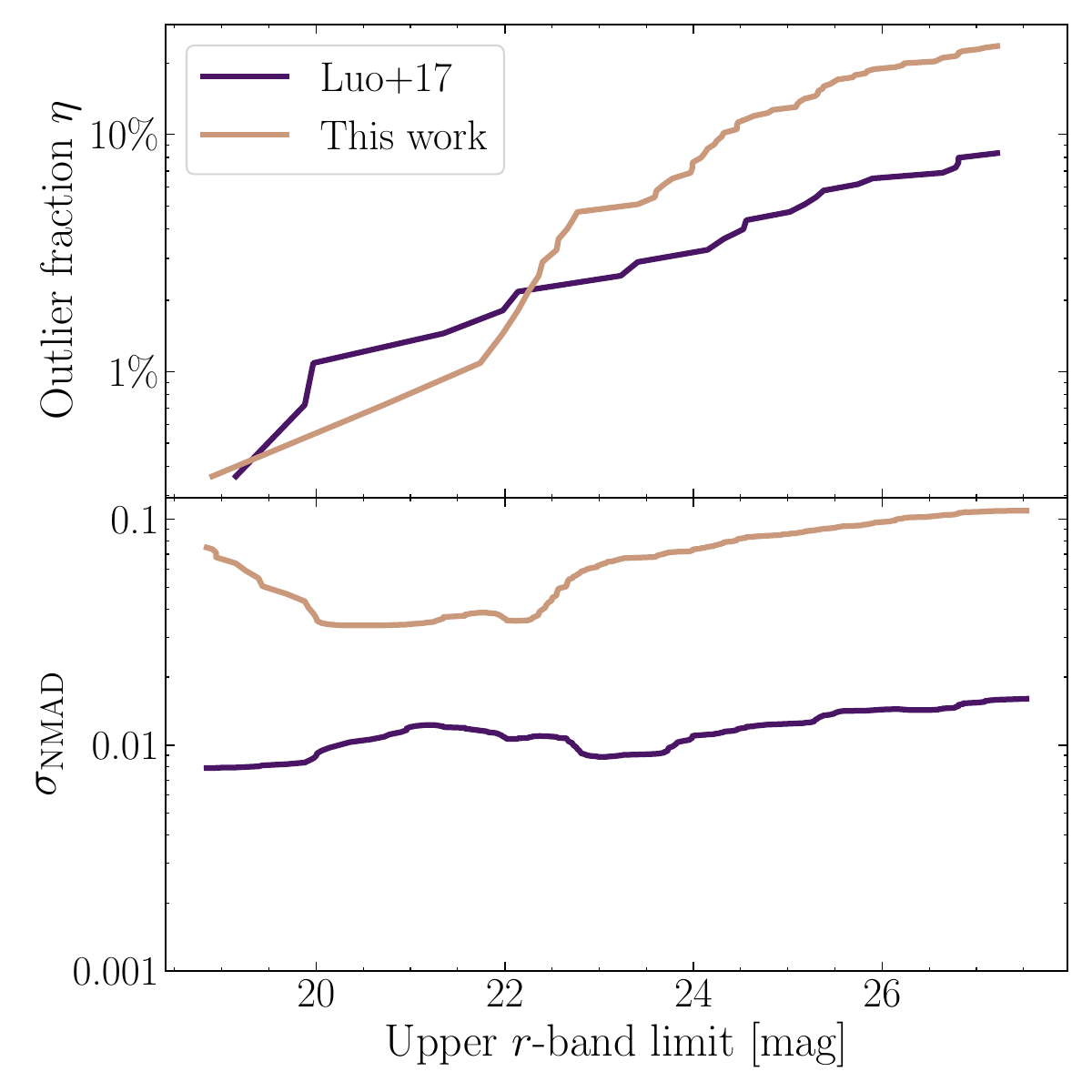}
\caption{Outlier fraction (top panel) and accuracy (bottom panel) as a function of limiting $r$-band magnitude for {\tt{PICZL}} and \citet{Luo_2017}. While the accuracy of \citet{Luo_2017} is superior to that reached by {\tt{PICZL}}, the fraction of outliers for bright sources ($r \lesssim 22.3$) is lower for {\tt{PICZL}}.}
\label{fig:A5}
\end{figure}

\section{HR to effective photon index \label{apdx:A3}}

\begin{figure}[H]
\centering
\includegraphics[angle=0,width=1\hsize]{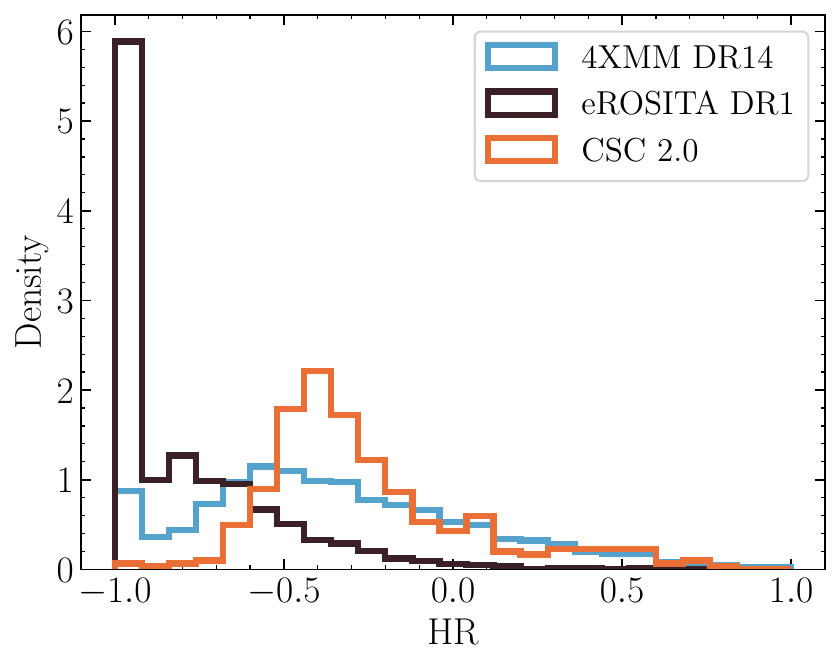}
\caption{Distribution of X-ray hardness ratio (HR) for eROSITA DR1$_{0.5-2\,\rm{keV}}^{2-5\,\rm{keV}}$, 4XMM DR14$_{0.5-2\,\rm{keV}}^{2-12\,\rm{keV}}$ and CSC2$_{0.5-2\,\rm{keV}}^{2-7\,\rm{keV}}$ sources. The distributions reflect differences in survey depth, energy band definitions, and source selection. While eROSITA shows a distribution skewed towards softer sources \mbox{($ \rm{HR} \rightarrow -1$)}, both 4XMM and CSC2 display peaks at harder HR values, consistent with their greater sensitivity to moderately obscured or intrinsically harder AGN.}
\label{fig:A6}
\end{figure}

\section{Column description of released CTP catalogue\label{apdx:C}}

\begin{enumerate}
    \item \textbf{Xray\_EUCLID\_ID}: Concatenation of X-ray\_ID and EUCLID\_ID (unique)
    \item \textbf{Xray\_ID}: X-ray source identifier (unique)
    \item \textbf{Xray\_RA}: J2000 Right Ascension of the X-ray source in the parent catalogue (deg)
    \item \textbf{Xray\_DEC}: J2000 Declination of the X-ray source in the parent catalogue (deg)
    \item \textbf{Xray\_CAT}: X-ray source catalogue
    \item \textbf{Xray\_pos\_err}: Positional uncertainty of X-ray source (arcsec)
    \item \textbf{Fx}: Mean 0.5--2\,keV X-ray flux from all detections of the X-ray source ($\textrm{erg} \, \textrm{cm}^{-2} \, \textrm{s}^{-1}$)
    \item \textbf{EFx}: Error on the mean 0.5--2\,keV X-ray flux from all detections of the X-ray source ($\textrm{erg} \, \textrm{cm}^{-2} \, \textrm{s}^{-1}$)
    \item \textbf{S\_Nx}: Signal-to-noise ratio for Fx/EFx
    \item \textbf{HR}: Count rate derived hardness ratio considering soft $0.5 - 2.0\,\rm{keV}$ and hard (4XMM DR14: $2.0 - 12.0\,\rm{keV}$, CSC 2.0: $2.0 - 7.0\,\rm{keV}$, and eROSITA: $2.0 - 5.0\,\rm{keV}$) bands. 
    \item \textbf{EDF}: \Euclid deep field
    \item \textbf{EUCLID\_ID}: EUCLID source identifier (unique)
    \item \textbf{EUCLID\_RA}: J2000 Right Ascension of the counterpart in Q1 catalogue (deg)
    \item \textbf{EUCLID\_DEC}: J2000 Declination of the counterpart in Euclid catalogue (deg)
    \item \textbf{EUCLID\_pos\_err}: Positional uncertainty of \Euclid source (arcsec)
    \item \textbf{EUCLID\_P\_Xray}: Probability to be an X-ray emitter based on the RF
    \item \textbf{EUCLID\_P\_Gal}: Probability of being a Galactic rather than extragalactic source based on the RF
    \item \textbf{Separation\_EUCLID\_Xray}: Distance between \Euclid and X-ray source (arcsec)
    \item \textbf{dist\_bayesfactor}: Logarithm of ratio between prior and posterior, from separation, positional error, and number density \citep[see Appendix in][]{Salvato_2018}
    \item \textbf{dist\_post}: Distance posterior probability comparing this association versus no association \citep[see Appendix in][]{Salvato_2018}
    \item \textbf{bias\_EUCLID\_Xray\_proba}: Probability weighting introduced by the \Euclid-based prior (1 indicates no change)
    \item \textbf{p\_single}: Same as dist\_post, but weighted by the prior \citep[see Appendix in][]{Salvato_2018}
    \item \textbf{p\_any}: Probability that there is a counterpart in LS10 for each X-ray entry \citep[see Appendix in][]{Salvato_2018}
    \item \textbf{p\_i}: Relative probability of the X-ray/\Euclid match \citep[see Appendix in][]{Salvato_2018}
    \item \textbf{match\_flag}: 1 for the most probable match; 2 for almost as good solutions (p\_i/p\_ibest > 0.5)
    \item \textbf{LS10\_FULLID}: Unique identifier of the LS10 source matched to the \Euclid CTP (concatenation of RELEASE, BRICKID, and OBJID)
    \item \textbf{LS10\_RA}: J2000 Right Ascension of the LS10 source (deg)
    \item \textbf{LS10\_DEC}: J2000 Declination of the LS10 source (deg)
    \item \textbf{ra\_spec}: J2000 Right Ascension of the spec-$z$ match to the \Euclid CTP (deg)
    \item \textbf{dec\_spec}: J2000 Declination of the spec-$z$ match to the \Euclid CTP (deg)
    \item \textbf{z\_spec}: Spectroscopic redshift
    \item \textbf{z\_err}: Error on the spectroscopic redshift
    \item \textbf{Cat\_spec}: Parent catalogue of the spec-$z$
    \item \textbf{phz}: {\tt{PICZL}} photometric redshift \citep[see][]{Roster24}
    \item \textbf{phz\_1sl}: {\tt{PICZL}} lower 1 sigma error
    \item \textbf{phz\_1su}: {\tt{PICZL}} upper 1 sigma error
    \item \textbf{phz\_Luo17}: Photometric redshift \citep[see][]{Luo_2017}
    \item \textbf{z\_final}: Spec-$z$ when available; otherwise \cite{Luo_2017} and {\tt{PICZL}} photo-$z$
    \item \textbf{Lx}: Rest-frame X-ray luminosity, assuming FlatLambdaCDM from \texttt{astropy.cosmology}

\end{enumerate}

\end{appendix}

\label{LastPage}
\end{document}